\documentclass[traditabstract]{aa}

\usepackage{graphicx}
\usepackage{afterpage}
\usepackage{amssymb}
\usepackage{amsmath}
\usepackage{rotating}
\usepackage{supertabular}
\usepackage[bookmarks=true, bookmarksnumbered=true, breaklinks=true, colorlinks=true, linkcolor=BrickRed, citecolor=Blue, urlcolor=Blue, filecolor=Blue]{hyperref}
\usepackage[all]{hypcap}
\usepackage{txfonts}
\usepackage{fixltx2e}
\usepackage{multirow}
\usepackage{pifont}
\usepackage[usenames,dvipsnames]{color}
\usepackage[all]{hypcap}

\def\aj{AJ}%
\def\araa{ARA\&A}%
\def\apj{ApJ}%
\def\apjl{ApJ}%
\def\apjs{ApJS}%
\def\ao{Appl.\ Opt.}%
\def\aap{A\&A}%
\def\nat{Nature}%
\newcommand{\ph}{\phantom{$-$}}

\newcommand{\oneDAV}{$\langle\mathrm{3D}\rangle$}
\newcommand{\oneDAVmath}{\langle\mathrm{3D}\rangle}
\newcommand{\marcs}{\textsc{marcs}}
\def\urltilda{\kern -.15em\lower .7ex\hbox{\~{}}\kern .04em}

\newcommand{\altcui}{Cu\,\textsc{i}}
\newcommand{\altcuii}{Cu\,\textsc{ii}}
\newcommand{\altzni}{Zn\,\textsc{i}}
\newcommand{\altznii}{Zn\,\textsc{ii}}
\newcommand{\altgai}{Ga\,\textsc{i}}
\newcommand{\altgaii}{Ga\,\textsc{ii}}
\newcommand{\altgei}{Ge\,\textsc{i}}
\newcommand{\altgeii}{Ge\,\textsc{ii}}
\newcommand{\altasi}{As\,\textsc{i}}
\newcommand{\altrbi}{Rb\,\textsc{i}}
\newcommand{\altrbii}{Rb\,\textsc{ii}}
\newcommand{\altsri}{Sr\,\textsc{i}}
\newcommand{\altsrii}{Sr\,\textsc{ii}}
\newcommand{\altyi}{Y\,\textsc{i}}
\newcommand{\altyii}{Y\,\textsc{ii}}
\newcommand{\altzri}{Zr\,\textsc{i}}
\newcommand{\altzrii}{Zr\,\textsc{ii}}
\newcommand{\altnbi}{Nb\,\textsc{i}}
\newcommand{\altnbii}{Nb\,\textsc{ii}}
\newcommand{\altmoi}{Mo\,\textsc{i}}
\newcommand{\altmoii}{Mo\,\textsc{ii}}
\newcommand{\altrui}{Ru\,\textsc{i}}
\newcommand{\altruii}{Ru\,\textsc{ii}}
\newcommand{\altrhi}{Rh\,\textsc{i}}
\newcommand{\altrhii}{Rh\,\textsc{ii}}
\newcommand{\altpdi}{Pd\,\textsc{i}}
\newcommand{\altpdii}{Pd\,\textsc{ii}}
\newcommand{\altagi}{Ag\,\textsc{i}}
\newcommand{\altagii}{Ag\,\textsc{ii}}
\newcommand{\altsni}{Sn\,\textsc{i}}
\newcommand{\altsnii}{Sn\,\textsc{ii}}
\newcommand{\altbai}{Ba\,\textsc{i}}
\newcommand{\altbaii}{Ba\,\textsc{ii}}
\newcommand{\altosi}{Os\,\textsc{i}}
\newcommand{\altosii}{Os\,\textsc{ii}}
\newcommand{\altiri}{Ir\,\textsc{i}}
\newcommand{\altirii}{Ir\,\textsc{ii}}
\newcommand{\altaui}{Au\,\textsc{i}}
\newcommand{\altauii}{Au\,\textsc{ii}}
\newcommand{\altpbi}{Pb\,\textsc{i}}
\newcommand{\altpbii}{Pb\,\textsc{ii}}
\newcommand{\altthi}{Th\,\textsc{i}}
\newcommand{\altthii}{Th\,\textsc{ii}}
\newcommand{\alteui}{Eu\,\textsc{i}}
\newcommand{\alteuii}{Eu\,\textsc{ii}}
\newcommand{\altsmi}{Sm\,\textsc{i}}
\newcommand{\altsmii}{Sm\,\textsc{ii}}
\newcommand{\alttbi}{Tb\,\textsc{i}}
\newcommand{\alttbii}{Tb\,\textsc{ii}}
\newcommand{\altwi}{W\,\textsc{i}}
\newcommand{\altwii}{W\,\textsc{ii}}
\newcommand{\altcdi}{Cd\,\textsc{i}}

\newcommand{\althfi}{Hf\,\textsc{i}}
\newcommand{\althfii}{Hf\,\textsc{ii}}
\newcommand{\altini}{In\,\textsc{i}}
\newcommand{\altinii}{In\,\textsc{ii}}
\newcommand{\altsbi}{Sb\,\textsc{i}}

\newcommand{\altlai}{La\,\textsc{i}}
\newcommand{\altlaii}{La\,\textsc{ii}}
\newcommand{\altcei}{Ce\,\textsc{i}}
\newcommand{\altceii}{Ce\,\textsc{ii}}
\newcommand{\altpri}{Pr\,\textsc{i}}
\newcommand{\altprii}{Pr\,\textsc{ii}}
\newcommand{\altdyi}{Dy\,\textsc{i}}
\newcommand{\altdyii}{Dy\,\textsc{ii}}
\newcommand{\altndi}{Nd\,\textsc{i}}
\newcommand{\altndii}{Nd\,\textsc{ii}}

\newcommand{\altgdi}{Gd\,\textsc{i}}
\newcommand{\altgdii}{Gd\,\textsc{ii}}
\newcommand{\althoi}{Ho\,\textsc{i}}
\newcommand{\althoii}{Ho\,\textsc{ii}}
\newcommand{\alteri}{Er\,\textsc{i}}
\newcommand{\alterii}{Er\,\textsc{ii}}
\newcommand{\altlui}{Lu\,\textsc{i}}
\newcommand{\altluii}{Lu\,\textsc{ii}}
\newcommand{\altybi}{Yb\,\textsc{i}}
\newcommand{\altybii}{Yb\,\textsc{ii}}
\newcommand{\alttmi}{Tm\,\textsc{i}}
\newcommand{\alttmii}{Tm\,\textsc{ii}}
\newcommand{\alttli}{Tl\,\textsc{i}}
\newcommand{\alttlii}{Tl\,\textsc{ii}}

\newcommand{\cui}{\altcui\ }
\newcommand{\cuii}{\altcuii\ }
\newcommand{\zni}{\altzni\ }
\newcommand{\znii}{\altznii\ }
\newcommand{\gai}{\altgai\ }
\newcommand{\gaii}{\altgaii\ }
\newcommand{\gei}{\altgei\ }
\newcommand{\geii}{\altgeii\ }
\newcommand{\asi}{\altasi\ }
\newcommand{\rbi}{\altrbi\ }
\newcommand{\rbii}{\altrbii\ }
\newcommand{\sri}{\altsri\ }
\newcommand{\srii}{\altsrii\ }
\newcommand{\yi}{\altyi\ }
\newcommand{\yii}{\altyii\ }
\newcommand{\zri}{\altzri\ }
\newcommand{\zrii}{\altzrii\ }
\newcommand{\nbi}{\altnbi\ }
\newcommand{\nbii}{\altnbii\ }
\newcommand{\moi}{\altmoi\ }
\newcommand{\moii}{\altmoii\ }
\newcommand{\rui}{\altrui\ }
\newcommand{\ruii}{\altruii\ }
\newcommand{\rhi}{\altrhi\ }
\newcommand{\rhii}{\altrhii\ }
\newcommand{\pdi}{\altpdi\ }
\newcommand{\pdii}{\altpdii\ }
\newcommand{\agi}{\altagi\ }
\newcommand{\agii}{\altagii\ }
\newcommand{\sni}{\altsni\ }
\newcommand{\snii}{\altsnii\ }
\newcommand{\bai}{\altbai\ }
\newcommand{\baii}{\altbaii\ }
\newcommand{\osi}{\altosi\ }
\newcommand{\osii}{\altosii\ }
\newcommand{\iri}{\altiri\ }
\newcommand{\irii}{\altirii\ }
\newcommand{\aui}{\altaui\ }
\newcommand{\auii}{\altauii\ }
\newcommand{\pbi}{\altpbi\ }
\newcommand{\pbii}{\altpbii\ }
\newcommand{\thi}{\altthi\ }
\newcommand{\thii}{\altthii\ }
\newcommand{\eui}{\alteui\ }
\newcommand{\euii}{\alteuii\ }
\newcommand{\smi}{\altsmi\ }
\newcommand{\smii}{\altsmii\ }
\newcommand{\tbi}{\alttbi\ }
\newcommand{\tbii}{\alttbii\ }
\newcommand{\wi}{\altwi\ }
\newcommand{\wii}{\altwii\ }
\newcommand{\cdi}{\altcdi\ }

\newcommand{\hfi}{\althfi\ }
\newcommand{\hfii}{\althfii\ }
\newcommand{\ini}{\altini\ }
\newcommand{\inii}{\altinii\ }
\newcommand{\sbi}{\altsbi\ }

\newcommand{\lai}{\altlai\ }
\newcommand{\laii}{\altlaii\ }
\newcommand{\cei}{\altcei\ }
\newcommand{\ceii}{\altceii\ }
\newcommand{\pri}{\altpri\ }
\newcommand{\prii}{\altprii\ }
\newcommand{\dyi}{\altdyi\ }
\newcommand{\dyii}{\altdyii\ }
\newcommand{\ndi}{\altndi\ }
\newcommand{\ndii}{\altndii\ }
\newcommand{\gdi}{\altgdi\ }
\newcommand{\gdii}{\altgdii\ }
\newcommand{\hoi}{\althoi\ }
\newcommand{\hoii}{\althoii\ }
\newcommand{\eri}{\alteri\ }
\newcommand{\erii}{\alterii\ }
\newcommand{\lui}{\altlui\ }
\newcommand{\luii}{\altluii\ }
\newcommand{\tmi}{\alttmi\ }
\newcommand{\tmii}{\alttmii\ }
\newcommand{\ybi}{\altybi\ }
\newcommand{\ybii}{\altybii\ }
\newcommand{\tli}{\alttli\ }
\newcommand{\tlii}{\alttlii\ }

\begin{document}

\title{The elemental composition of the Sun}
\subtitle{III. The heavy elements Cu to Th}

\titlerunning{Solar abundances III. The heavy elements (Cu to Th)}
\authorrunning{Grevesse et al.}

\author{Nicolas Grevesse\inst{1,2}
\and
Pat Scott\inst{3}
\and
Martin Asplund\inst{4}
\and
A.~Jacques Sauval\inst{5}}

\institute{Centre Spatial de Li{\`e}ge, Universit{\'e} de Li{\`e}ge, avenue Pr{\'e} Aily, B-4031 Angleur-Li{\`e}ge, Belgium
\and
Institut d'Astrophysique et de G{\'e}ophysique, Universit{\'e} de Li{\`e}ge, All{\'e}e du 6 ao{\^u}t, 17, B5C, B-4000 Li{\`e}ge, Belgium\\
\email{nicolas.grevesse@ulg.ac.be}
\and
Department of Physics, Imperial College London, Blackett Laboratory, Prince Consort Road, London SW7 2AZ, UK\\
\email{p.scott@imperial.ac.uk}
\and
Research School of Astronomy and Astrophysics, Australian National University, Cotter Rd., Weston Creek, ACT 2611, Australia\\
\email{martin.asplund@anu.edu.au}
\and
Observatoire Royal de Belgique, avenue circulaire, 3, B-1180 Bruxelles, Belgium\\
\email{jacques.sauval@oma.be}
}

\date{Received 1 May 2014 / Accepted 1 Sep 2014}

\abstract{We re-evaluate the abundances of the elements in the Sun from copper ($Z=29$) to thorium ($Z=90$).  Our results are mostly based on neutral and singly-ionised lines in the solar spectrum.  We use the latest 3D hydrodynamic solar model atmosphere, and in a few cases also correct for departures from local thermodynamic equilibrium (LTE) using non-LTE (NLTE) calculations performed in 1D.  In order to minimise statistical and systematic uncertainties, we make stringent line selections, employ the highest-quality observational data and carefully assess oscillator strengths, hyperfine constants and isotopic separations available in the literature, for every line included in our analysis.   Our results are typically in good agreement with the abundances in the most pristine meteorites, but there are some interesting exceptions.  This analysis constitutes both a full exposition and a slight update of the relevant parts of the preliminary results we presented in Asplund, Grevesse, Sauval \& Scott (2009), including full line lists and details of all input data that we have employed.}

\maketitle
%

\section{Introduction}

Due to the Coulomb barrier and the fact that nuclear binding energy peaks at iron, the elements beyond the Fe peak ($Z\geq 29$, i.e. Cu and heavier elements) are not produced by exothermic charged-particle fusion reactions in stable nuclear burning inside stars. Instead, most of these elements are forged by successive neutron capture followed by $\beta$-decay.\footnote{In some cases the origins may also include contributions from nuclear statistical equilibrium, proton capture or interactions with neutrinos. For example, roughly half of the solar Zn is $^{64}$Zn produced in nuclear statistical equilibrium (possibly in connection with hypernovae), but the neutron-rich isotopes stem from neutron capture (e.g. Kobayashi \& Nakasoto \cite{2011ApJ...729...16K}).}  There are two main channels through which the isotopes of the heavy elements are produced: slow and rapid neutron capture (the $s$- and $r$-processes, respectively).  The distinction between $s$ and $r$ depends on the neutron density, such that the timescale for $\beta$-decay is shorter ($s$-process) or longer ($r$-process) than the timescale for neutron capture.  Once the neutron flux recedes, the neutron-rich isotopes from the $r$-process decay back to the valley of stability (Burbidge et al.\ \cite{1957RvMP...29..547B}; Cameron \cite{1957AJ.....62....9C}). 
As a result, most elements have a mixed origin, although a few isotopes and elements are produced almost exclusively by one or the other neutron-capture process (e.g. Sneden et al.\ \cite{2008ARA&A..46..241S}). Roughly half of the isotopes come primarily from the $s$-process and the other half from the $r$-process (see Table 3 in Anders \& Grevesse \cite{ag89}, for a full inventory). 
Observationally, the heavy elements show large abundance variations between stars, especially at low metallicity (e.g. Roederer et al.\ \cite{2014ApJ...784..158R}). 
The $r$-process produces abundance peaks mainly at \mbox{$Z\approx 32$}, 54 and 78 (Ge, Xe and Pt), whereas a typical $s$-process abundance pattern has peaks at $Z\approx 38$, 56 and 82 (Sr, Ba and Pb). 

The main site of the $s$-process is believed to be thermally pulsing asymptotic giant branch (AGB) stars (e.g. Sneden et al.\ \cite{2008ARA&A..46..241S}).  The main neutron source in these stars is thought to be $^{13}$C($\alpha$,n)$^{16}$O, although $^{22}$Ne($\alpha$,n)$^{25}$Mg can also operate, especially in more massive AGB stars (e.g. Busso et al.\ \cite{2001ApJ...557..802B}; Karakas et al.\ \cite{2012ApJ...751....8K}). More recently, it has been realised that the $s$-process can also be efficient in rapidly-rotating massive stars at low metallicity, through the production of primary N later converted to $^{22}$Ne (Pignatari et al.\ \cite{2008ApJ...687L..95P}).

The site(s) of the $r$-process have not yet been unanimously agreed upon. Numerous possibilities have been proposed, none altogether convincing. Currently the main contenders are core-collapse supernovae (e.g. in connection with a neutrino wind, Wanajo \cite{2013ApJ...770L..22W}) or mergers of neutron stars with either black holes or other neutron stars (Goriely et al.\ \cite{2011ApJ...738L..32G}), but achieving the necessary neutron fluxes and entropies is challenging in either case. A handful of stars at low metallicity are strongly enriched in the $r$-process elements, with an abundance pattern remarkably similar to the $r$-process signature in the solar system, suggesting the dominance of some universal process active throughout cosmic time (e.g. Cayrel et al.\ \cite{2001Natur.409..691C}, Sneden et al.\ \cite{2003ApJ...591..936S}, Frebel et al.\ \cite{2007ApJ...660L.117F}). In a few cases, the $r$-process enhancement is so high that it has enabled the detection of radioactive elements like thorium and uranium in the stellar spectra, which has made age-dating of the nucleosynthesis sites possible ({\it cosmo-chronology}, Butcher \cite{1987Natur.328..127B}). Because the $r$-process elements can have a primary origin (being produced from Fe-nuclei seeds) in contrast to the secondary origin of the $s$-process elements (which require first the production of $^{13}$C or $^{22}$Ne nuclei as neutron sources), the $r$-process should dominate in the early Universe (Truran \cite{1981A&A....97..391T}). 

A proper understanding of the nuclear processes involved in the production of the heavy elements requires a detailed knowledge of the solar system abundances.  In a recent review (Asplund et al.\ \cite{asp8}: AGSS09), we presented a summary of our redetermination of nearly all available elements in the solar photosphere. In this work (Paper III), we give detailed and homogeneous results for the elements Cu to Th, updating AGSS09 where necessary.  We use the most recent 3D atmospheric model together with the best atomic and solar data.  Previous papers in this series have focussed on Na -- Ca (Scott et al.\ \cite{AGSS_NaCa}: Paper I) and the Fe group (Scott et al.\ \cite{AGSS_Fegroup}: Paper II) with remaining elements to be dealt with in future studies.  Amongst the heavy elements, so far zirconium (Ljung et al.\ \cite{ljung}; Caffau et al.\ \cite{caf1}), osmium (Caffau et al.\ \cite{caf3}), europium (Mucciarelli et al.\ \cite{mucc}), hafnium and thorium (Caffau et al.\ \cite{caf2}) have been subjected to 3D solar analysis.

Sect.~\ref{observations} describes the observations we use and how we employ them. Sect.~\ref{3Dmodel} quickly recaps the salient details of the model atmospheres used in this series, and Sect. \ref{atomicdata} describes the atomic data and NLTE corrections we have employed in this paper.  In Sect. \ref{results} we give our abundance results for all elements, followed by a discussion of their dependence on the chosen model atmosphere in Sect. \ref{discussion}. We compare our results with some previous compilations of the solar chemical composition for the heavy elements in Sect. \ref{compilations}, before concluding in Sect. \ref{conclusions}.

\section{Observations}
\label{observations}

We have employed the Jungfraujoch (Delbouille et al.\ \cite{delb1}) and Kitt Peak (visual: Neckel \& Labs \cite{neck}; near-infrared: Delbouille et al.\ \cite{delb2}) disc-centre solar spectral atlases to carefully measure the equivalent widths of a large number of lines of the elements Cu to Th. Wherever possible, we have been very demanding in the quality of the lines we retained. We carefully examined the shape and full width of each line, to detect any trace of blending lines.  We gave each line a weight from 1 to 3, depending on the uncertainty that we estimated for the measured equivalent width, factoring in potential pitfalls such as blends and continuum placement. We integrated observed and theoretical line profiles over the same wavelength regions when measuring the equivalent widths. 

For most of the elements we used the measured equivalent widths to derive abundances with the new model atmosphere described in Sec.~\ref{3Dmodel}.  In some cases however (Nb, rare earths, Hf, W), we instead update results obtained by other authors with spectral synthesis and the Holweger \& M\"{u}ller (\cite{hol}) photospheric model, to account for the effect of the new 3D hydrodynamic solar model (Sect. \ref{3Dmodel}).  We included isotopic and hyperfine structure of lines as a series of blending features, as described in detail in Paper I, using the adopted atomic data detailed below for each relevant species (Sect. \ref{atomicdata}).

\section{Solar model atmospheres and spectral line formation}
\label{3Dmodel}

We follow the same procedure as in Papers I and II. We use the new 3D, time-dependent, hydrodynamic simulation of the solar surface convection first employed in AGSS09 as a realistic model of the solar photosphere. The reader is referred to Paper I (Scott et al.\ \cite{AGSS_NaCa}) for further details of the 3D solar model. As demonstrated extensively by  Pereira et al.\ (\cite{pereira_models}), this 3D model performs extremely well against an arsenal of key observational tests, indicating that the model is highly realistic. In particular, we note that our 3D solar model perfectly predicts the continuum centre-to-limb variation, which is a sensitive probe of the mean temperature structure in the typical line-forming region of the photosphere. The corresponding theoretical absolute intensities agree extremely well with observations as well. 

For comparison, and to quantify the systematic errors of our abundance results, we have also again performed identical calculations with four different 1D models of the solar photosphere: the widely-used, hydrostatic, semi-empirical model of Holweger \& M\"{u}ller (\cite{hol}; HM) which has been pressure-integrated to be consistent with our continuous opacities and chemical compositions, the mean 3D model \oneDAV\ obtained by spatially and temporally averaging the 3D model over about 45\,min of solar time, the solar model from the grid of {\sc marcs} theoretical models  extensively used for abundance analysis of solar-type stars  (Asplund et al.\ \cite{asp0}, Gustafsson et al.\ \cite{marcs08}), and the {\sc miss} semi-empirical model obtained from 1D LTE spectral inversion of Fe lines (Allende Prieto et al.\ \cite{alle}).  

For all spectral line formation calculations we assumed local thermodynamic equilibrium (LTE). Non-LTE (NLTE) analyses performed in a 1D framework for the Sun exist to our knowledge only for \altcui, \altzni, \altsri, \altsrii, \altzri, \altzrii,  \altbaii, \alteuii, \pbi and \thii (see below for individual references for each element). When such NLTE data are available, we correct the 3D LTE and 1D LTE results with those NLTE abundance corrections. We note that this is not fully self-consistent as the departures from LTE may be different in 3D than in 1D, indeed often depending on the particular 1D model atmosphere used.  As a rule of thumb, NLTE effects are more pronounced in theoretical model atmospheres than in semi-empirical ones due to their steeper temperature gradients, a phenomenon dubbed `NLTE masking' by Rutten \& Kostik (\cite{1982A&A...115..104R}). For completeness, we also mention that NLTE calculations are often performed for flux spectra, but we employ disc-centre intensity spectra; NLTE effects tend to be less severe in intensity than flux because the absorption lines are formed at lower heights in intensity than in flux. As is clear from the few NLTE studies available, much work still remains to be done in this area for the heavy elements. The shortage of NLTE investigations is partially due to the relatively small number of experts in the field.  For these elements however, perhaps the biggest reason for the dearth of NLTE analyses is a lack of necessary atomic data such as photo-ionisation cross-sections, transition probabilities and collisional rates for the many relevant atomic processes in what are often highly complicated atoms (Asplund \cite{asp7}).  NLTE calculations with incomplete model atoms and/or less precise atomic data can often be rather misleading -- but so can LTE.  The assumption of LTE is an extreme one, and not at all a cautious middle ground.  Caution should therefore be exercised with elements lacking NLTE corrections.  Fortunately, for the heavy elements, in most cases we rely on lines from the dominant ionisation stage, which typically exhibit less severe departures from LTE than minority species (Asplund \cite{asp7}); in the solar case the once-ionised species are typically in the majority. 

As explained in Paper I, with a 3D solar model there is no need to invoke any micro- or macroturbulent velocities to obtain correct line broadening (Asplund et al.\ \cite{asp1}).
For all 1D models, we adopted a microturbulence of 1\,km\,s$^{-1}$.  We carried out 3D line formation calculations for 45 snapshots from the full time sequence of the 3D solar model, spaced 1\,min apart,  and averaged them before carrying out the continuum normalisation.

\section{Atomic data and line selection}
\label{atomicdata}

The atomic data (transition probabilities, isotopic structure and hyperfine structure: HFS) we adopt are discussed in detail below for each element.  We give our adopted lines, oscillator strengths, NLTE corrections, equivalent widths, excitation potentials and derived abundances for all elements in Table~\ref{table:lines}, and isotopic and HFS data in Table~\ref{table:hfs}.  We have also re-examined the ionisation energies and partition functions in detail, updating some data relative to AGSS09; this is not trivial, as even many recent analyses are still based on quite erroneous data.  These data are given in Table~\ref{table:partition}.  For \hoii our partition functions come from from Bord \& Cowley (\cite{bord1}), and for all other species, from Barklem \& Collet (in preparation); these are in good agreement with values computed from NIST atomic energy levels.  Our ionisation energies come from the NIST data tables.

Modern line-broadening data for collisions with hydrogen atoms are available from Barklem, Piskunov \& O'Mara (\cite{bark}) for all the neutral lines we consider in this paper, but only \baii and \srii amongst the ionised lines.  We use this data wherever it exists; otherwise, we use the classical recipe of Uns\"old (\cite{unsold}) with an enhancement factor of 2.0.

\subsection{Copper}
Although Cu is essentially once ionised in the solar photosphere, only \cui lines have been identified in the solar spectrum. From the works of Kock \& Richter (\cite{kock}) and Sneden \& Crocker (\cite{sned2}), we retained the five \cui lines given in Table~\ref{table:lines}. For the first three lines, we have used the experimental $gf$-values of Kock \& Richter, renormalised to the lifetimes of Carlsson et al.\ (\cite{carl}; the renormalisation amounts to a change of only $-0.006$\,dex).  For the last two lines, we took oscillator strengths from Bielski (\cite{biel}), based on the measurements of Meggers et al.\ (\cite{Meggers61}). 

Cu is approximately 69.2\% $^{63}$Cu and 30.8\% $^{65}$Cu (Rosman \& Taylor \cite{IUPAC98}), both of which have nuclear spin $I=\frac32$.  The strongest solar lines are affected by isotopic broadening and HFS.  We accounted for isotopic splitting using the data of Fischer, H\"{u}nermann \& Kollath (\cite{fisch}).  We took HFS constants from the same paper as well as from Hannaford \& McDonald (\cite{hann}).

NLTE \cui line formation in the Sun was investigated by Shi et al.\ (\cite{Shi14}), who found positive NLTE corrections (\mbox{$\approx +0.02$}\,dex) for our first three lines, and a larger negative correction (\mbox{$\approx -0.05$}\,dex)  for the last line (\cui  809.3\,nm).  We adopt their results for $S_\mathrm{H}=0.1$, as recommended in their paper.  Because the \cui 793.3\,nm and \cui 809.3\,nm lines differ only in the $J$ values of their lower levels, and Shi et al.\ (\cite{Shi14}) explain the large negative NLTE offset in the abundance returned by \cui 809.3\,nm as mainly due to underpopulation of the upper level of this transition, we assume that the NLTE correction for \cui 793.3\,nm is the same as for \cui 809.3\,nm.  The addition of NLTE corrections for \cui is a new feature of the analysis here, compared to AGSS09. We note that the NLTE study of Shi et al.\ is for flux rather than disc-centre intensity spectra, which means that the adopted NLTE effects may be slightly exaggerated. 

\subsection{Zinc}
Zinc is mostly neutral in the photosphere. We retain the five \zni lines given in Table~\ref{table:lines}. The $gf$-values we use come from Bi\'{e}mont \& Godefroid (\cite{biem8}), whose theoretical results agree well with measured lifetimes. For two lines (472.2\,nm and 481.0\,nm), we renormalise these data to the accurate lifetimes measured by Kerkhoff et al.\ (\cite{Kerkhoff80}) using laser spectroscopy, resulting in an increase of 0.01\,dex.

Y.~Takeda (private communication, 2011) used the model atom from Takeda et al.\ (\cite{take}) to compute NLTE corrections of between $-0.01$ and $-0.04$\,dex for our lines at the centre
of the solar disc, for different values of $S_\mathrm{H}$. For $S_\mathrm{H}$ values between 0.1 and 1 the NLTE corrections vary very little.  We chose to use the corrections at $S_\mathrm{H}=0.3$.

\subsection{Gallium}
Although essentially once ionised, only the near-UV \gai resonance line at 417.2\,nm (see Table~\ref{table:lines}) has been used by Ross \& Aller (\cite{ross}) and Lambert et al.\ (\cite{lamb}) to derive the solar Ga abundance.  This line is rather heavily perturbed and can only be extracted by spectrum synthesis; doing so, these authors found $\epsilon_{\mathrm{Ga}}= 2.80$ and $\epsilon_{\mathrm{Ga}}= 2.84$ respectively. Lambert et al.\ also suggested using a very faint unidentified infrared line at 1194.915\,nm.  They argued that this line could be due to \altgai, as in their analysis it led to an abundance in agreement with the near-UV line.  Oscillator strengths have been discussed by Lambert et al., and we adopt their chosen value for the near-UV line.  This comes from Cunningham \& Link (\cite{Cunningham67}), who set a theoretical branching fraction to an absolute scale with their own accurate experimental lifetime.

We rechecked the IR line: it is suspiciously wide (suggesting unidentified blends), and its equivalent width
is too inaccurate to keep as a good abundance indicator. We therefore synthesised the spectrum around
the near-UV resonance line and derived an accurate estimate of the \gai
contribution: 5.22$\pm$0.30\,pm.

\subsection{Germanium}
Ge is also mostly once ionised in the solar photosphere, but only very few \gei lines have been identified. Accurate $gf$-values have been measured by Bi\'emont et al.\ (\cite{biem2}).  With the only usable \gei line (326.9\,nm; Table~\ref{table:lines}), they applied their oscillator strength to derive a solar abundance of $\log\epsilon_{\mathrm{Ge}}=3.58$.

We also investigated the intercombination line at 468.583\,nm suggested by Lambert et al.\ (\cite{lamb}) as a possible indicator of the Ge abundance. It is however blended by a Co\,\textsc{i} line, which contributes 0.155\,pm to the total equivalent width of the feature when computed with the new solar Co abundance derived in Paper II.  With this blend removed, the equivalent width is about 0.4\,pm, and the abundance of Ge derived from this line is about a factor of four too large; this line is definitely blended by another unknown line.

\subsection{Arsenic}
No As line is definitively identified in the solar spectrum. Gopka et al.\ (\cite{gopka}) used two lines in the near-UV, at 299.0984 and 303.2846\,nm, believed to be due to \asi to derive the solar abundance of As.  They found $\log \epsilon_{\mathrm{As}}=2.33$, although the $gf$-values for these two \asi lines are extremely uncertain. As explained below, we argue that no reliable As abundance can be derived for the Sun with the available information. 

\subsection{Rubidium}
Rubidium is very much once-ionised in the solar photosphere, but only two faint, near-IR resonance lines  of \rbi at 780.0 and 794.7\,nm can be identified (Table~\ref{table:lines}).  Both are strongly broadened by isotopic and hyperfine structure, and perturbed by stronger neighbouring lines (e.g. the 794.7\,nm line is heavily
perturbed by a water vapour line).  We very carefully measured the equivalent widths of the two \rbi lines, eventually using spectral synthesis to derive the best values. We measured the faintest line only on the Jungfraujoch solar spectrum, which shows by far the smallest water vapour content of the available solar atlases.

Accurate $gf$-values are available for both lines, from lifetime measurements of the upper levels by Volz \& Schmoranzer (\cite{vol}) and Simsarian et al.\ (\cite{sim}).  As advocated by Morton \cite{Morton00}), we adopted the mean from these two studies, weighted according 
to their uncertainties.

Natural Rb shows HFS because it is about 72.2\% $^{85}$Rb, with $I=5/2$, and 27.8\% $^{87}$Rb, with $I=3/2$ (Rosman \& Taylor \cite{IUPAC98}).  We took HFS constants from a range of different sources.  For $^{85}$Rb, we adopted values from Nez et al.\ (\cite{nez}), Rapol et al.\ (\cite{rapol}) and Beacham \& Andrew (\cite{beach}).  We used data from Beacham \& Andrew also for $^{87}$Rb, as well as from Ye et al.\ (\cite{ye}) and Bize et al.\ (\cite{bize}).  Isotopic splitting in these lines is small compared to the separation of HFS components (Banerjee et al.\ \cite{banerjee}) and thus neglected here.

Relative to AGSS09, in this analysis we have added HFS data and updated the oscillator strengths for both lines.

\subsection{Strontium}
We analysed the few \sri and \srii lines previously studied by Gratton \& Sneden (\cite{grat}) and Barklem \& O'Mara (\cite{bark2}), listed in Table~\ref{table:lines}. These authors also discuss the accuracy of available $gf$-values in detail.  For the two \sri lines, we adopted experimental oscillator strengths from García \& Campos (\cite{garc}), who normalised their own relative values using existing accurate lifetimes, and Migdalek \& Baylis (\cite{Migdalek}). For \altsrii, we derived $gf$-values by setting the relative data of Gallagher (\cite{Gallagher}) to an absolute scale using the mean of lifetimes from Kuske et al.\ (\cite{Kuske}) and Pinnington et al.\ (\cite{Pinnington95}).  The \sri $gf$-values are substantially more precise overall ($\pm$0.02--0.03\,dex) than the \srii ones ($\pm$0.08\,dex). 

The broadening parameters for the rather strong IR lines of \srii have been calculated by Barklem \& O'Mara (\cite{bark2}).

We treat each Sr line as a single component in our abundance calculations, as isotopic splitting is small for Sr lines (Hauge \cite{hauge2}), and HFS exists only for $^{87}$Sr (which accounts for just 7\% of Sr).

M.~Bergemann (private communication, 2011) has computed NLTE corrections for disc-centre intensity both for our \sri and \srii lines at $S_\mathrm{H}=0.05$ using the {\sc marcs}, HM and \oneDAV\ models; for the 3D case we adopt the \oneDAV\ results, which should closely approximate the full 3D results.  The NLTE effects are substantial, and in opposite directions for \sri and \srii (Table~\ref{table:lines}). Bergemann's results are in good agreement with those of Mashonkina \& Gehren (\cite{mash}) for the three \srii lines.

Relative to AGSS09, we have updated the NLTE corrections and the $gf$-value of the \sri 707.0\,nm line.  We also discarded the \srii line at 416.1\,nm, because no accurate $gf$-value is available for this line.

\subsection{Yttrium}
The most recent solar analysis is from Hannaford et al.\ (\cite{hann1}). They obtained accurate $gf$-values, both for \yi and \yii lines, combining lifetimes with branching fraction measurements.  They derived the solar abundance of Y, $\log\epsilon_{\mathrm{Y}}=2.24\pm0.03$ from eight \yi and 41 \yii solar lines. In the present work we carefully examined all these solar lines, deciding to discard all \yi lines because they are very weak, with large uncertainties arising from the measurement of equivalent widths. Adopting the same demanding selection criteria as for other elements, we reduced the \yii sample to the ten best lines.  We adopted $gf$-values from Hannaford et al.\ (\cite{hann1}) directly for seven of these. For the other three (490.0, 547.3 and 572.8\,nm), we updated the $gf$-values of Hannaford et al.\ using new accurate lifetimes by W\"{a}nnstr\"{o}m et al.\ (\cite{wann}) and Bi\'{e}mont et al.\ (\cite{biem7}).  We calculated new lifetimes as means of lifetimes from as many of these three sources as possible in each case.  We took unweighted means, as differences between the three sets of lifetimes indicate a systematic error that is not quantified in any of the stated lifetime errors.

Y consists entirely of $^{89}$Y, which has $I=\frac12$.  We included HFS data for \yii lines where available, drawing on HFS constants measured by W\"{a}nnstr\"{o}m et al.\ (\cite{wann}) and Dinneen et al.\ (\cite{Dinneen}), as well as theoretical calculations by Beck (\cite{Beck}).

Relative to AGSS09, we have added HFS data for all lines, and renormalised the oscillator strengths of the 490.0, 547.3 and 572.8\,nm lines.

\subsection{Zirconium}
Zr is essentially in the form of \zrii in the solar photosphere.  However, a
large number of faint \zri lines are also present in the photospheric
spectrum.  Bi\'emont et al.\ (\cite{biem3}) measured $gf$-values by the
lifetime+branching fraction technique for a large number of \zri and
\zrii transitions, and applied these data to determine the solar
abundance of Zr, using equivalent widths and the HM model. Thirty-four \zri and 24 \zrii lines
led to the same result (\altzri: $\log\epsilon_{\mathrm{Zr}}= 2.57\pm0.21$, \altzrii: $\log\epsilon_{\mathrm{Zr}}= 2.56\pm0.14$), but the dispersions are uncomfortably large. 
This indicates that many solar \zri and \zrii lines are blended.
Ljung et al.\ (\cite{ljung}) measured new branching fractions and combined 
them with previously-measured lifetimes to obtain a new set of accurate \zrii $gf$-values. 
These authors also analysed a few solar \zrii lines, using their new data, equivalent widths and an older version of the 3D model that we employ in this series (they used the same model as used in AGS05) to derive a 3D LTE solar abundance of $\log\epsilon_{\mathrm{Zr}}= 2.58\pm0.02$ (standard deviation). 

As for other elements, we carefully selected the lines to be used in the
present abundance analysis. Although we give \zri results for comparison in Tables~\ref{table:lines} 
and \ref{table:abuns}, using oscillator strengths from Bi\'emont et al.\ (\cite{biem3}), we do not retain \zri as a good indicator
of the solar Zr abundance.  This is because the lines are all very weak and many are blended.  Because \zri is the minor species, 
it is also prone to very large departures from LTE (\mbox{$\approx +0.3$}\,dex for $S_{\rm H}=1.0$), whereas low excitation \zrii lines are essentially formed in LTE (Velichko et al.\ \cite{veli}); note though that Velichko et al.\ (\cite{veli2}) suspect that their published NLTE corrections are overestimated due to their incomplete \zri model atom.  We finally retained the ten low-excitation \zrii lines given in Table~\ref{table:lines}. The $gf$-values are mean values (on a log scale) from Bi\'emont et al.\ (\cite{biem3}) and Ljung et al.\ (\cite{ljung}), except for 402.4\,nm, which comes only from Ljung et al., as it was not measured by Bi\'emont et al.\  For our lines, the differences between these two data sets are within the claimed uncertainties, and the uncertainties of the mean $gf$-values we adopt are of order 5--10$\%$.

\subsection{Niobium}
Because of the rather large ratio \altnbii/\altnbi, as discussed in
Hannaford et al.\ (\cite{hann2}), only \nbii is a reliable indicator
of the solar abundance of Nb.  All useful \nbii lines are unfortunately strongly blended 
in the solar photospheric spectrum.  We therefore choose to update the recent value obtained 
by Nilsson et al.\ (\cite{nils}), who derived the solar Nb abundance from spectral synthesis
using the HM model.  They used accurate new experimental $gf$-values and HFS data from Nilsson
\& Ivarsson (\cite{nils2}), supplemented by some of their own oscillator strengths.  We adopt 
the same data here.  Nb is entirely $^{93}$Nb ($I=\frac92$), so HFS is important but isotopic 
structure is not.

In AGSS09 we performed our own analysis without HFS, whereas here we update the result of Nilsson et al.\ (\cite{nils}), using HFS data.

\subsection{Molybdenum}
Mo is essentially once ionised in the solar photosphere. Unfortunately all
the lines identified as \moii are too blended to be useful, except for one
at 329.2\,nm. However, no $gf$-value exist for this line apart
from the value from Corliss \& Bozman (\cite{corl}), which suffer from large
uncertainties. The abundance result from this \moii line appears to be
orders of magnitude larger than the meteoritic value, indicating a problem with the transition probability. We therefore
rely on the minor indicator, \altmoi, as done by
Bi\'emont et al.\ (\cite{biem4}).  They used accurate $gf$-values from Whaling
et al.\ (\cite{whal1}), finding $\log\epsilon_{\mathrm{Mo}}=1.92\pm0.05$ with twelve very
faint lines, all difficult to measure. After a careful analysis of
these lines, we retain just two as reliable indicators of the Mo abundance (Table~\ref{table:lines}).
For these lines, we employ the slightly improved oscillator strengths of Whaling \& Brault (\cite{whal2}).

Mo consists of a pot pouri of different isotopes, but the two lines we use are very weak, and their isotopic splitting is 
small (Hughes \cite{Hughes}; Golovin \& Striganov \cite{Golovin}).  We can therefore safely ignore isotopic and hyperfine structure for these lines.

Compared to AGSS09, we use the $gf$-values of Whaling \& Brault (\cite{whal2}) instead of those of Whaling et al.\ (\cite{whal1}), which translates to a mean abundance change of $-0.02$\,dex.

\subsection{Ruthenium}
As for the previous elements, the ratio \altruii/\rui is large in the solar photosphere.
Unfortunately none of the \ruii lines identified in the solar
spectrum could be used as they are all hopelessly blended. New $gf$
values have recently been measured and computed by Fivet et al.\ (\cite{five}) 
for a large number of lines of \altrui. These new data
are in reasonable agreement with the accurate $gf$-values measured by
Wickliffe et al.\ (\cite{wick}) using the lifetime+branching fraction
technique. We rely on the purely experimental $gf$-values rather than using
the data from Fivet et al.\ (\cite{five}), because for our lines, the latter are 
based only on theoretical values.

Fivet et al.\ (\cite{five}) applied their new data to determine the
solar Ru abundance from six good \rui lines: $\log\epsilon_{\mathrm{Ru}}=1.72\pm0.12$. 
We adopt the same six lines in our analysis (Table~\ref{table:lines}).  Whilst Ru consists of
many different isotopes, the only isotopic splits available are for the 408.1\,nm and 
455.5\,nm lines, which are so weak that isotopic structure has no effect for abundance determinations.

Relative to AGSS09, here we use $gf$-values from Wickliffe et al.\ (\cite{wick}) instead of Fivet et al.\ (\cite{five}).

\subsection{Rhodium}
The ratio \altrhii/\rhi is again large in the solar photosphere. Unfortunately, none of the few
\rhii lines identified in the UV can be measured: they are all very
heavily blended and could not be analysed even by spectrum synthesis.
Kwiatkowski et al.\ (\cite{kwia1}) measured  lifetimes for \rhi and
derived $gf$-values by combining these lifetimes with branching
fractions from Corliss \& Bozman (\cite{corl}). They used these
new $gf$-values with the HM model and a sample of solar \rhi lines 
in the near UV, to derive an abundance of $\log\epsilon_{\mathrm{Rh}}=1.12\pm0.12$. As many of these lines
are quite difficult to measure with accuracy, we only retain two of
them as reliable indicators of the Rh abundance.  We adopt the $gf$-values of Kwiatkowski et al.\ (\cite{kwia1})
for both these lines, as later data (Duquette \& Lawler \cite{Duquette}) suffers from radiation trapping for the only one of our lines measured.  

Rhodium is 100\% $^{103}$Rh, which has nuclear spin $I=\frac12$.  Where possible, we include HFS data for our two lines, from Chan et al.\ (\cite{Chan}).  The analysis in AGSS09 did not include HFS.

\subsection{Palladium}
The ratio \altpdii/\pdi is about 4 in the solar photosphere but no \pdii lines have been identified in the solar spectrum. The most recent Pd abundance is from Xu et al.\ (\cite{xu}), who combined new lifetimes and branching fractions to derive accurate $gf$-values for \pdi lines. They used these new data to refine an earlier analysis by Bi\'emont et al.\ (\cite{biem5}), using equivalent widths, five of the eight \pdi lines employed in the earlier study, and the same HM model.  As the lines are very perturbed, we only retain two of those used in these earlier works.

Isotopic and hyperfine structure is small for \pdi (Engleman et al.\ \cite{Engleman}) and can be ignored for our purposes.

\subsection{Silver}
There is no \agii line in the solar spectrum. The only two \agi lines are in the near-UV (see Table~\ref{table:lines}), and very difficult to measure because of blends and uncertainty in the continuum placement. Grevesse (\cite{grev}) revised an older analysis by Ross \& Aller (\cite{ross2}; $\log\epsilon_{\mathrm{Ag}}=0.85\pm0.15$), using $gf$-values from Hannaford \& Lowe (\cite{hann3}).  We adopt newer $gf$-values here, based on the total lifetime of the resonance doublet as measured by Carlsson et al.\ (\cite{carl2}), and the relative contribution of each line calculated by Civi\v{s} et al.\ (\cite{civis}).

Silver is approximately 51.8\% $^{107}$Ag and 48.2\% $^{109}$Ag (both $I=\frac12$), and exhibits substantial isotopic and hyperfine structure.  We calculated the isotopic separations of our \agi lines using data from Crawford et al.\ (\cite{craw}) and Jackson \& Kuhn (\cite{jack}), relying on component identifications from Brix et al.\ (\cite{brix}) and Wessel \& Lew (\cite{wess}).  We took HFS constants from the experiments of Wessel \& Lew (\cite{wess}) and Carlsson et al.\ (\cite{carl2}).

Compared to AGSS09, here we employ both updated HFS data and oscillator strengths; in AGSS09 we used the $gf$-values of Hannaford \& Lowe (\cite{hann3}).

\subsection{Cadmium}
There is a \cdi line identified in the solar spectrum at 326.1065\,nm, but it is awfully blended. Even spectrum synthesis by Youssef et al.\ (\cite{yous}) was not very successful; we have chosen to discard this line. Another line (508.6\,nm; Table~\ref{table:lines}) was attributed to \cdi by Lambert et al.\ (\cite{lamb}). Youssef et al.\ derived the Cd abundance from this very weak line using the HM model and an accurate $gf$-value, measured by the lifetime+branching fraction technique by Veer et al.\ (\cite{veer}). We adopt this line and the same oscillator strength for our analysis.

We carefully measured the 508.6\,nm line, finding an equivalent width of $0.100\pm0.015$\,pm.  This line is however also blended, by a faint Fe\,\textsc{i} line.  The lower level of the Fe\,\textsc{i} line has an excitation energy of 3.88\,eV, but the line has no accurately-known oscillator strength. The only $gf$-value available comes from the calculations of Kurucz (\cite{kur}; $\log gf=-4.325$);  the accuracies of Kurucz's semi-empirical $gf$-values degrade markedly as one moves to weaker transitions such as this.  In order to empirically estimate the contribution of this blend, we identified five other weak Fe\,\textsc{i} lines with similar excitation potentials, near to the 508.6\,nm line in wavelength.  We measured their equivalent widths, and used the Fe abundance derived in Paper II together with the relative $gf$-values to estimate the equivalent width of the Fe\,\textsc{i} contribution to the feature at 508.6\,nm.  We found an implied Fe\,\textsc{i} equivalent width of $0.027\pm0.014$\,pm, giving an overall \cdi contribution of $0.073\pm0.021$\,pm.

\subsection{Indium}
Although indium is essentially \inii because of its very low ionisation energy (5.78\,eV; the ratio \altinii/\ini is $>$$100$ in the solar photosphere), there is only one identified \ini line in the solar photospheric spectrum, at 451.13\,nm.  This line has been analysed by many authors: Lambert et al.\ (\cite{lamb}), Bord \& Cowley (\cite{bord}, \cite{bord1}) and Gonzalez (\cite{gonz}). Their results for the solar abundance of indium cluster around $\log\epsilon_{\mathrm{In}}= 1.60$, which is 0.84\,dex larger than the meteoritic value: $\log\epsilon_{\mathrm{In}}= 0.76\pm0.03$ (Lodders et al.\ \cite{lodd}). To explain this large difference, Bord \& Cowley and Gonzalez suggested that because its condensation temperature is just 536\,K, indium did not fully condense in the solar nebula, leading 
to a relative depletion in meteorites.  However, other elements with similarly low condensation temperatures do not exhibit the same differences in their solar and meteoritic abundances.

Vitas et al.\ (\cite{vita}) recently made a detailed analysis of this \ini line by spectrum synthesis, including HFS and many known blends.  They concluded that the photospheric line at 451.13\,nm is in fact not \altini, because they could not reproduce the observed spectrum, regardless of their adopted indium abundance.  They also analysed the same region in a sunspot spectrum, finding that it could be reproduced nicely by spectrum synthesis with the meteoritic indium abundance, if Zeeman splittings were properly taken into account. Vitas et al.\ therefore suggest that the solar indium abundance is the meteoritic value, and that the photospheric line at 451.13\,nm is due to an unidentified high-excitation ionic line, which disappears in the much cooler sunspot.

Based on the ratio Sn/In, Vitas et al.\ (\cite{vita}) also make the argument that the observed meteoritic ratio is perfectly reproduced by models of $r$- and $s$-process nucleosynthesis, but that the high solar In abundance is not.

\subsection{Tin}
The ratio \altsnii/\sni is about 10.  The only reasonable Sn line available is a very weak \sni line in the near-UV (see Table~\ref{table:lines}).  We adopt the $gf$-value obtained by Lotrian et al.\ (\cite{lotri}) for this line, from lifetime and branching fraction measurements.  The equivalent width of this very faint line (0.12\,pm) has an uncertainty of order 20$\%$, i.e.\ 0.08\,dex.  Although tin has many significant isotopes, for this line isotopic structure is neither available nor, given its weakness, necessary.

\subsection{Antimony}
Only one very faint and perturbed line has been
identified as \sbi in the solar photospheric spectrum, at 323.2547\,nm. The
abundance of Sb has not been re-analysed for many decades: the
latest result was from Grevesse et al.\ (\cite{grev2}), slightly revised
by Ross \& Aller (\cite{ross3}) to $\log\epsilon_{\mathrm{Sb}}=1.00$.
This value had a very large uncertainty, difficult to estimate because of the
uncertainty of the equivalent width and the fact that the only available $gf$-value came from
the notoriously uncertain data of Corliss \& Bozman (\cite{corl}). We measured 
an equivalent width of 0.045$\pm$0.015\,pm, with a large uncertainty of 33$\%$.
Spectral synthesis would not substantially reduce this error, given how weak and
badly blended this line is.

Guern \& Lotrian (\cite{guer}) and Gonzales et al.\ (\cite{gonz2}) measured the 
oscillator strength of this line by combining lifetimes and branching fractions, 
but the values differ by 0.18\,dex: $\log gf=-0.72$ and $-0.90$, respectively.

\subsection{Barium}
Because of its low ionisation energy, Ba is essentially \baii in the photosphere. We retained the three \baii lines of Table~\ref{table:lines}.  These lines are sensitive to NLTE effects, and substantially broadened by HFS.  M.\ Bergemann (private communication, 2011) computed the NLTE effects for these lines.  Her values at $S_\mathrm{H}=0.05$ (Table~\ref{table:lines}) are in excellent agreement with the results of Mashonkina et al.\ (\cite{mash3}) and Mashonkina \& Gehren (\cite{mash2}).  We adopt the accurate $gf$-values available from Davidson et al.\ (\cite{davi}) and Kurz et al.\ (\cite{Kurz}).

The many stable isotopes of Ba are present in the approximate ratio $^{130}$Ba:$^{132}$Ba:$^{134}$Ba:$^{135}$Ba:$^{136}$Ba:$^{137}$Ba:$^{138}$Ba = 0.1:0.1:2.4:6.6:7.9:11.2:71.7 (Rosman \& Taylor \cite{IUPAC98}), with only the odd-$A$ nuclei possessing nuclear spin ($I=\frac32$ in both cases).  We obtained isotopic separations for the 455.4\,nm line from Wendt et al.\ (\cite{Wendt84}), from van Hove et al.\ (\cite{vanhove}) for the 585.4\,nm line, and from Villemoes et al.\ (\cite{villemoes}) for the 649.7\,nm line.  We also drew on the latter for HFS data, along with the papers of Trapp et al.\ (\cite{trapp}) and van Hove et al.\ (\cite{vanhove2}).

In AGSS09 we did not include the 455.4\,nm line.  Here we have also updated the HFS, isotopic and NLTE data, as well as the oscillator strength of the 585.4\,nm line.

\subsection{Rare Earths (La to Lu) and Hafnium}
\label{rare_gfs}
In recent years, the Wisconsin group have systematically measured the atomic data required for accurate abundance analyses of all the once-ionised species of the rare Earth elements. They have derived accurate $gf$-values by measuring lifetimes and branching fractions, as well as HFS constants and isotopic splits wherever necessary. With the HM model, spectral synthesis and/or equivalent widths, and as many lines as possible, they have systematically applied their very accurate atomic data to the determination of the solar abundances of all the rare Earth elements (La: Lawler et al.\ \cite {law1}, Ce: Lawler et al.\ \cite{law2}, Nd: Den Hartog et al.\ \cite{den1}, Sm: Lawler et al.\ \cite{law3}, Eu: Lawler et al.\ \cite{law4}, Gd: Den Hartog et al.\ \cite{den2}, Tb: Lawler et al.\ \cite{law5}, Ho: Lawler et al.\ \cite{law6}, Er: Lawler et al.\ \cite{law7}), Pr, Dy, Tm, Yb and Lu: Sneden et al.\ \cite{sned1}), as well as Hf (Lawler et al.\ \cite{law8}).

We have not repeated the careful spectral synthesis computations done by Sneden, Lawler and collaborators, but rather corrected them for the abundance differences we see between the results with our 3D model and the HM model. For this purpose we have selected a sample of representative lines of each species, and have derived the 3D$-$HM abundance differences. We included all HFS and isotopic splittings summarised by Sneden et al.\ (\cite{sned1}) in our calculations.

For Nd, new $gf$-values are also available from Li et al.\ (\cite{li}).  Taking the mean of $gf$-values from Li et al.\ (\cite{li}) and from Den Hartog et al.\ (\cite{den1}) produces exactly the same abundance as using only the values of Den Hartog et al, however.

For \altsmii, new accurate $gf$-values have been measured by Rehse et al.\ (\cite{reh}), which agree quite well with the values of Lawler et al.\ (\cite{law3}). We chose to adopt the mean of the $gf$-values from these two sources.  In AGSS09 we simply adopted the values of Lawler et al.\ (\cite{law3}).

As in the analysis of AGSS09, for Eu we also applied a small NLTE correction of $+0.03$\,dex for every line, as derived by Mashonkina \& Gehren (\cite{mash2}).
We are not aware of any other NLTE study of rare Earth elements in the Sun.

For \alttbii, we only kept one of the three lines (365.9\,nm) used by Lawler et al.\ (\cite{law5}), because of the difficulty in fitting the profiles of the other two, due to blends and very wide HFS.

\subsection{Tungsten}
\label{w_gfs}
The ratio \altwii/\wi is of order ten in the solar photosphere but unfortunately no \wii lines are available in the solar spectrum.  Holweger \& Werner (\cite{hol2}) used spectral synthesis to analyse the two weak \wi lines at 400.9 and 484.4\,nm with the HM model. New $gf$-values have become available from den Hartog et al.\ (\cite{denh}) for both lines, and from Kling \& Koch (\cite{klin}) for the 484.4\,nm line.  Here we update the results of Holweger \& Werner for the new $gf$-values (where we take the mean of the two new values for the 484.4\,nm line); we did not previously apply this update in AGSS09.  

The lines are too faint for isotopic or hyperfine structure to matter.  This is in fact true of all lines we use from elements heavier than Hf, so we will discuss neither HFS nor isotopic effects any further in this Section.

\subsection{Osmium}
Only very few faint lines of \osi have been identified in the solar spectrum. Kwiatkowsky et al.\ (\cite{kwia2}) measured accurate lifetimes and used branching fractions from Corliss \& Bozman (\cite{corl}) to derive $gf$-values for a few \osi lines of solar interest.  Quinet et al.\ (\cite{quin}) recently made new measurements of lifetimes and combined them with modern theoretical branching fractions, leading to values in good agreement with those of Kwiatkowsky et al.\ (\cite{kwia2}).  

Kwiatkowsky et al.\ (\cite{kwia2}) used a series of nine \osi lines to derive the solar abundance. Quinet et al.\ (\cite{quin}) however showed that many of these lines are too faint and blended to be reliably measured. We agree with Quinet et al.\ (\cite{quin}) that very few \osi lines are realistically usable (see their Table 7). We are ultimately even more demanding than them in our line selection: we only retain the 330.2\,nm line as the unique tracer of the solar Os abundance. We adopt the accurate experimental oscillator strength of Ivarsson et al.\ (\cite{ivar}) for this line, but note that the  value differs from that of Quinet et al.\ (\cite{quin}) by only 0.003\,dex.  The two other lines considered by Quinet et al.\ (327.0 and 442.0\,nm; see their figure 6), are very difficult to analyse accurately, even by spectral synthesis.

\subsection{Iridium}
Drake \& Aller (\cite{drake}) analysed the \iri line at 322.1\,nm by spectrum synthesis.  Youssef \& Khalil (\cite{yous2}) analysed three \iri lines in the near-UV, including the 322.1\,nm line, also by spectrum synthesis using the HM model.

We retain only the 322.1\,nm \iri line, as the other two used by Youssef \& Khalil (\cite{yous2}) are too heavily perturbed. Even the 322.1\,nm line is itself heavily blended. Rechecking this line in the Jungfraujoch solar tracing (Delbouille et al.\ \cite{delb1}; no tracing is available from Kitt
Peak at these wavelengths) we noticed that Youssef \& Khalil's adopted continuum was too high in this spectral region. We therefore directly remeasured the equivalent width of this blended line: 0.975$\pm$0.125\,pm.  We confirmed this value using spectral synthesis with the HM solar photospheric model.

We have been able to derive a very accurate $gf$-value for 322.1\,nm line, using lifetime measurements by Gough et al.\ (\cite{gough}; uncertainty 3.6$\%$) and Xu et al.\ (\cite {xu2}; uncertainty 6.8$\%$).  We weighted these lifetimes by their respective uncertainties and took the mean, then paired the resulting value with the branching fraction measured by Gough et al.\  We did not use the theoretical branching fraction of Xu et al.\ because of its uncertain accuracy.

\subsection{Gold}
Youssef (\cite{yous3}) used spectral synthesis with the HM model to analyse the much-perturbed \aui 312.3\,nm line (Table~\ref{table:lines}), the only useful Au feature in the solar spectrum.  We carefully remeasured this \aui line on the Jungfraujoch disc-centre solar spectrum (Delbouille et al.\ \cite{delb1}); no Kitt Peak spectrum is available in this wavelength region. We found an equivalent width of 0.29\,pm, with an uncertainty of 10$\%$.

A very weak Fe\,\textsc{i} line is known to exist coincident with the \aui line, with parameters \mbox{$\lambda=312.2775$\,nm}, \mbox{E$_\mathrm{exc}=2.45$\,eV}, \mbox{$\log gf = -4.144$} (Kurucz \cite{kur}).  Unfortunately, the oscillator strength is not reliable enough to make any serious estimate of the contribution to the equivalent width of the \aui line, and the spectrum is far too crowded in this region to employ the same strategy as we did for \altcdi, where we relied on a number of other nearby faint Fe\,\textsc{i} lines.

The best available $gf$-value for the \aui line comes from Hannaford et al.\ (\cite{hann4}), who obtained log $gf$ $=-0.95\pm0.06$ with lifetime and branching fraction measurements.  The lifetime of Hannaford et al.\ for the upper level agrees perfectly with that obtained by Gaarde et al.\ (\cite{Gaarde}) using a similar technique.

\subsection{Lead}
Bi\'emont et al.\ (\cite{biem6}) used one \pbi line, for which they had accurately derived the $gf$-value, to revise the solar Pb abundance.  We use the same line and oscillator strength (Table~\ref{table:lines}), but our equivalent width is $0.855\pm0.035$\,pm instead of 0.91\,pm, derived anew for this paper by both direct measurement and spectrum synthesis.  This line is situated in the outer red wing of an extremely strong line with contributions from Co\,\textsc{i}, Fe\,\textsc{i} and V\,\textsc{i}, and in the outer blue wing of a somewhat less strong Fe\,\textsc{i} line. The continuum is very difficult to determine in this crowded region, but a viable reference point can be found at 3683.8\,\AA.  We rule out the large equivalent width of Bi\'emont et al.; our spectral synthesis calculations confirm that their continuum placement is incorrect.

Mashonkina et al.\ (\cite{mash12}) computed the effect of departures from LTE on the formation of this \pbi line at the centre of the solar disc, which are substantial. Between $S_\mathrm{H}=0$ and $S_\mathrm{H}=1$, the NLTE correction with the \marcs\ model atmosphere varies from $+0.15$ to $+0.07$\,dex.  We adopt Mashonkina et al.'s preferred value of $+$0.12\,dex, derived with $S_\mathrm{H}=0.1$.  At the time of AGSS09, no NLTE correction was available.

\subsection{Thorium}
The only reliable faint \thii line lies at 401.9\,nm, in the red wing of a much stronger Fe\,\textsc{i} line. It is also blended by Co\,\textsc{i} and V\,\textsc{i} lines, both with about the same wavelengths.

The best oscillator strength for the \thii line comes from the accurate lifetime+branching fraction results of Nilsson et al.\ (\cite{nils3}).  Atomic data (wavelengths, HFS and $gf$-values) concerning the blends are known from the works of Learner et al.\ (\cite{learn}), Lawler et al.\ (\cite{law9}) and Pickering \& Semeniuk (\cite{pick}).  Although the $gf$-value for the Co\,\textsc{i} line ($\log gf=-2.27\pm0.04$, $E_\mathrm{exc}=2.28$\,eV) is very accurate (Lawler et al.\ \cite{law9}), the value for the V\,\textsc{i} blend ($E_\mathrm{exc}=1.80$\,eV) is probably less well known, even though Pickering \& Semeniuk give $\log gf= -1.30\pm0.04$.  The stated uncertainty is for the relative value with respect to the $gf$-values of two stronger V\,\textsc{i} lines, to which Pickering \& Semeniuk compared this line. The absolute $gf$-values of these stronger lines come from from Kurucz (\cite{kur}), and are only known to about $\pm$0.08\,dex.

We measured the total equivalent width of the feature containing \altthii.  We also computed the contributions of the Co\,\textsc{i} and V\,\textsc{i} blends from the $gf$-values above, and the abundances derived in Paper II, employing a number of different solar photospheric model atmospheres.  The equivalent widths of the two blends predicted in this way are almost model-independent. We found predicted equivalent widths of 0.208\,pm for the Co\,\textsc{i} blend, and 0.038\,pm for the V\,\textsc{i} blend. As the total measured equivalent width is 0.56\,pm, this leaves 0.314\,pm for \altthii. We estimate the uncertainty of the \thii contribution to be of order 20\%, based on the uncertainty of the total measured equivalent width and the uncertainties of the Co\,\textsc{i} and V\,\textsc{i} contributions.

Mashonkina et al.\ (\cite{mash12}) estimated small NLTE corrections of between $+0.06$\,dex ($S_\mathrm{H}=0$) and $0.00$\,dex ($S_\mathrm{H}=1$) for this \thii line, using the \marcs\ model atmosphere.  We adopt $\Delta_{\rm NLTE}=+0.01$\,dex, corresponding to $S_\mathrm{H}=0.1$ (Mashonkina et al.'s preferred value).  This NLTE correction was not available for AGSS09.

\begin{figure*}[p]
\centering
\begin{minipage}[t]{0.39\textwidth}
\centering
\includegraphics[width=\linewidth]{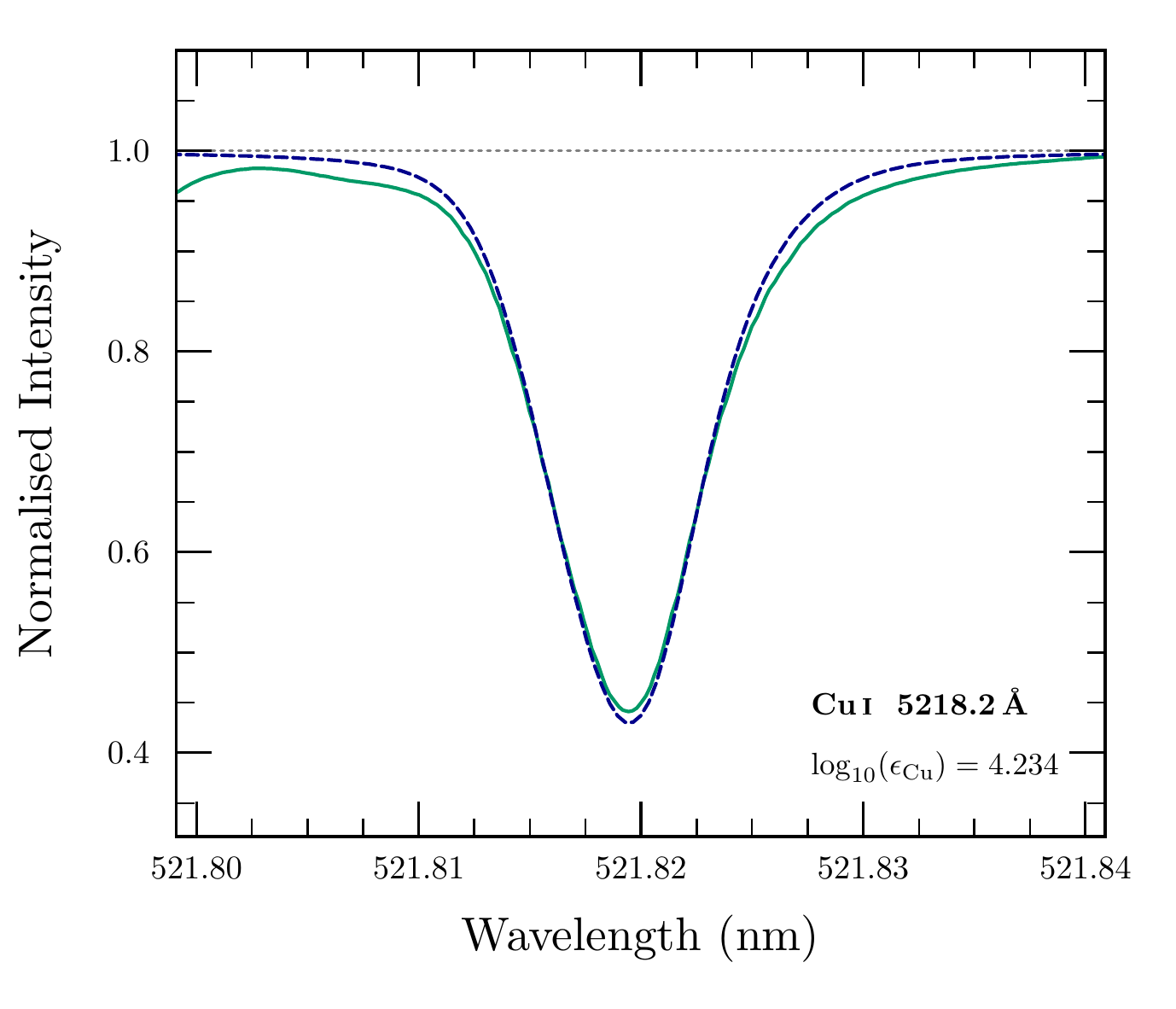}
\includegraphics[width=\linewidth]{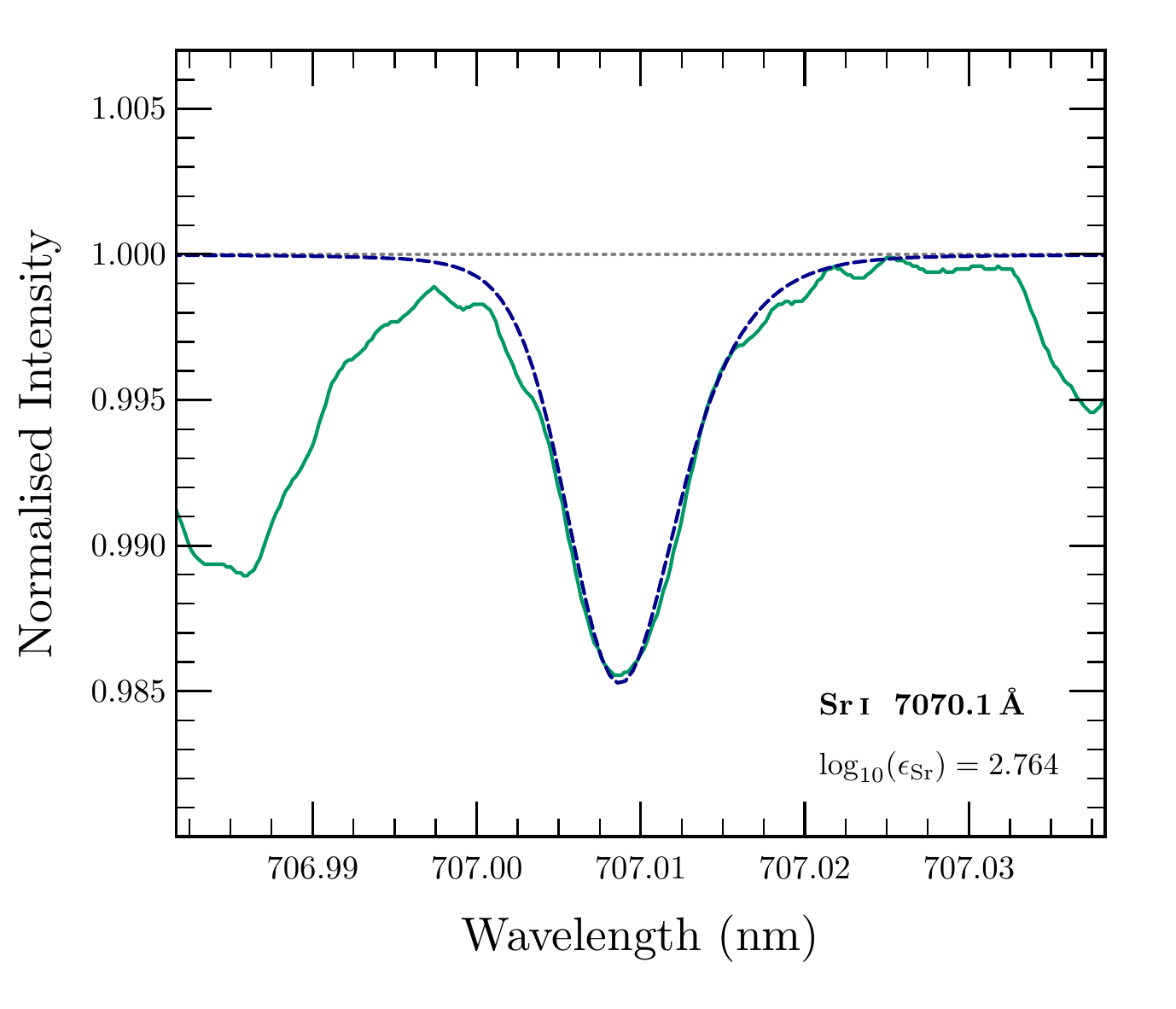}
\includegraphics[width=\linewidth]{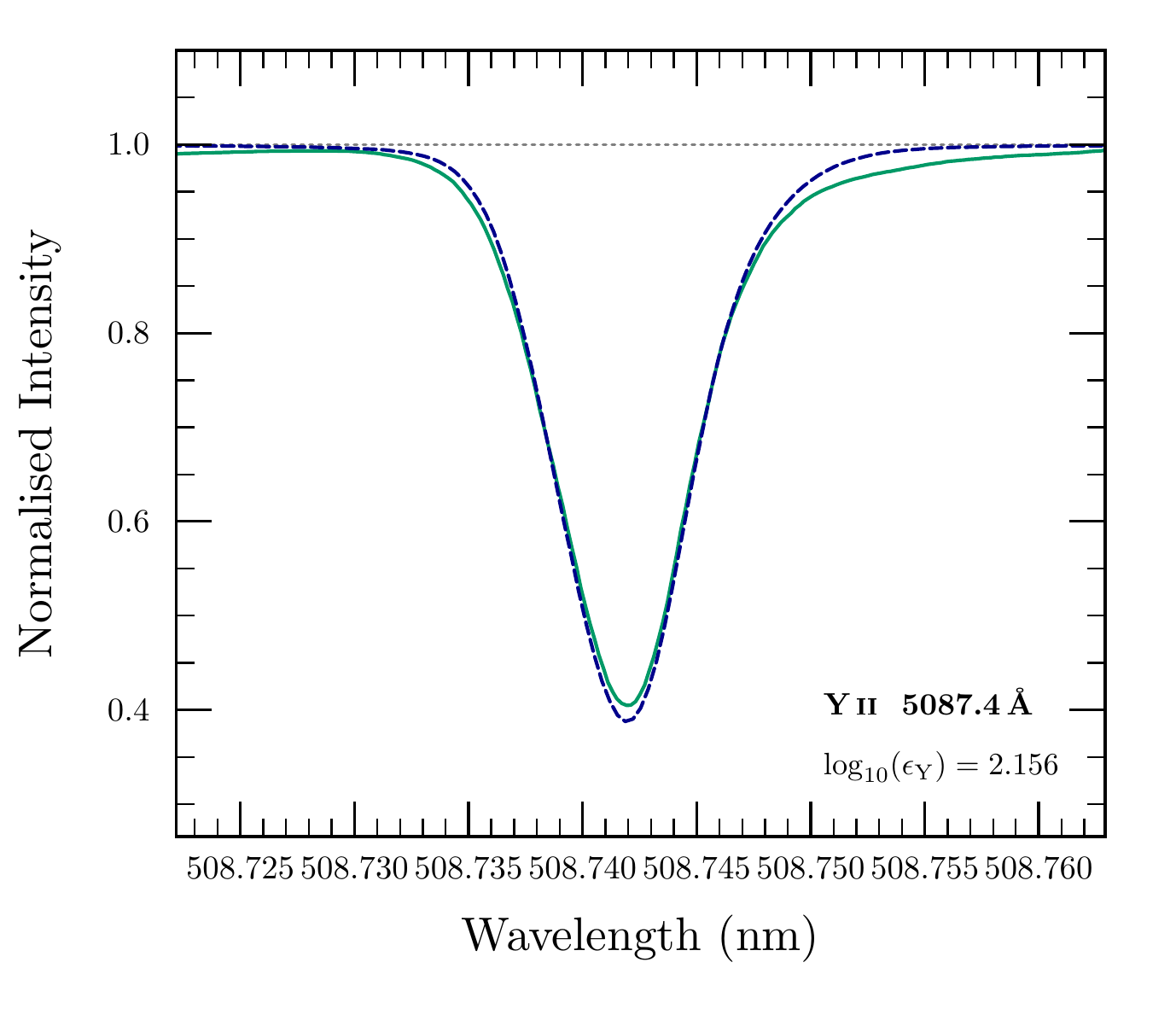}
\end{minipage}
\hspace{0.05\textwidth}
\begin{minipage}[t]{0.39\textwidth}
\centering
\includegraphics[width=\linewidth]{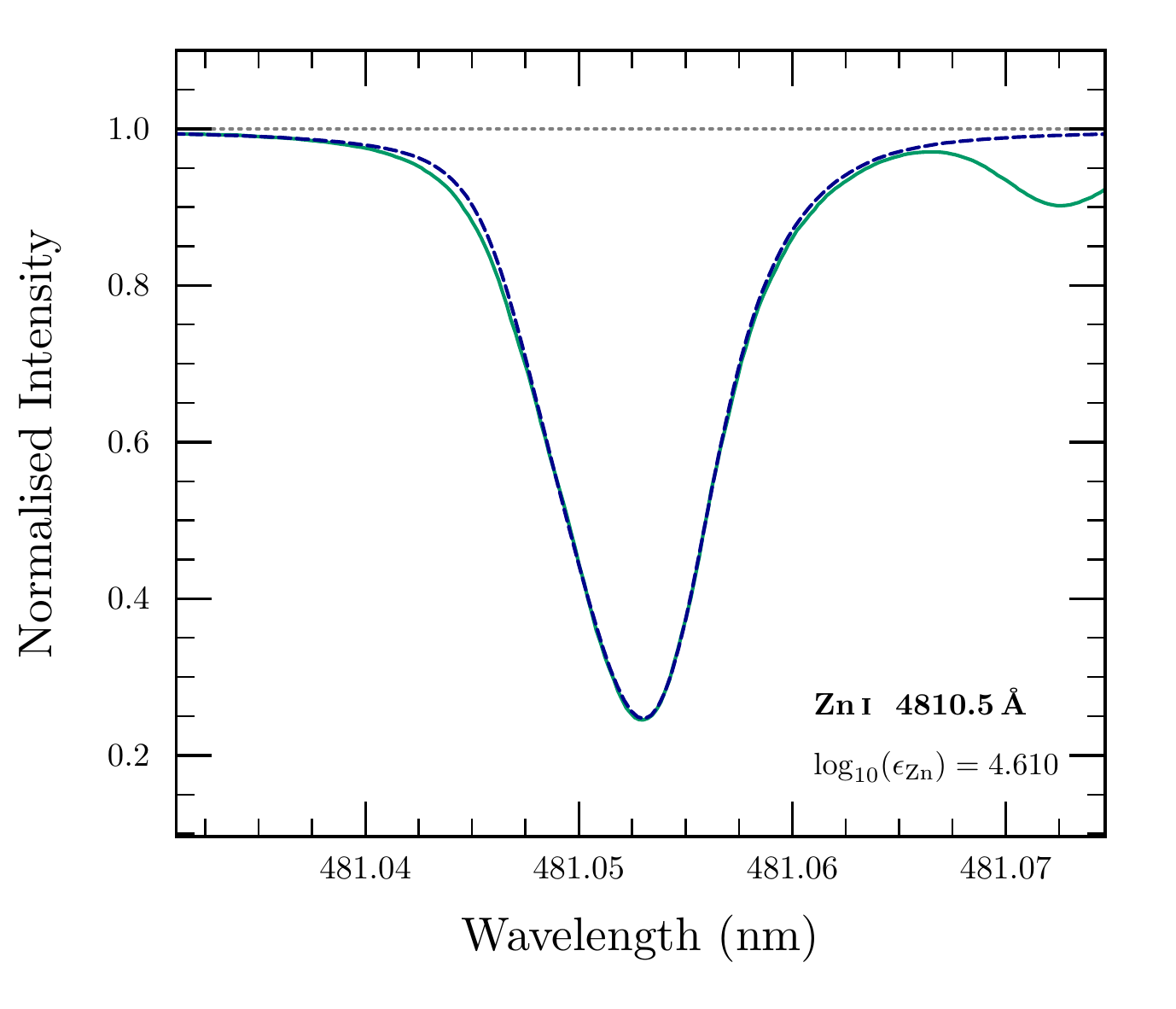}
\includegraphics[width=\linewidth]{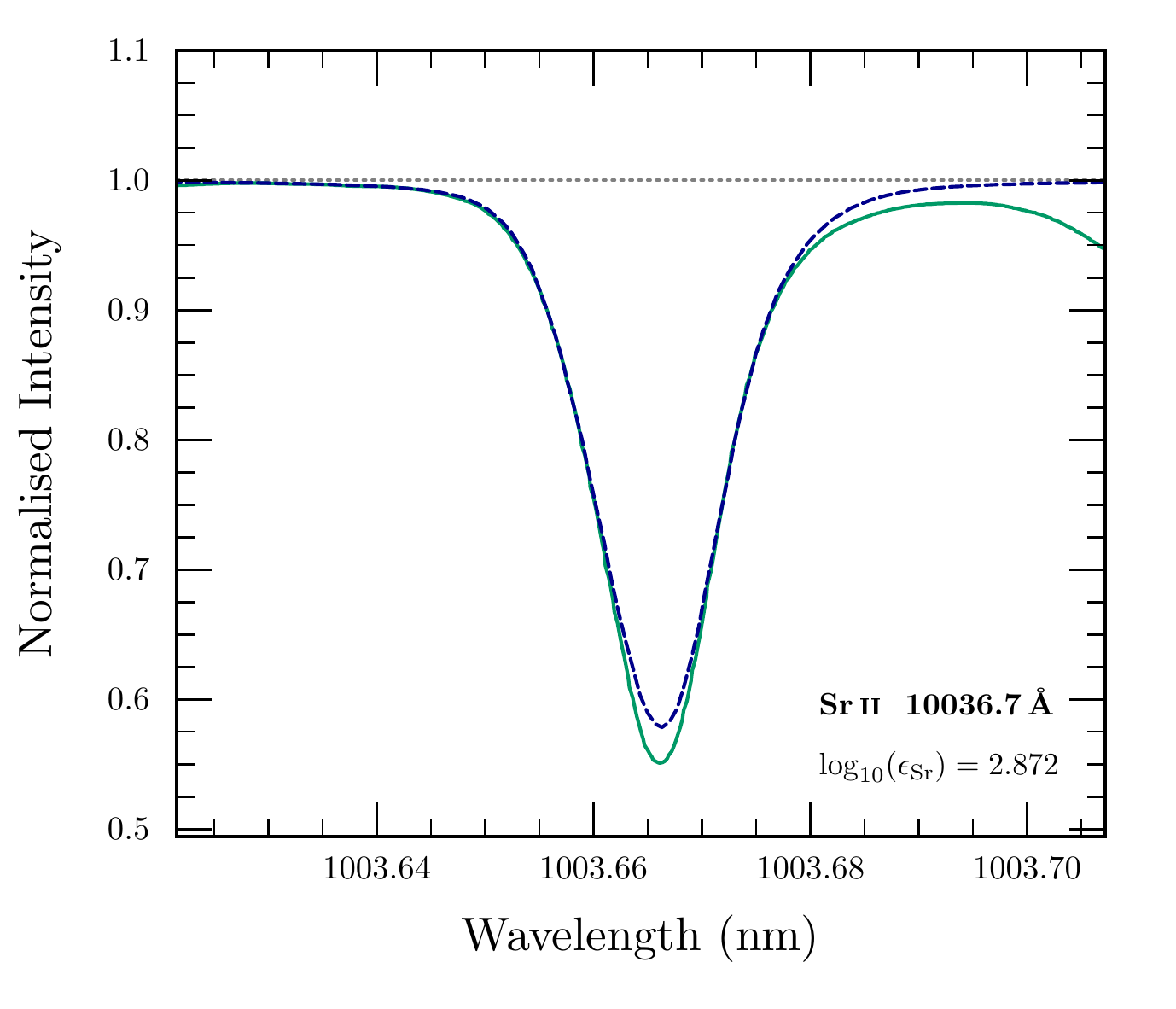}
\includegraphics[width=\linewidth]{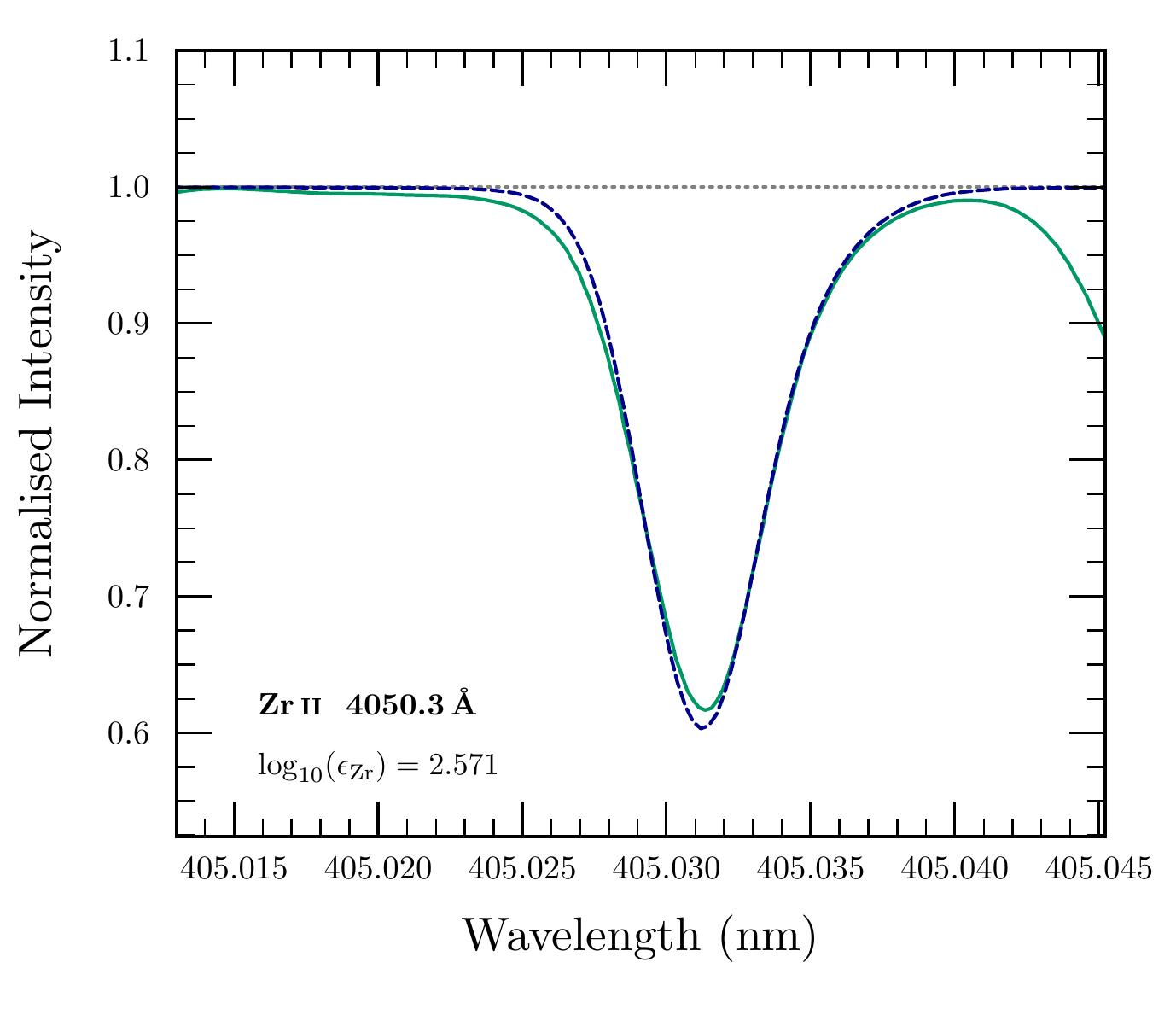}
\end{minipage}
\includegraphics[width=0.39\linewidth]{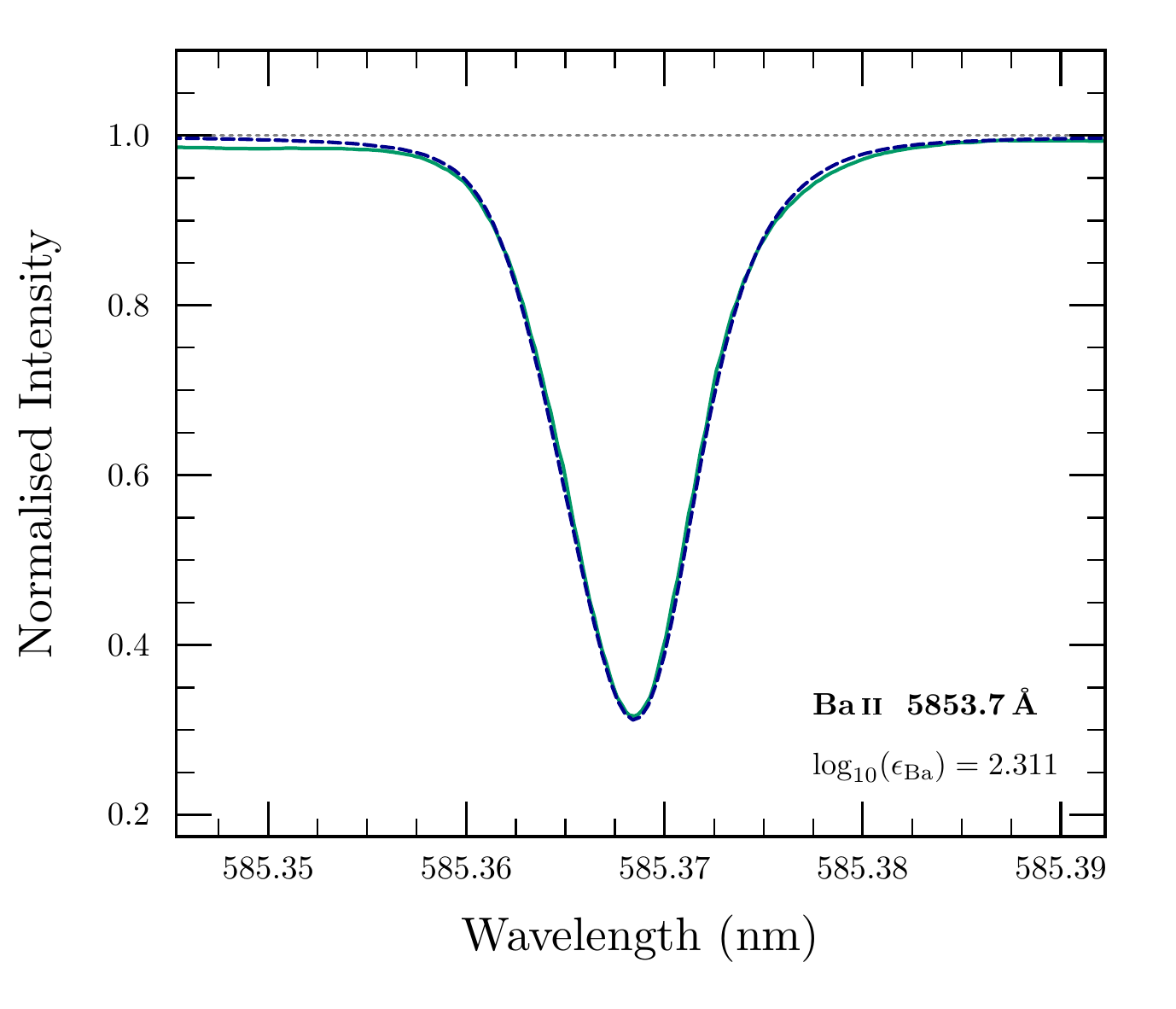}
\caption{Example spatially and temporally averaged, disk-centre synthesised line profiles (blue dashed), shown in comparison to the observed FTS profile (solid green).  The solar gravitational redshift was removed from the FTS spectrum.  The synthesised profiles have been convolved with an instrumental sinc function and fitted in abundance; wavelength and intensity normalisations have been adjusted for display purposes. Note that the plotted profiles are the 3D LTE ones, whereas the quoted abundance in each panel accounts also for the predicted 1D NLTE correction (if available).}
\label{fig:profiles}
\end{figure*}

\begin{figure*}[p]
\centering
\begin{minipage}[t]{0.4\textwidth}
\centering
\includegraphics[width=\linewidth]{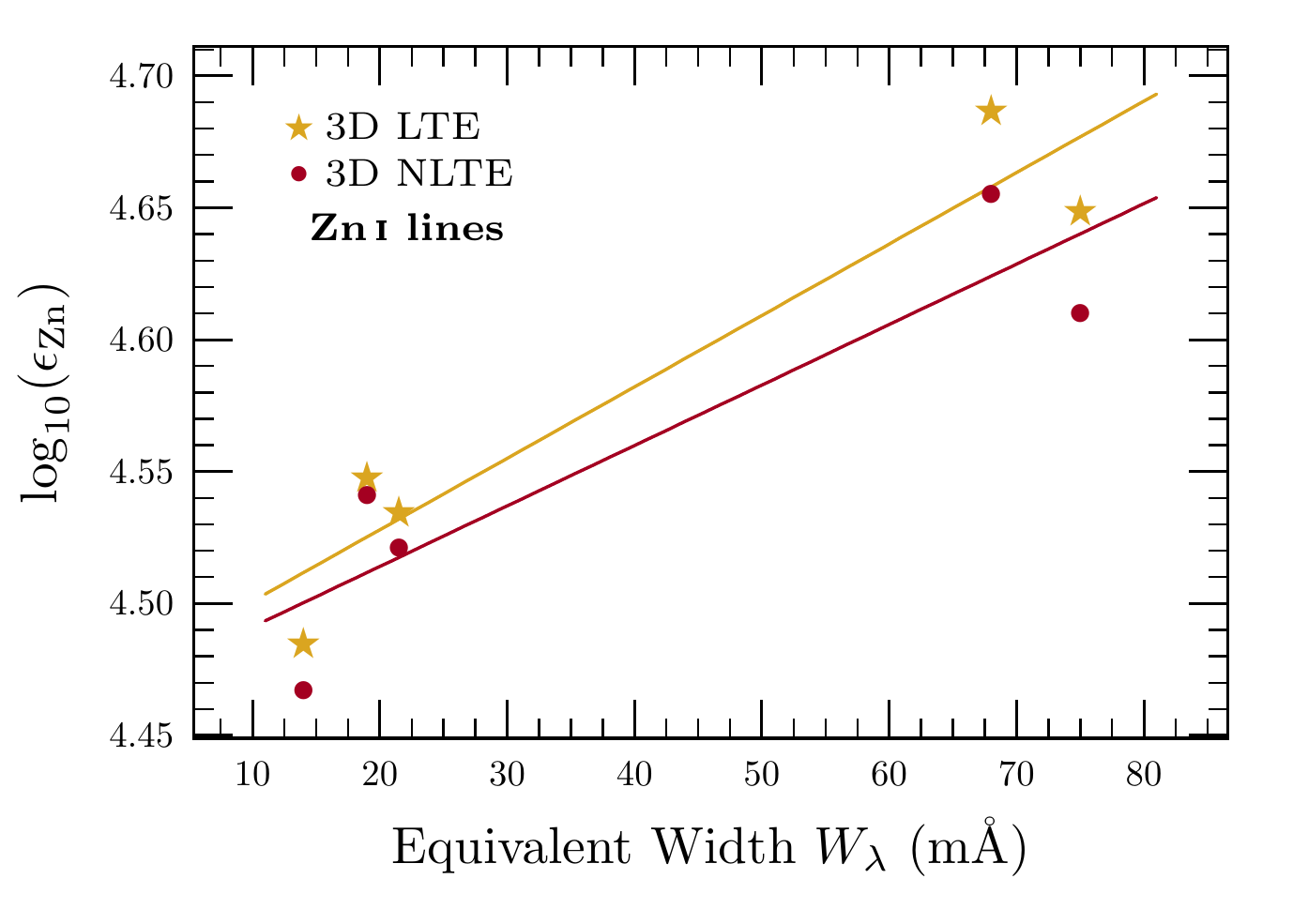}
\includegraphics[width=\linewidth]{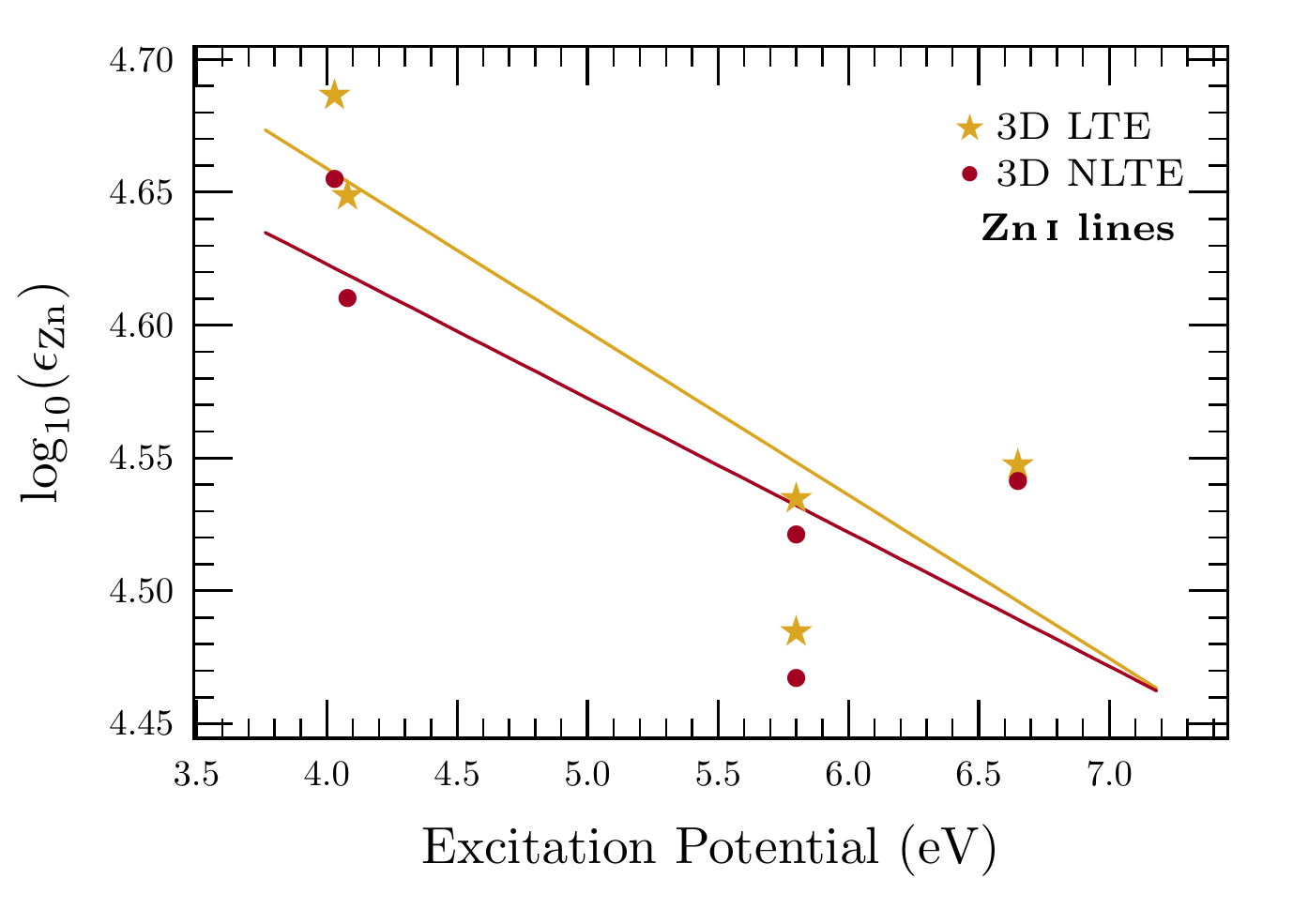}
\end{minipage}
\hspace{0.05\textwidth}
\begin{minipage}[t]{0.4\textwidth}
\centering
\includegraphics[width=\linewidth]{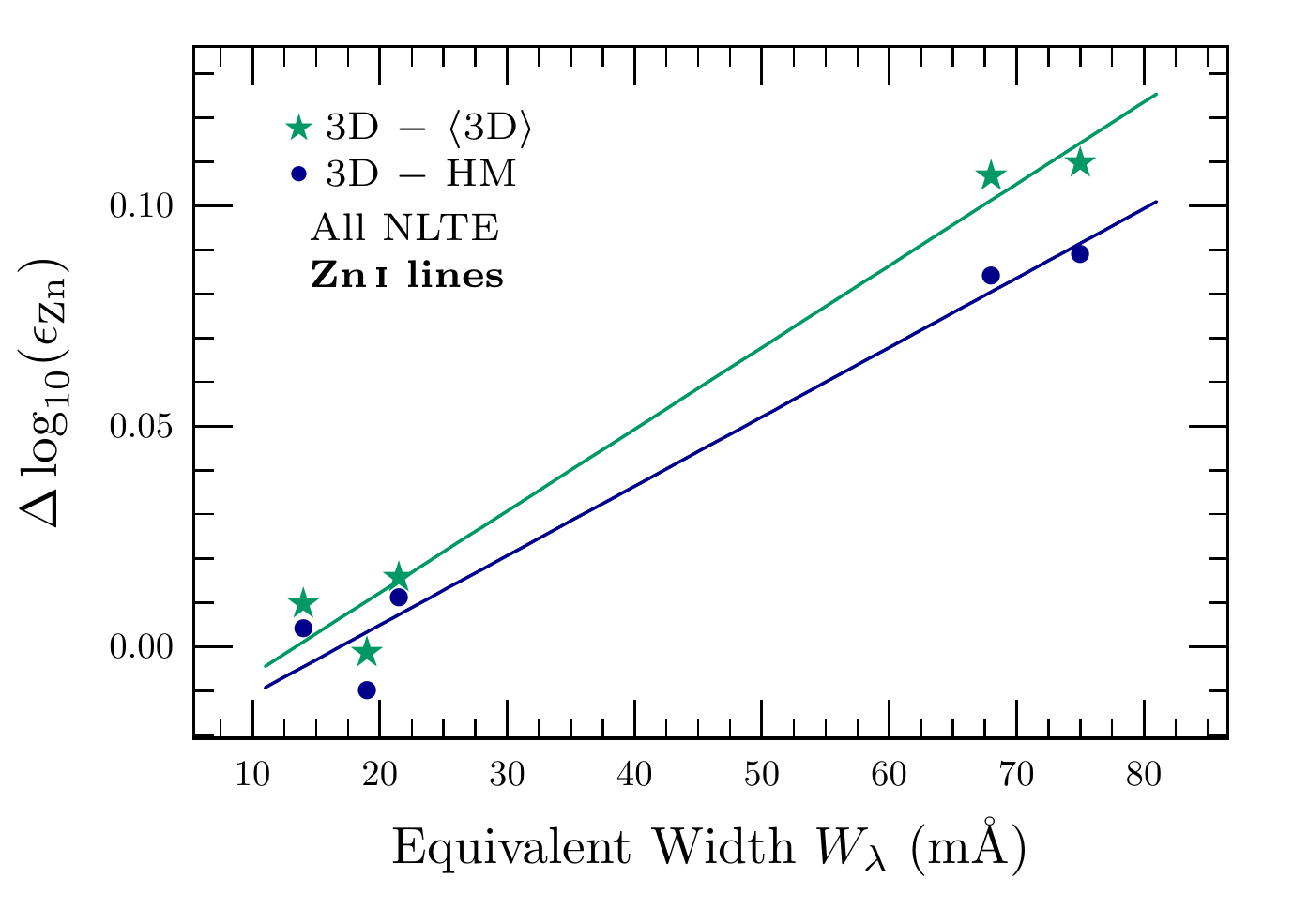}
\includegraphics[width=\linewidth]{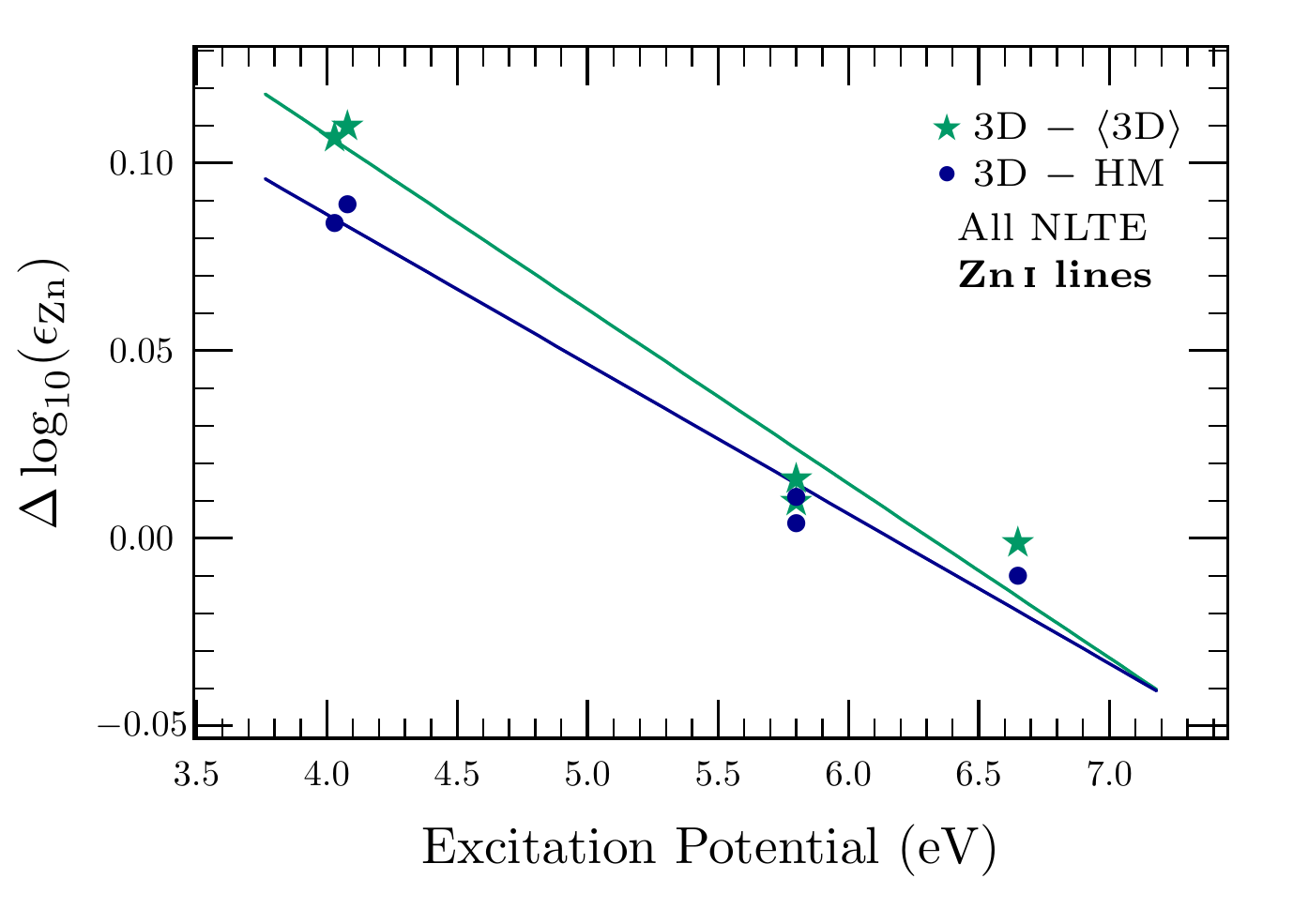}
\end{minipage}
\caption{\textit{Left}: Zn abundances derived from \zni lines with the 3D model, shown as a function of equivalent width and lower excitation potential.  \textit{Right}: Line-by-line differences between Zn abundances obtained with the 3D and \oneDAV\ models, and between those obtained with the 3D and HM models.  Trendlines give equal weight to each line (unlike our mean abundances, where we give larger weights to better lines).}
\label{fig:zn}
\end{figure*}

\begin{figure*}[p]
\centering
\begin{minipage}[t]{0.4\textwidth}
\centering
\includegraphics[width=\linewidth]{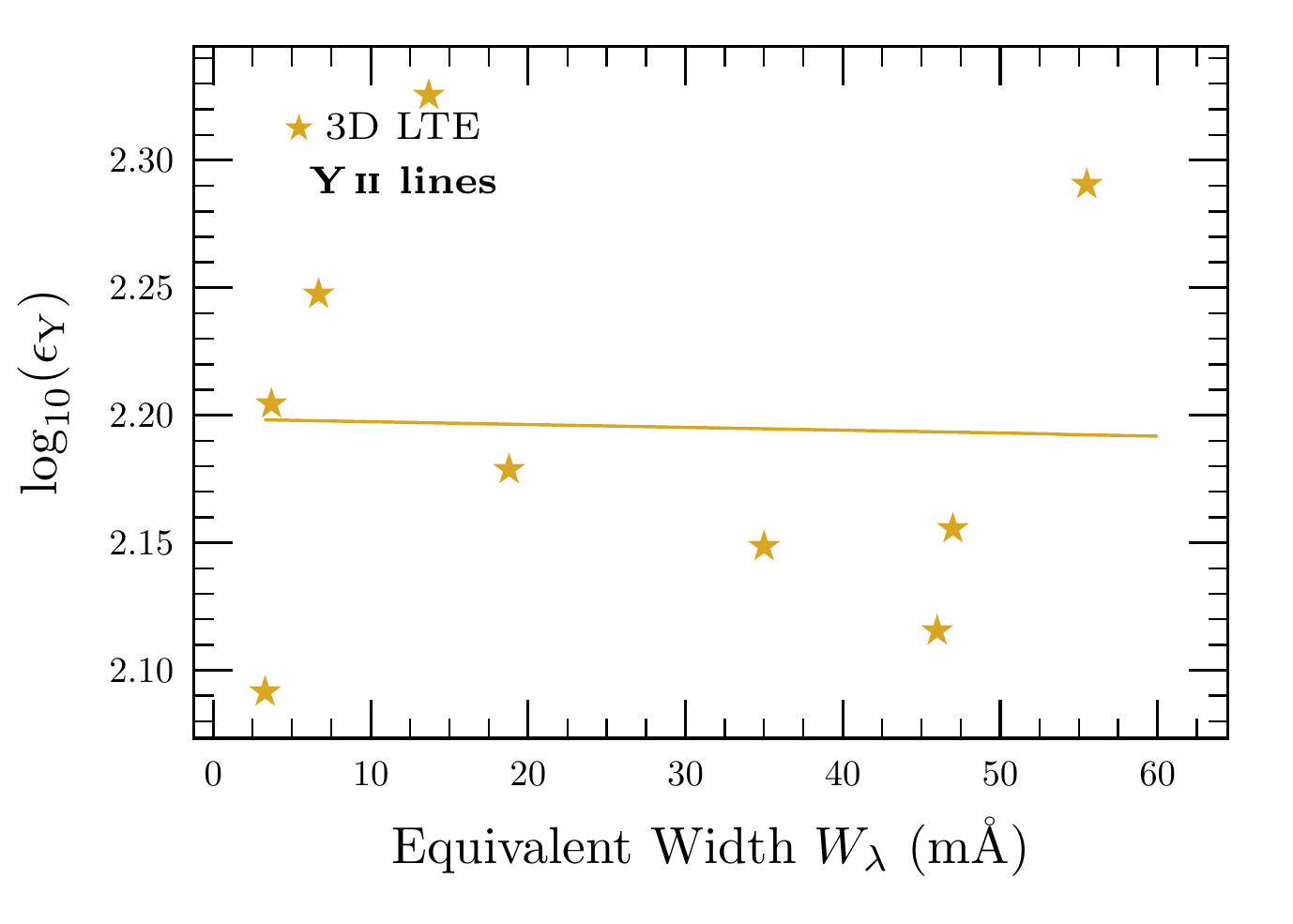}
\includegraphics[width=\linewidth]{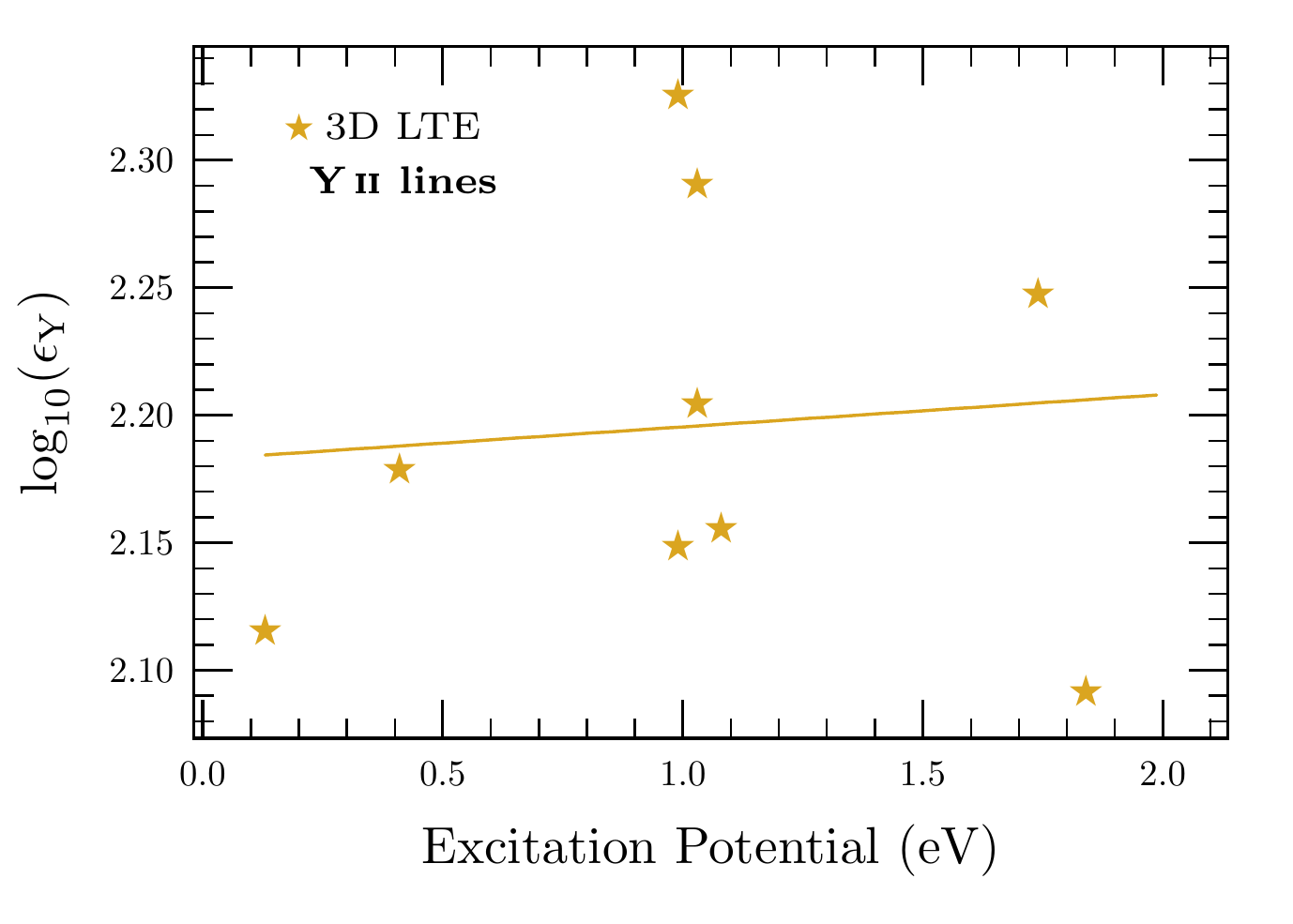}
\end{minipage}
\hspace{0.05\textwidth}
\begin{minipage}[t]{0.4\textwidth}
\centering
\includegraphics[width=\linewidth]{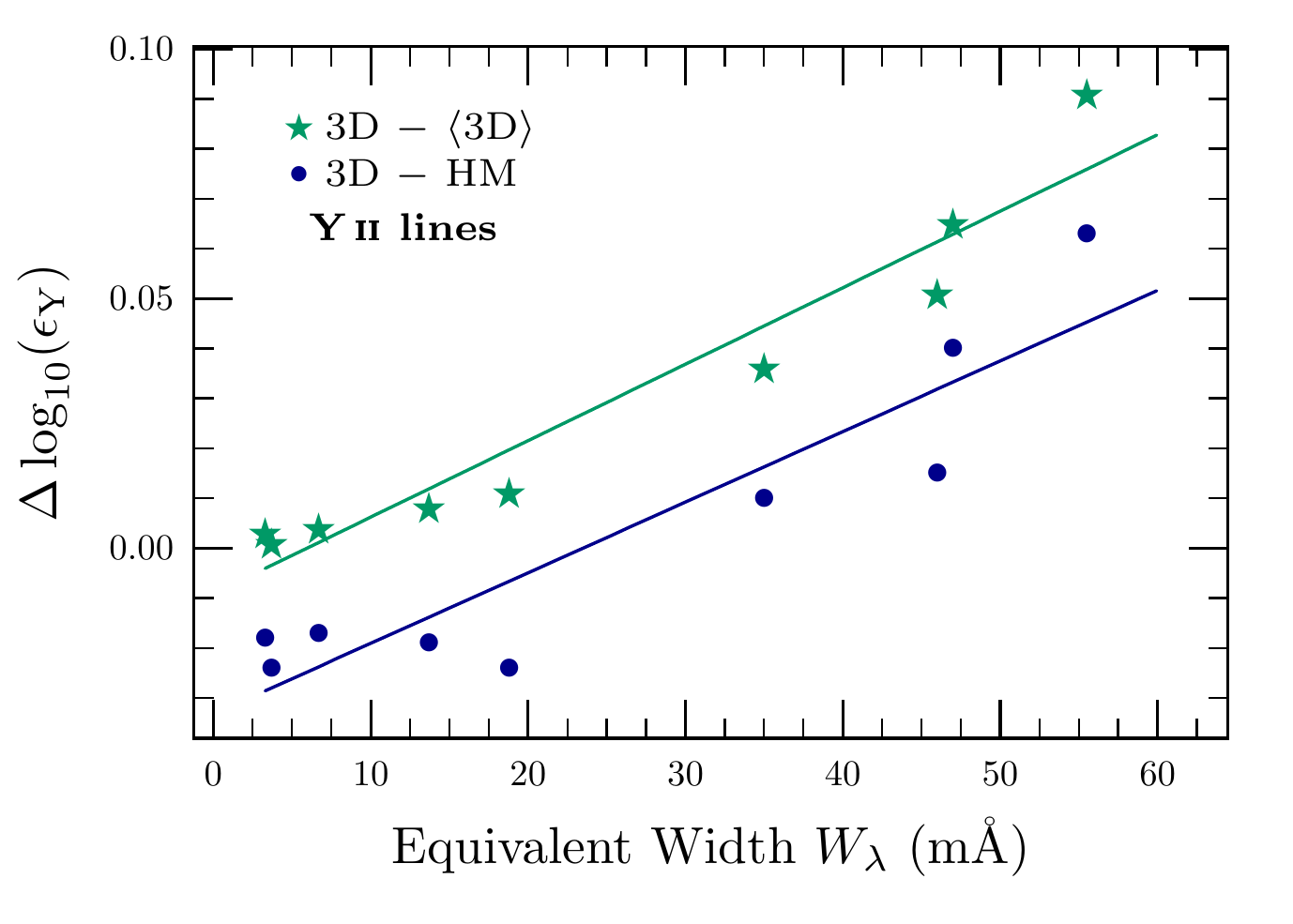}
\includegraphics[width=\linewidth]{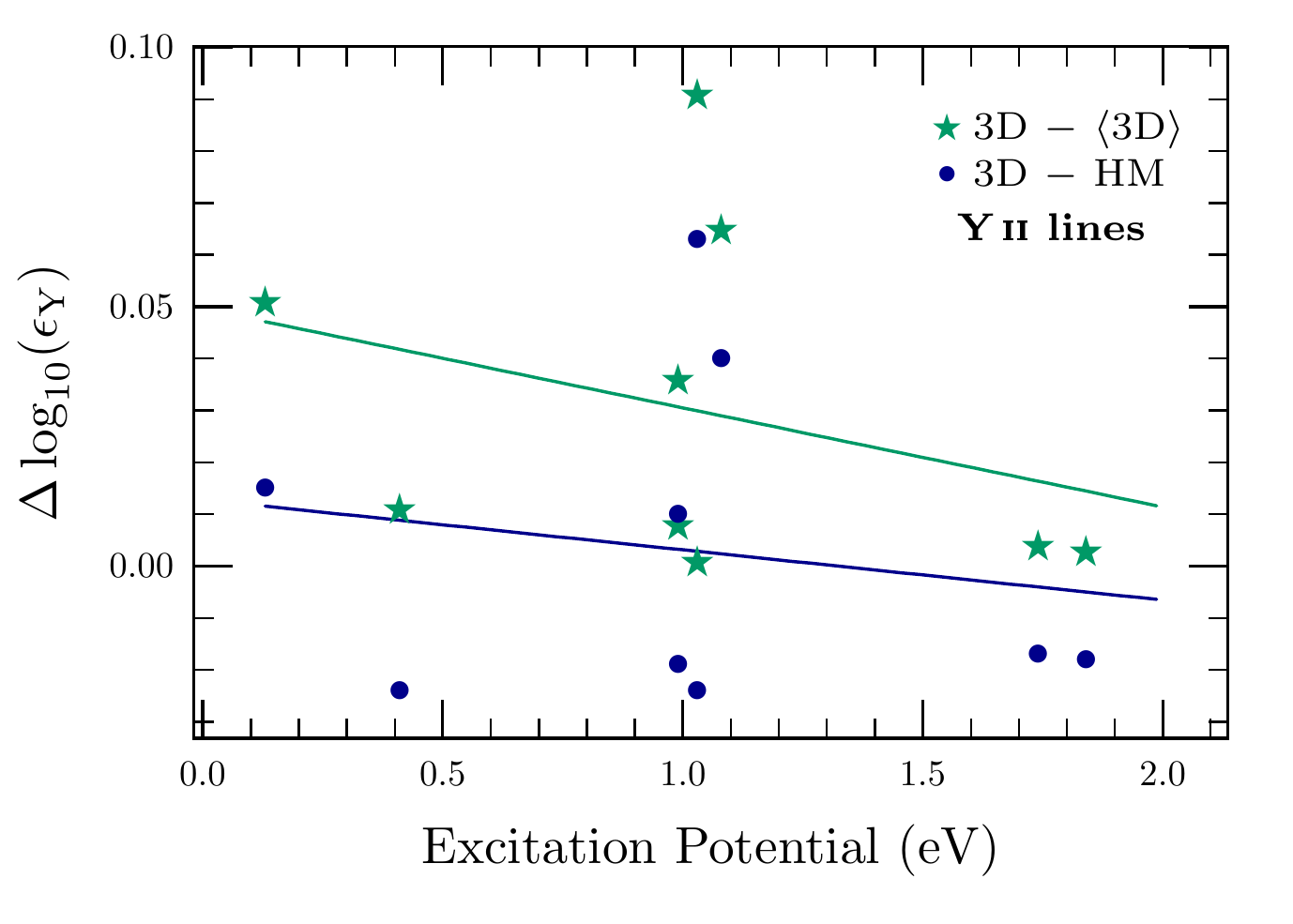}
\end{minipage}
\caption{\textit{Left}: Y abundances from \yii lines in our 3D LTE analysis, as a function of equivalent width and lower excitation potential.  \textit{Right}: Line-by-line differences between abundances obtained with the 3D and \oneDAV\ models, and between those obtained with the 3D and HM models.}
\label{fig:y}
\end{figure*}

\begin{figure*}
\centering
\begin{minipage}[t]{0.4\textwidth}
\centering
\includegraphics[width=\linewidth]{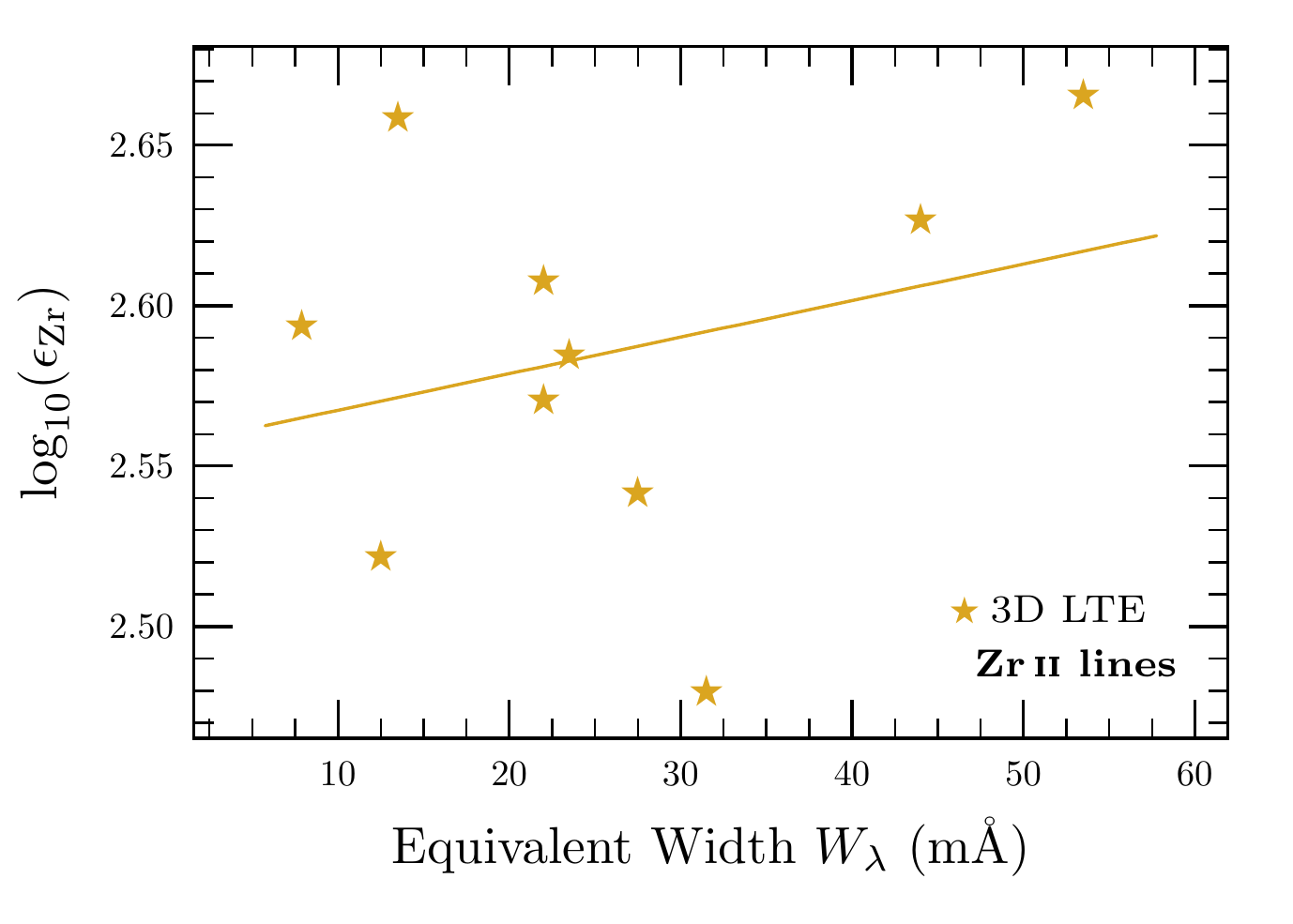}
\end{minipage}
\hspace{0.05\textwidth}
\begin{minipage}[t]{0.4\textwidth}
\centering
\includegraphics[width=\linewidth]{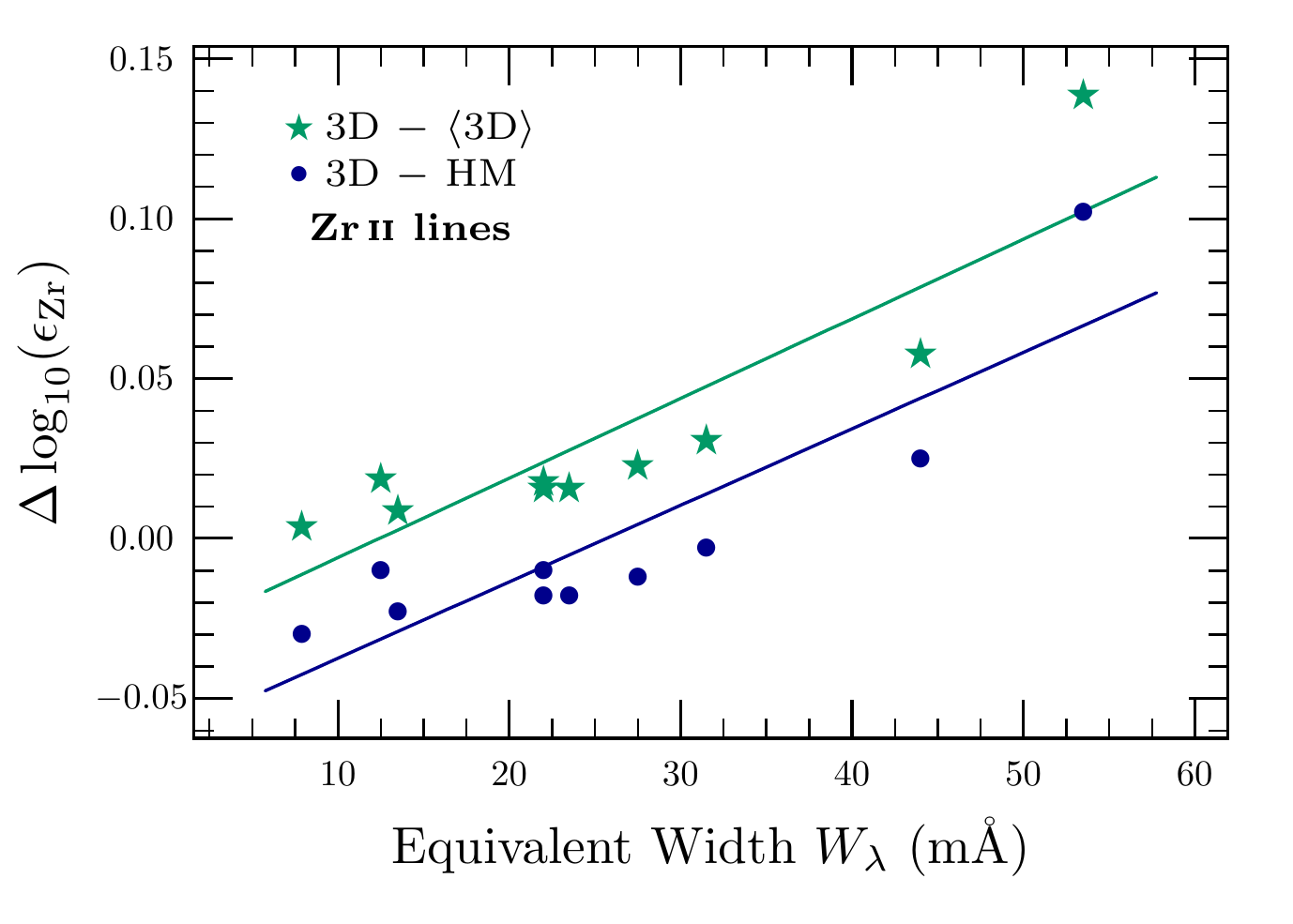}
\end{minipage}
\caption{\textit{Left}:  Zr abundances from \zrii lines for the 3D LTE case, as a function of equivalent width.  \textit{Right}: Line-by-line differences between abundances obtained with the 3D and \oneDAV\ models, and between those obtained with the 3D and HM models.}
\label{fig:zr}
\end{figure*}

\section{Derived solar elemental abundances}
\label{results}

The detailed results for the different model atmospheres we used are presented in Table~\ref{table:lines}, with each element individually discussed below.  Line profiles produced in 3D generally show very good agreement with the observed spectrum; some examples are given in Fig.~\ref{fig:profiles}.

Besides statistical errors in our derived abundances (which we specify as the standard deviation of the mean), we quantify three possible systematic errors arising from potential problems in the atmospheric and atomic modelling: departures from LTE, the mean photospheric temperature structure and atmospheric inhomogeneities.  We add these in quadrature to the statistical errors to estimate overall uncertainties. Full details can be found in Paper I.  

In the following we compare our derived abundances to CI chondritic abundances taken from the recent careful compilation and analysis of Lodders et al.\ (\cite{lodd}).  As we did in AGSS09, we renormalise these data to the photospheric abundance of silicon determined in Paper I ($\log\epsilon_\mathrm{Si}=7.51$).

Below we often refer to the ``3D effect'' on the derived elemental abundances. We define this 3D effect as $\log\epsilon_{\rm 3D}-\log\epsilon_{\rm HM}$, i.e.\ the correction to bring the abundance computed with the HM model into agreement with the abundance obtained with our 3D model.  Although this is not strictly just the effect of using a 3D model atmosphere rather than a 1D one (see Sec.\ \ref{discussion} for a discussion), we use this definition because the HM model has been the de facto standard for abundance determinations in past decades. We note that other authors prefer to define the 3D effect as $\log\epsilon_{\rm 3D}-\log\epsilon_{\oneDAVmath}$ to isolate the impact of the atmospheric inhomogeneities. We argue however that differences in the mean stratification arising from the treatment of convection should be considered when evaluating the full impact of using 3D models rather than 1D ones.

\subsection{Copper}
Our 3D+NLTE copper abundance is 
\begin{displaymath}
\log\epsilon_{\mathrm{Cu}}= 4.18\pm 0.05\ (\pm0.02\ \mathrm{stat},\ \pm0.04\ \mathrm{sys}).
\end{displaymath}
The 3D effect on these lines is rather small; with the HM model the abundance is 4.21. Interestingly, accounting for departures from LTE as computed by Shi et al.\ (\cite{Shi14}) slightly increases the line-to-line scatter  ($\sigma = 0.03$ and $0.05$\,dex, respectively) but the net effect on the Cu abundance is small ($-0.01$\,dex). Our Cu abundances in 3D and with HM are not too different from the results of Shi et al.\ (\cite{Shi14}; $4.19 \pm 0.10$, from ten lines), Kock \& Richter (\cite{kock}; $4.16 \pm 0.08$, from six lines) and Sneden \& Crocker (\cite{sned2}; 4.12, from two lines).  We do not reproduce the rather large difference in abundance found by Shi et al.\ (\cite{Shi14}) between the low and high excitation lines with the HM model. However, our sample only includes one of the low excitation lines used by those authors.  Our adopted value agrees reasonably well with the meteoritic abundance ($\log\epsilon_{\mathrm{Cu}}=4.25\pm0.04$; Lodders et al.\ \cite{lodd}).

\subsection{Zinc}
Our 3D+NLTE zinc abundance is 
\begin{displaymath}
\log \epsilon_{\mathrm{Zn}}=4.56\pm0.05\ (\pm0.03\ \mathrm{stat},\ \pm0.04\ \mathrm{sys}). 
\end{displaymath}
We note that the abundance only varies by 0.026\,dex when we go from $S_\mathrm{H}=0.1$ to 1. The mean NLTE correction is also very small: $-0.02$\,dex. \zni is one of the few species where the HM model gives a slightly lower abundance than the 3D model: $\log \epsilon_{\mathrm{Zn}}=4.53$.  Referring to Fig.~\ref{fig:zn}, we see that this is essentially due to two strong lines only, indicating that the microturbulence adopted in 1D is probably too high, or possibly that Doppler broadening due to convection is insufficient in the 3D case.  Curiously, these two lines return an abundance approximately $0.1$\,dex larger with the 3D model than any 1D model, including \oneDAV.  This leads to pronounced trends with line strength and excitation potential, although the significance of the trends is debatable given the small number of lines.  These are precisely the two lines for which we renormalised oscillator strengths to the scale of Kerkhoff et al.\ (\cite{Kerkhoff80}), but this operation amounts to a change of just $0.01$\,dex, so cannot explain the $0.1$\,dex offset. Another possibility is that the NLTE corrections may have been underestimated for these high-forming lines, given that they were computed using a 1D model atmosphere, but it seems difficult to believe that this could be a $0.1$\,dex effect.

Our results are in good agreement with the LTE result of Bi\'{e}mont \& Godefroid (\cite{biem}; $\log \epsilon_{\mathrm{Zn}}=4.56$ with the HM model), and with Sneden \& Crocker's result (\cite{sned2}; $\log\epsilon_{\mathrm{Zn}}=4.62$, based on two lines only).  Our final abundance also overlaps the meteoritic one ($\log \epsilon_{\mathrm{Zn}}=4.63\pm0.04$; Lodders et al.\ \cite{lodd}).

\subsection{Gallium}
For Ga we find 
\begin{displaymath}
\log\epsilon_{\mathrm{Ga}}=3.02\pm0.05\ \mathrm{(pure\ systematic\ error)}
\end{displaymath}
from the sole \gai\ 417.2\,nm line.
This value is about $0.2$\,dex higher than the earlier estimates by Ross \& Aller (\cite{ross}) and Lambert et al.\ (\cite{lamb}), but agrees with the meteoritic abundance ($\log\epsilon_{\mathrm{Ga}}=3.08\pm0.02$; Lodders et al.\ \cite{lodd}) to within the mutual uncertainties.

\subsection{Germanium}
For Ge we derive 
\begin{displaymath}
\log \epsilon_{\mathrm{Ge}}= 3.63\pm0.07\ \mathrm{(pure\ systematic\ error)},
\end{displaymath}
in reasonable agreement with Bi\'emont et al.\ (\cite{biem2}) and the meteoritic abundance ($\log\epsilon_{\mathrm{Ge}}=3.58\pm0.04$; Lodders et al.\ \cite{lodd}).

\subsection{Arsenic}
We rechecked the two lines at 299.1 and 303.2\,nm used previously by Gopka et al.\ (\cite{gopka}) and could not derive any meaningful abundance, because of the very heavy blending. For this reason, and the unreliable nature of the $gf$-values for these lines, we conclude that the As abundance cannot be reliably derived from the solar photospheric spectrum.

\subsection{Rubidium}
With our chosen data, we find
\begin{displaymath}
\log\epsilon_{\mathrm{Rb}}=2.47\pm0.07\ (\pm0.06\ \mathrm{stat},\ \pm0.05\ \mathrm{sys}).
\end{displaymath}
This value is 0.10\,dex lower than the abundance obtained with the HM model. We note however that the 780.0\,nm line leads to an abundance 0.11\,dex larger than the 794.7\,nm line, suggesting that an unknown blend might contribute about 25$\%$ of its equivalent width.

Lambert \& Mallia (\cite{lamb3}) and Hauge (\cite{hauge}) analysed these same two lines.  Hauge found $\log\epsilon_{\mathrm{Rb}}=2.60$, in pretty good agreement with Lambert \& Mallia's $\log\epsilon_{\mathrm{Rb}}=2.63$.  Our derived Rb abundance is quite a bit lower than both of these values but still significantly above the CI meteoritic value ($\log\epsilon_{\mathrm{Rb}}=2.36\pm0.03$; Lodders et al.\ \cite{lodd}).

\subsection{Strontium}
Gratton \& Sneden (\cite{grat}) concluded that the \sri lines lead to an abundance of $\log\epsilon_{\mathrm{Sr}}=2.75$, much smaller than the \srii lines ($\log\epsilon_{\mathrm{Sr}}=2.97$).  However, Barklem \& O'Mara (\cite{bark2}) found good agreement between \sri and \srii with the same HM atmospheric model, using their new damping parameters.  Their final abundance was $\log\epsilon_{\mathrm{Sr}}=2.92\pm0.05$. 

Our 3D+NLTE results are $\log\epsilon_{\mathrm{Sr}}=2.80\pm0.06$ ($1\sigma$ scatter, \altsri) and $\log\epsilon_{\mathrm{Sr}}=2.85\pm0.08$ ($1\sigma$ scatter, \altsrii).  Abundances from the two ionisation stages now agree very well in NLTE, whereas in LTE they differ by a factor of two: 2.69 (\altsri) versus 3.02 (\altsrii). The mean 3D+NLTE Sr abundance, from all \sri and \srii lines together, becomes
\begin{displaymath}
\log\epsilon_{\mathrm{Sr}}=2.83\pm0.06\ (\pm0.02\ \mathrm{stat},\ \pm0.05\ \mathrm{sys}).
\end{displaymath}
This result is substantially lower (by 0.09\,dex) than obtained by Barklem \& O'Mara, which is also true for our HM-based value. Our recommended value still overlaps the meteoritic Sr abundance ($\log \epsilon_{\mathrm{Zn}}=2.88\pm0.03$; Lodders et al.\ \cite{lodd}).

\subsection{Yttrium}
The new 3D LTE abundance is 
\begin{displaymath}
\log\epsilon_{\mathrm{Y}}=2.21\pm0.05\ (\pm0.03\ \mathrm{stat},\ \pm0.04\ \mathrm{sys}).
\end{displaymath}
As seen in Fig.~\ref{fig:y}, the 3D effect is quite small except for stronger lines, which is as expected for a dominant species like \altyii.  No appreciable trend can be seen in the 3D results, either in equivalent width or excitation potential.  Our result agrees with both that of Hannaford et al.\ (\cite{hann1}, $\log \epsilon_{\mathrm{Y}}=2.24\pm0.03$) and the meteoritic abundance ($\log \epsilon_{\mathrm{Y}}=2.17\pm0.04$; Lodders et al.\ \cite{lodd}).

\begin{figure*}
\centering
\begin{minipage}[t]{0.4\textwidth}
\centering
\includegraphics[width=\linewidth]{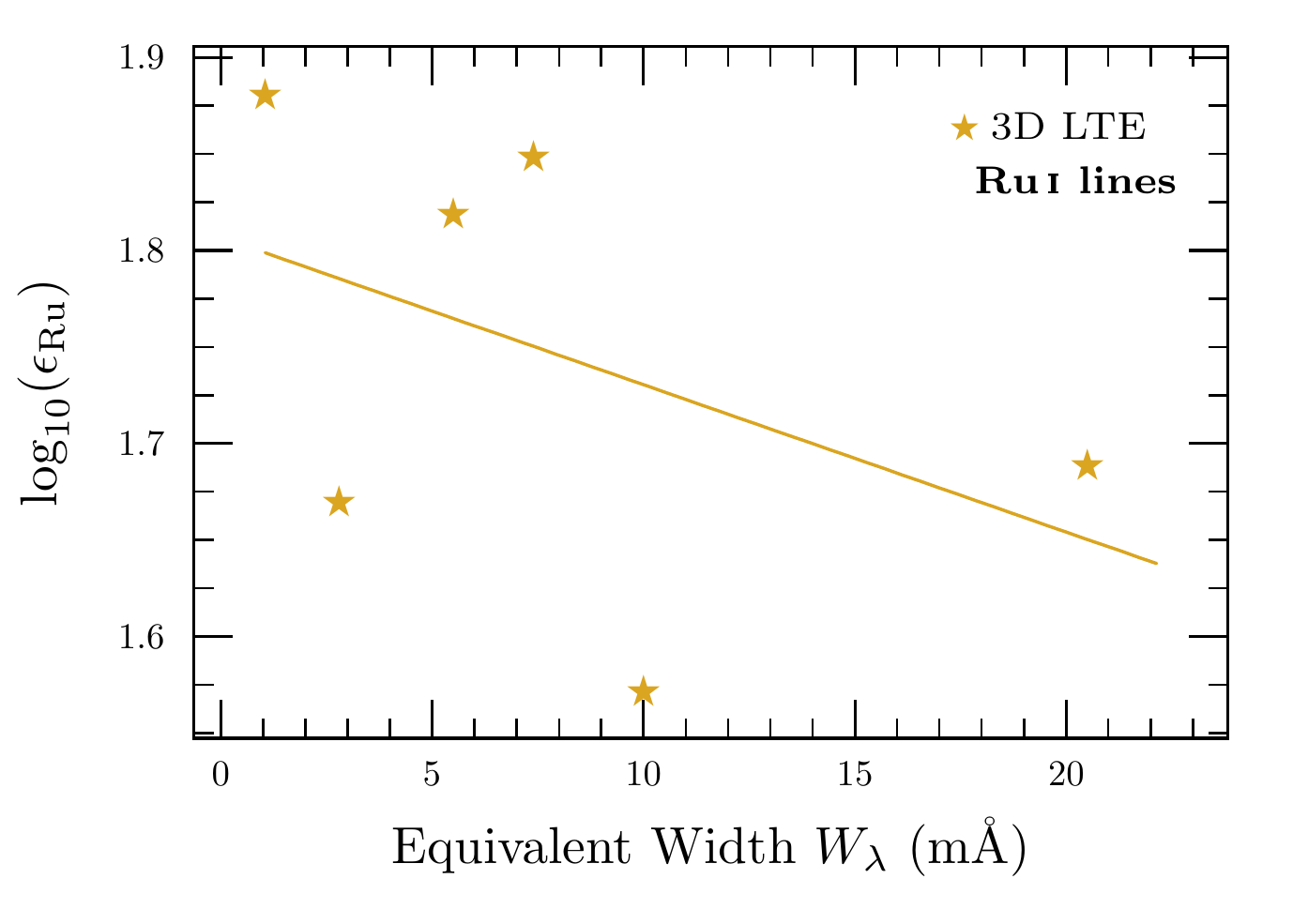}
\end{minipage}
\hspace{0.05\textwidth}
\begin{minipage}[t]{0.4\textwidth}
\centering
\includegraphics[width=\linewidth]{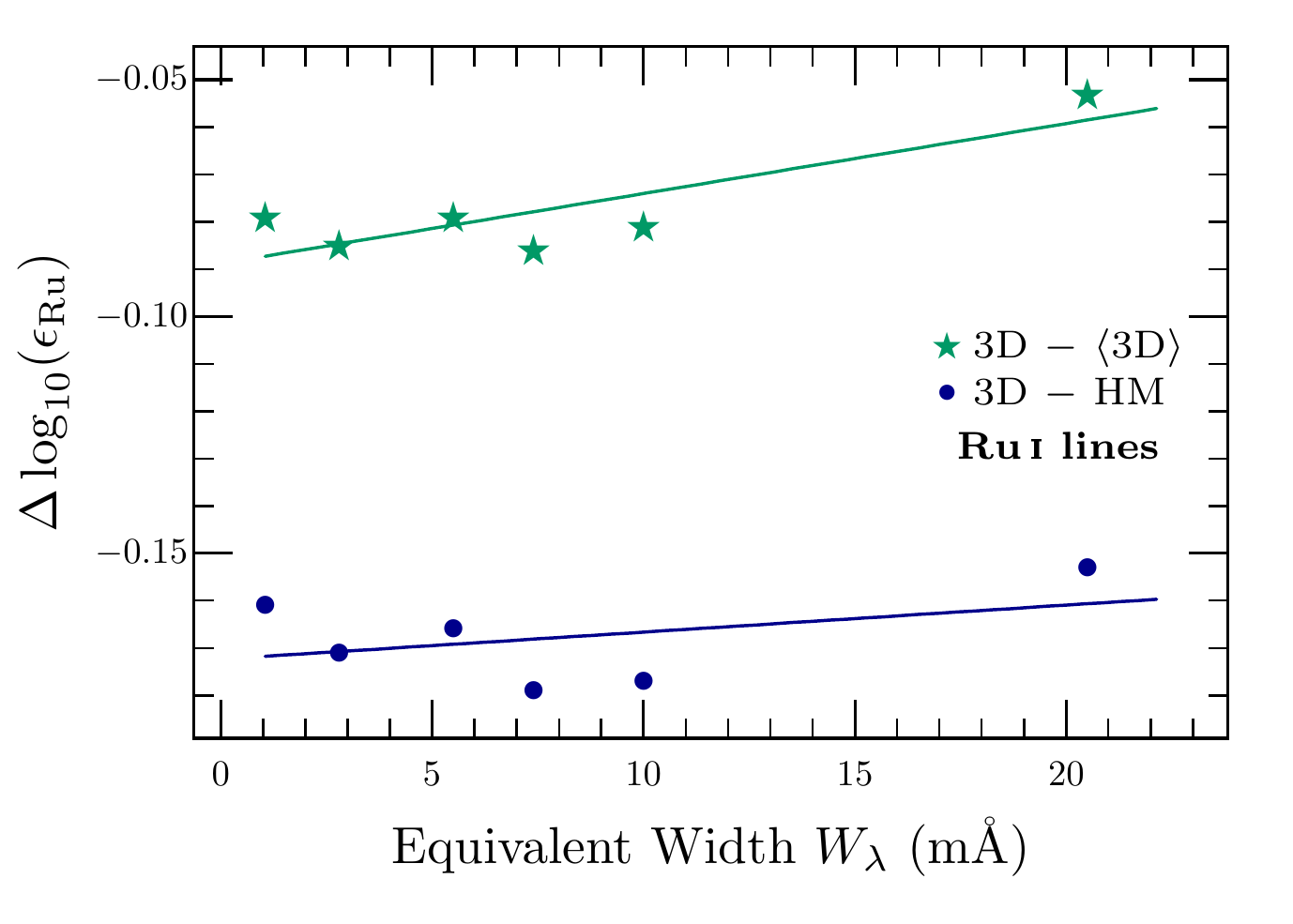}
\end{minipage}
\caption{\textit{Left}: Ru abundances in 3D LTE from \rui lines, as a function of equivalent width.  \textit{Right}: Line-by-line differences between abundances obtained with the 3D and \oneDAV\ models, and between those obtained with the 3D and HM models.}
\label{fig:ru}
\end{figure*}

\subsection{Zirconium}
\label{zrresults}
The 3D LTE result from our \zrii lines is 
\begin{displaymath}
\log\epsilon_{\mathrm{Zr}}=2.59\pm0.04\ (\pm0.01\ \mathrm{stat},\ \pm0.04\ \mathrm{sys}).
\end{displaymath}
We see from Fig.~\ref{fig:zr} and Table~\ref{table:lines} that abundances from \zrii lines, as expected for relatively high excitation lines of a dominant ionisation stage, are quite independent of the adopted photospheric model.  The use of the HM model for example leads to exactly the same value of the mean abundance.  From Table~\ref{table:lines}, we also see a very large scatter and strong model-dependence in the \zri results, as expected for a minor contributor.  If we postulate that the four very weak \zri lines that lead to the smallest abundance are the only \zri lines that are not blended, these lines return a 3D abundance of $\log\epsilon_{\mathrm{Zr}}=2.32\pm0.07$ (1$\sigma$), 0.27\,dex smaller than the value derived from the \zrii lines.  This difference may be explained by the very large NLTE effects on \zri lines (Velichko et al.\ \cite{veli}, \cite{veli2}).

Our new result is in perfect agreement with the Zr abundance found by Ljung et al.\ (\cite{ljung}), from a smaller sample of seven \zrii lines with wavelengths $>400$\,nm analysed in LTE with an older version of our 3D model.  It also overlaps the meteoritic value ($\log \epsilon_{\mathrm{Zr}}=2.53\pm0.04$; Lodders et al.\ \cite{lodd}) to within the mutual uncertainties.  Caffau et al.\ (\cite{caf1}) recently redetermined the solar abundance of Zr using their 3D model. Their 3D LTE result,
$\log\epsilon_{\mathrm{Zr}}=2.62\pm0.06$, from a sample of 15 \zrii lines, is slightly larger than ours. This can be explained by their use of the $gf$-values from Ljung et al.\ (\cite{ljung}) rather than the mean values we employ, by the different 3D models and by their use of a somewhat larger sample of \zrii lines, including a number of doubtful lines that we suspect might be blended.

\subsection{Niobium}
We give the raw line-by line results of Nilsson et al.\ (\cite{nils}) in the HM column of Table~\ref{table:lines}, then give in the 3D column the implied 3D abundance for each line, when the 3D$-$HM correction we have derived here is applied.  The final Nb abundance is 
\begin{displaymath}
\log\epsilon_{\mathrm{Nb}}=1.47\pm0.06,
\end{displaymath}
where we have retained the stated error of Nilsson et al.\ (\cite{nils}), as it is clear that systematics in the spectral synthesis fitting dominate the error budget in this case.  In their paper Nilsson et al.\ (\cite{nils}) give $\log\epsilon_{\mathrm{Nb}}=1.47$ as the mean of their HM results, but from the results in their Table 8 for the four lines they recommend to use, the mean should be $\log\epsilon_{\mathrm{Nb}}=1.49$.  They gave a final recommended abundance of $\log\epsilon_{\mathrm{Nb}}=1.44\pm0.06$ after applying a $-0.03$\,dex correction for 3D effects, based on the typical difference 3D$-$HM we reported for rare earth elements in AGSS09.  The mean difference 3D$-$HM we see here for \nbii is indeed $-0.026$\,dex.  The final abundance agrees with earlier HM-based one of Hannaford et al.\ (\cite{hann2}; $\log\epsilon_{\mathrm{Nb}}=1.42\pm0.06$), and with the meteoritic abundance ($\log \epsilon_{\mathrm{Nb}}=1.41\pm0.04$; Lodders et al.\ \cite{lodd}).

\subsection{Molybdenum}
We obtain a 3D abundance
\begin{displaymath}
\log\epsilon_{\mathrm{Mo}}=1.88\pm0.09\ (\pm0.06\ \mathrm{stat},\ \pm0.06\ \mathrm{sys}),
\end{displaymath}
in agreement with Bi\'emont et al.\ (\cite{biem4}) and the meteoritic abundance ($\log \epsilon_{\mathrm{Mo}}=1.94\pm0.04$; Lodders et al.\ \cite{lodd}).  The 3D effect is very large for these low-excitation lines of a minority neutral species: $-0.16$\,dex, giving an abundance of $\log\epsilon_{\mathrm{Mo}}=2.04$ with the HM model.

\subsection{Ruthenium}
Using our new 3D model and the experimental $gf$-values of Wickliffe et al.\ (\cite{wick}), the solar Ru abundance becomes
\begin{displaymath}
\log\epsilon_{\mathrm{Ru}}=1.75\pm0.08\ (\pm0.05\ \mathrm{stat},\ \pm0.07\ \mathrm{sys}).
\end{displaymath}
This is our recommended value, in good agreement with the result of Fivet et al.\ (\cite{five}) and the meteoritic value ($\log \epsilon_{\mathrm{Ru}}=1.76\pm0.03$; Lodders et al.\ \cite{lodd}).  As can be seen in Fig.~\ref{fig:ru}, 3D effects on \rui line formation are quite large: approximately $-0.08$ each from the mean temperature structure and the presence of inhomogeneities, for a total effect of $-0.17$ dex.  No significant trends with equivalent width or excitation potential are evident.

\subsection{Rhodium}
From our two selected \rhi lines, we get
\begin{displaymath}
\log\epsilon_{\mathrm{Rh}}=0.89\pm0.08\ (\pm0.04\ \mathrm{stat},\ \pm0.07\ \mathrm{sys}),
\end{displaymath}
well below both the result of Kwiatkowski et al.\ (\cite{kwia1}) and the meteoritic abundance ($\log \epsilon_{\mathrm{Ru}}=1.06\pm0.04$; Lodders et al.\ \cite{lodd}).
The 3D effect is very large ($-0.17$\,dex on average), not surprisingly for a minority neutral species. The photospheric value we recommend is $0.02$\,dex smaller than in AGSS09.

\subsection{Palladium}
Our 3D LTE result for Pd is 
\begin{displaymath}
\log\epsilon_{\mathrm{Pd}}=1.55\pm0.06\ (\pm0.02\ \mathrm{stat},\ \pm0.06\ \mathrm{sys}).
\end{displaymath}
The 3D effect, $-0.06$\,dex, is smaller than for the other minority species Rb, Zr, Mo, Ru and Rh,
as the ratio ion/neutral is smaller for Pd. The Pd abundance derived with the HM model is $\log \epsilon_{\mathrm{Pd}}=1.61$, somewhat lower than found by Xu et al.\ (\cite{xu}; $\log\epsilon_{\mathrm{Pd}}=1.66\pm0.04$), due in part to our more stringent line selection.  Interestingly, our 3D abundance is substantially lower than the meteoritic one ($\log \epsilon_{\mathrm{Pd}}=1.65\pm0.02$; Lodders et al.\ \cite{lodd}).

\subsection{Silver}
Our 3D LTE solar abundance of silver is
\begin{displaymath}
\log\epsilon_{\mathrm{Ag}}=0.96\pm0.10\ (\pm0.09\ \mathrm{stat},\ \pm0.06\ \mathrm{sys}),
\end{displaymath}
in agreement with the earlier value from Grevesse (\cite{grev}; $\log\epsilon_{\mathrm{Ag}}=0.94\pm0.25$), but well below the meteoritic abundance ($\log \epsilon_{\mathrm{Ag}}=1.20\pm0.02$; Lodders et al.\ \cite{lodd}). The HM-based abundance is significantly larger ($1.04$), but the presence of atmospheric inhomogeneities, as opposed to the mean stratification, matters little.

\subsection{Cadmium}
Taking into account the large uncertainty on the measured equivalent width ($W_\lambda = 0.073\pm 0.021$\,pm) of the one \cdi line available, our 3D solar Cd abundance is
\begin{displaymath}
\log\epsilon_{\mathrm{Cd}}=1.77\pm0.15\ \mathrm{(pure\ systematic\ error)}.
\end{displaymath}
The 3D effect is small: $-0.03$\,dex.  Our result is coincidentally in perfect agreement with that of Youssef et al.\ (\cite{yous}; $\log\epsilon_{\mathrm{Cd}}=1.77\pm0.11$), and also easily overlaps the meteoritic abundance ($\log \epsilon_{\mathrm{Cd}}=1.71\pm0.03$; Lodders et al.\ \cite{lodd}).

\subsection{Indium}
We adopt the solar abundance of In suggested by Vitas et al.\ (\cite{vita}):
\begin{displaymath}
\log\epsilon_{\mathrm{In}}=0.80\pm0.20.
\end{displaymath}

Using the meteoritic abundance, the equivalent width of the \ini contribution to the observed 
photospheric feature at 451.13\,nm should be 0.055\,pm, i.e.\ \ini contributes $<20\%$ of the
observed faint photospheric line, which has $W_\lambda=0.33$\,pm.  In principle, the problems 
in the photosphere might be related to potentially large NLTE and/or 3D effects on \ini line formation.
If we assume (purely for the sake of argument) that the entire line is \altini, then we confirm the high abundance
value with the HM model: $\log\epsilon_{\mathrm{In}}=1.61$.  With the 3D model, we would instead find 
$\log\epsilon_{\mathrm{In}}=1.46$, meaning that 3D effects are indeed large ($-0.14$\,dex after rounding), but certainly
not large enough to reconcile the abundance with the meteoritic value.  Combining NLTE and 3D effects
could never successfully explain the 0.8\,dex difference between the photospheric and meteoritic abundances, 
confirming the conclusion of Vitas et al.\ (\cite{vita}) that the photospheric line is severely blended.

\subsection{Tin}
Taking into account the error on the equivalent width of the single \sni line we use (in addition to our standard error budget), the 3D result is
\begin{displaymath}
\log\epsilon_{\mathrm{Sn}}=2.02\pm0.10\ \mathrm{(pure\ systematic\ error)}.
\end{displaymath}
With the HM model we obtained $\log\epsilon_{\mathrm{Sn}}=2.13$, so the 3D effect is rather large: $-0.11$\,dex.  A very old solar abundance by Grevesse et al.\ (\cite{grev2}; $\log\epsilon_{\mathrm{Sn}}=1.32$) was revised by Lambert et al.\ (\cite{lamb}) with a new, more accurate $gf$-value, and then by Ross \& Aller (\cite{ross3}), resulting in $\log\epsilon_{\mathrm{Sn}}=2.0\pm0.4$.  Our result is very close to that of Ross \& Aller (\cite{ross3}), and is entirely consistent with the meteoritic abundance ($\log \epsilon_{\mathrm{Sn}}=2.07\pm0.06$; Lodders et al.\ \cite{lodd}).

\subsection{Antimony}
Taking into account the large uncertainties of the equivalent width and the experimental $gf$-values, 
we find a solar Sb abundance of $\log\epsilon_{\mathrm{Sb}}\approx1.5\pm0.3$. For
this reason, we do not recommend any photospheric abundance of antimony.

\subsection{Barium}
For the Sun, the few available \baii lines have been analysed by Holweger \& M\"uller (\cite{hol}), Rutten (\cite{rutt}), Gratton \& Sneden (\cite{grat}), Mashonkina et al.\ (\cite{mash3}) and Mashonkina \& Gehren (\cite{mash2}), among others.  The most recent estimate of the solar abundance is $\log\epsilon_{\mathrm{Ba}}=2.21$ (Mashonkina \& Gehren \cite{mash2}).

We find a 3D+NLTE Ba abundance of
\begin{displaymath}
\log\epsilon_{\mathrm{Ba}}=2.25\pm0.07\ (\pm0.03\ \mathrm{stat},\ \pm0.07\ \mathrm{sys}),
\end{displaymath}
Our HM-based value is actually 0.07\,dex lower. Both the HM and 3D results overlap the recommended value of Mashonkina \& Gehren (\cite{mash2}), as well as the meteoritic value ($\log \epsilon_{\mathrm{Ba}}=2.18\pm0.03$; Lodders et al.\ \cite{lodd}).
 
\subsection{Rare Earths (La to Lu) and Hafnium}
As we explain in Sect.\ \ref{rare_gfs}, we do not attempt a detailed 3D-based analysis of all the often-blended lines of rare Earth elements and Hf. Instead we rely on the careful work done by Lawler, Sneden and collaborators using 1D spectrum synthesis with the HM model, and simply correct their derived abundance with our predicted 3D-HM corrections for a subsample of lines (for each element, the lines behave very similarly in this respect so there is no need to compute the 3D effect for all lines).
As expected for low excitation lines of a dominant species, the 3D$-$HM effect on all the rare Earth lines we investigated is small, varying from $-0.01$ to $-0.04$\,dex. The 3D results for all elements La--Hf are summarised in Table~\ref{table:abuns}.

Updating the original Sm abundance of Lawler et al.\ for the new $gf$-values we describe in Sect.\ \ref{rare_gfs} leads to a small change in the HM abundance: $\log\epsilon_{\mathrm{Sm}}=0.99$ instead of 1.00.  

Our 3D LTE result for Eu, $\log\epsilon_{\mathrm{Eu}}=0.49$, is 0.03\,dex lower than the recent LTE Eu solar abundance found by Mucciarelli et al.\ (\cite{mucc}; $\log\epsilon_{\mathrm{Eu}}=0.52$), using their own 3D model. Our recommended Eu abundance however also includes NLTE corrections of $+0.03$\,dex from Mashonkina \& Gehren (\cite{mash2}), which cancel the 3D correction of $-0.03$\,dex to bring our final result into (coincidentally) perfect agreement with those of both Mucciarelli et al.\ (\cite{mucc}) and Lawler et al.\ (\cite{law4}).

With the 3D model, the one \tbii line we retained returns $\log\epsilon_{\mathrm{Tb}}=0.31\pm0.10$. If we had kept all three lines analysed by Lawler et al.\ (\cite{law5}), we would have obtained an abundance of $\log\epsilon_{\mathrm{Tb}}=0.24\pm0.08$.

Our new Hf abundance, $\log\epsilon_{\mathrm{Hf}}=0.85\pm0.05$, is in very good agreement with a recent result by Caffau et al.\ (\cite{caf2}) using their own 3D model: $\log\epsilon_{\mathrm{Hf}}=0.87\pm0.04$.  Although the solar abundances of all elements from La to Lu are consistent with the meteoritic values (to within the mutual uncertainties), the solar abundance of Hf is quite a bit higher than the meteoritic one ($\log \epsilon_{\mathrm{Hf}}=0.71\pm0.02$; Lodders et al.\ \cite{lodd}). Perhaps the most natural explanation would be erroneous $gf$-values for the \hfii lines, but the experimental data from Lawler et al.\ (\cite{law8}) appear reliable. 

\subsection{Tungsten}
Updating the results of Holweger \& Werner (\cite{hol2}) for the new $gf$-values we discuss in Sect.\ \ref{w_gfs}, the mean abundance becomes $\log\epsilon_{\mathrm{W}}=1.03$.  We further updated this result by computing the 3D$-$HM abundance corrections for these two lines: $-0.19$\,dex. This is not surprising for these low excitation lines of a minor species ($\mathrm{\altwii/\wi}\sim10$). The 3D LTE abundance of W thus becomes
\begin{displaymath}
\log\epsilon_{\mathrm{W}}=0.83\pm0.11\ (\pm0.03\ \mathrm{stat},\ \pm0.11\ \mathrm{sys}), 
\end{displaymath}
where the systematic error includes both the regular sources (0.08\,dex) and the error in the equivalent widths (0.07\,dex), due to the fact that these lines are very weak (0.35 and 0.035\,pm respectively, cf.\ Table~\ref{table:lines}). Even with spectral synthesis these lines are extremely difficult to reproduce accurately.  Despite the large 3D effects, the solar tungsten abundance is substantially larger than seen in meteorites ($\log \epsilon_{\mathrm{W}}=0.65\pm0.04$; Lodders et al.\ \cite{lodd}), similar to the case of Hf mentioned above.

\subsection{Osmium}
We find with our 3D model an Os abundance of
\begin{displaymath}
\log\epsilon_{\mathrm{Os}}=1.40\pm0.05\ \mathrm{(pure\ systematic\ error)},
\end{displaymath}
in reasonable agreement with the meteoritic value ($\log \epsilon_{\mathrm{Os}}=1.35\pm0.03$; Lodders et al.\ \cite{lodd}).  The 3D effect is quite large: $-0.09$\,dex.

An example of the problem with the \osi lines used in past analyses is seen in the results of Kwiatkowsky et al.\ (\cite{kwia2}): the dispersion is very large ($\sigma=0.15$\,dex), and the difference between the lowest and largest abundances is 0.42\,dex.  Recent results by Caffau et al.\ (\cite{caf3}), who used the lines and solar data of Kwiatkowsky et al.\ with their own 3D model, show a similarly unacceptable dispersion.  Their result, $\log\epsilon_{\mathrm{Os}}=1.36\pm0.19$\,dex, includes a striking difference of 0.71\,dex between the lowest and largest abundance.  We note that Caffau et al.\ used an obsolete value for the ionisation potential of \osi (8.7\,eV), whereas the correct value is $8.4382\pm0.0002$\,eV (Colarusso et al.\ \cite{cola}).  We estimate that their abundance results from \osi lines should therefore be increased by about 0.15\,dex, to $\log\epsilon_{\mathrm{Os}}=1.51\pm0.19$\,dex.  This result is much larger than what we find here, as expected from the severely blended nature of some of their chosen lines.

\subsection{Iridium}
With our adopted equivalent width and $gf$-value, we find a 3D abundance of 
\begin{displaymath}
\log\epsilon_{\mathrm{Ir}}=1.42\pm0.07\ \mathrm{(pure\ systematic\ error)},
\end{displaymath}
where the error includes the uncertainty in the measured equivalent width (0.06\,dex) and the combined contributions of the usual three systematics (0.04\,dex). This abundance is in rather poor agreement with the meteoritic value ($\log \epsilon_{\mathrm{Ir}}=1.32\pm0.02$; Lodders et al.\ \cite{lodd}).  Our result is however consistent with previous determinations: Youssef \& Khalil (\cite{yous2}) found $\log\epsilon_{\mathrm{Ir}}=1.38\pm0.05$, and the result of Drake \& Aller (\cite{drake}) goes from $\log\epsilon_{\mathrm{Ir}}=0.82$ to $\log\epsilon_{\mathrm{Ir}}=1.37$ when updated to the $gf$-value we adopt.

\subsection{Gold}
We find a 3D solar gold abundance of
\begin{displaymath}
\log\epsilon_{\mathrm{Au}}=0.91\pm0.08\ \mathrm{(pure\ systematic\ error)}, 
\end{displaymath}
where we have incorporated the uncertainty in the equivalent width and $gf$-value into the final uncertainty.  The 3D effect is small: $-$0.02\,dex.  Updating the HM spectrum synthesis result of Youssef (\cite{yous3}; $\log\epsilon_{\mathrm{Au}}=0.95$) with the 3D$-$HM effect would lead to $\log\epsilon_{\mathrm{Au}}$=0.93, in excellent agreement with the value we obtain here.  Our adopted gold abundance remains substantially higher than the meteoritic value ($\log \epsilon_{\mathrm{Au}}=0.80\pm0.04$; Lodders et al.\ \cite{lodd}).

\subsection{Lead}
\label{pbresults}
The LTE abundances of lead with the 3D and HM models are $\log\epsilon_{\mathrm{Pb}}=1.80$ and 1.93, respectively; the 3D$-$HM effect is rather large, $-$0.14\,dex after rounding. This is expected because \pbi is a minority species compared to \pbii in the solar photosphere. 

Taking into account the significant NLTE correction ($+0.12$\,dex) of Mashonkina et al.\ (\cite{mash12}), our final 3D abundance of Pb becomes
\begin{displaymath}
\log\epsilon_{\mathrm{Pb}}=1.92\pm0.08\ \mathrm{(pure\ systematic\ error)}.
\end{displaymath}
This is somewhat smaller than obtained by Bi\'emont et al.\ (\cite{biem6};  $\log\epsilon_{\mathrm{Pb}}=2.00\pm0.06$) with the HM model and an erroneously large equivalent width. Our photospheric Pb abundance is also substantially smaller than the corresponding CI meteoritic value ($\log \epsilon_{\mathrm{Pb}}=2.04\pm0.03$; Lodders et al.\ \cite{lodd}).  

\subsection{Thorium}
The 3D+NLTE solar abundance of Th is
\begin{displaymath}
\log\epsilon_{\mathrm{Th}}=0.03\pm0.10\ \mathrm{(pure\ systematic\ error)},
\end{displaymath}

\setcounter{table}{3}
\begin{table*}[p]
\centering
\caption[Abundances implied by all lines]{Average abundances implied by all lines of elements from Cu to Th.  Abundances are given as the weighted mean across all lines in the given list, taking into account NLTE corrections for \altzni, \altsri, \altsrii, \altbaii, \alteuii, \pbi and \altthii.  \zri is shown in brackets because we do not consider this result reliable enough to include in our final adopted abundance.  Note that because all means were computed using abundances accurate to three decimal places, entries in the columns 8 and 9 differ in some cases from the differences between the entries in columns 3--5.  We also give our final recommended solar photospheric abundance of each element, compared with the abundance in CI chondritic meteorites (Lodders et al.\ \cite{lodd}, normalised to the silicon abundance determined in Paper I).}
\label{table:abuns}
\begin{tabular}{l l c c c c c c c c c}
\hline
\hline
& Species & 3D & \oneDAV & HM & \textsc{marcs} & {\sc miss} & 3D$-$HM & 3D$-$\oneDAV & Recommended & Meteoritic\\
\hline
$\log \epsilon_\mathrm{Cu}$     & \cui    & $4.18\pm0.05$ & 4.16 & 4.21 & 4.11 & 4.18 & $-$0.03 & \ph0.02 & $4.18\pm0.05$ & $4.25\pm0.04$\vspace{1mm}\\
$\log \epsilon_\mathrm{Zn}$     & \zni    & $4.56\pm0.05$ & 4.52 & 4.53 & 4.46 & 4.54 & \ph0.03 & \ph0.04 & $4.56\pm0.05$ & $4.63\pm0.04$\vspace{1mm}\\
$\log \epsilon_\mathrm{Ga}$     & \gai    & $3.02\pm0.05$ & 3.00 & 3.09 & 2.96 & 3.02 & $-$0.06 & \ph0.02 & $3.02\pm0.05$ & $3.08\pm0.02$\vspace{1mm}\\
$\log \epsilon_\mathrm{Ge}$     & \gei    & $3.63\pm0.07$ & 3.55 & 3.62 & 3.51 & 3.57 & \ph0.01 & \ph0.09 & $3.63\pm0.07$ & $3.58\pm0.04$\vspace{1mm}\\
$\log \epsilon_\mathrm{Kr}$     && \multicolumn{7}{c}{interpolated $s$-process production rate (AGSS09)}  & $3.25\pm0.06$ & $-2.27$      \vspace{1mm}\\
$\log \epsilon_\mathrm{Rb}$     & \rbi    & $2.47\pm0.07$ & 2.52 & 2.57 & 2.48 & 2.54 & $-$0.10 & $-$0.04 & $2.47\pm0.07$ & $2.36\pm0.03$\vspace{1mm}\\
$\log \epsilon_\mathrm{Sr}$     & \sri    & $2.80\pm0.06$ & 2.82 & 2.84 & 2.77 & 2.79 & $-$0.03 & $-$0.01 & &            \\
                                & \srii   & $2.85\pm0.11$ & 2.75 & 2.86 & 2.70 & 2.86 & $-$0.01 & \ph0.09 &               &                          \\
                                & Sr all  & $2.83\pm0.06$ & 2.78 & 2.85 & 2.73 & 2.83 & $-$0.02 & \ph0.05 &  $2.83\pm0.06$              &    $2.88\pm0.03$           \vspace{1mm}\\
$\log \epsilon_\mathrm{Y}$      & \yii    & $2.21\pm0.05$ & 2.17 & 2.20 & 2.14 & 2.20 & \ph0.01 & \ph0.04 & $2.21\pm0.05$ & $2.17\pm0.04$\vspace{1mm}\\
$\log \epsilon_\mathrm{Zr}$     & (\zri)  & $2.57\pm0.12$ & 2.67 & 2.75 & 2.64 & 2.69 & $-$0.18 & $-$0.10 & &            \\
                                & \zrii   & $2.59\pm0.04$ & 2.55 & 2.59 & 2.53 & 2.59 & $-$0.00 & \ph0.03 &  $2.59\pm0.04$              &     $2.53\pm0.04$          \vspace{1mm}\\
$\log \epsilon_\mathrm{Nb}$     & \nbii   & $1.47\pm0.06$ &      & 1.49 &      &      & $-$0.03 & \ph0.00 & $1.47\pm0.06$ & $1.41\pm0.04$\vspace{1mm}\\
$\log \epsilon_\mathrm{Mo}$     & \moi    & $1.88\pm0.09$ & 1.96 & 2.04 & 1.93 & 1.98 & $-$0.16 & $-$0.08 & $1.88\pm0.09$ & $1.94\pm0.04$\vspace{1mm}\\
$\log \epsilon_\mathrm{Ru}$     & \rui    & $1.75\pm0.08$ & 1.82 & 1.91 & 1.80 & 1.84 & $-$0.17 & $-$0.08 & $1.75\pm0.08$ & $1.76\pm0.03$\vspace{1mm}\\
$\log \epsilon_\mathrm{Rh}$     & \rhi    & $0.89\pm0.08$ & 0.98 & 1.07 & 0.95 & 0.99 & $-$0.17 & $-$0.08 & $0.89\pm0.08$ & $1.06\pm0.04$\vspace{1mm}\\
$\log \epsilon_\mathrm{Pd}$     & \pdi    & $1.55\pm0.06$ & 1.52 & 1.61 & 1.49 & 1.54 & $-$0.06 & \ph0.03 & $1.55\pm0.06$ & $1.65\pm0.02$\vspace{1mm}\\
$\log \epsilon_\mathrm{Ag}$     & \agi    & $0.96\pm0.10$ & 0.95 & 1.04 & 0.92 & 0.96 & $-$0.08 & \ph0.01 & $0.96\pm0.10$ & $1.20\pm0.02$\vspace{1mm}\\
$\log \epsilon_\mathrm{Cd}$     & \cdi    & $1.77\pm0.15$ & 1.76 & 1.79 & 1.73 & 1.79 & $-$0.03 & \ph0.01 & $1.77\pm0.15$ & $1.71\pm0.03$\vspace{1mm}\\
$\log \epsilon_\mathrm{In}$     & \ini    & \multicolumn{7}{c}{sunspot (Vitas et al.\ \cite{vita})}       & $0.80\pm0.20$ & $0.76\pm0.03$\vspace{1mm}\\
$\log \epsilon_\mathrm{Sn}$     & \sni    & $2.02\pm0.10$ & 2.06 & 2.13 & 2.04 & 2.09 & $-$0.11 & $-$0.05 & $2.02\pm0.10$ & $2.07\pm0.06$\vspace{1mm}\\
$\log \epsilon_\mathrm{Xe}$     && \multicolumn{7}{c}{interpolated $s$-process production rate (AGSS09)}  & $2.24\pm0.06$ & $-1.95$      \vspace{1mm}\\
$\log \epsilon_\mathrm{Ba}$     & \baii   & $2.25\pm0.07$ & 2.15 & 2.18 & 2.10 & 2.17 & \ph0.07 & \ph0.10 & $2.25\pm0.07$ & $2.18\pm0.03$\vspace{1mm}\\
$\log \epsilon_\mathrm{La}$     & \laii   & $1.11\pm0.04$ &      & 1.14 &      &      & $-$0.03 & \ph0.00 & $1.11\pm0.04$ & $1.17\pm0.02$\vspace{1mm}\\
$\log \epsilon_\mathrm{Ce}$     & \ceii   & $1.58\pm0.04$ &      & 1.61 &      &      & $-$0.03 & \ph0.01 & $1.58\pm0.04$ & $1.58\pm0.02$\vspace{1mm}\\
$\log \epsilon_\mathrm{Pr}$     & \prii   & $0.72\pm0.04$ &      & 0.76 &      &      & $-$0.04 & $-$0.01 & $0.72\pm0.04$ & $0.76\pm0.03$\vspace{1mm}\\
$\log \epsilon_\mathrm{Nd}$     & \ndii   & $1.42\pm0.04$ &      & 1.45 &      &      & $-$0.03 & \ph0.00 & $1.42\pm0.04$ & $1.45\pm0.02$\vspace{1mm}\\
$\log \epsilon_\mathrm{Sm}$     & \smii   & $0.95\pm0.04$ &      & 0.99 &      &      & $-$0.04 & $-$0.01 & $0.95\pm0.04$ & $0.94\pm0.02$\vspace{1mm}\\
$\log \epsilon_\mathrm{Eu}$     & \euii   & $0.52\pm0.04$ &      & 0.55 &      &      & $-$0.03 & \ph0.00 & $0.52\pm0.04$ & $0.51\pm0.02$\vspace{1mm}\\
$\log \epsilon_\mathrm{Gd}$     & \gdii   & $1.08\pm0.04$ &      & 1.11 &      &      & $-$0.03 & \ph0.01 & $1.08\pm0.04$ & $1.05\pm0.02$\vspace{1mm}\\
$\log \epsilon_\mathrm{Tb}$     & \tbii   & $0.31\pm0.10$ &      & 0.28 &      &      & $-$0.04 & $-$0.00 & $0.31\pm0.10$ & $0.32\pm0.03$\vspace{1mm}\\
$\log \epsilon_\mathrm{Dy}$     & \dyii   & $1.10\pm0.04$ &      & 1.13 &      &      & $-$0.03 & \ph0.00 & $1.10\pm0.04$ & $1.13\pm0.02$\vspace{1mm}\\
$\log \epsilon_\mathrm{Ho}$     & \hoii   & $0.48\pm0.11$ &      & 0.51 &      &      & $-$0.03 & \ph0.00 & $0.48\pm0.11$ & $0.47\pm0.03$\vspace{1mm}\\
$\log \epsilon_\mathrm{Er}$     & \erii   & $0.93\pm0.05$ &      & 0.96 &      &      & $-$0.03 & \ph0.01 & $0.93\pm0.05$ & $0.92\pm0.02$\vspace{1mm}\\
$\log \epsilon_\mathrm{Tm}$     & \tmii   & $0.11\pm0.04$ &      & 0.14 &      &      & $-$0.03 & \ph0.01 & $0.11\pm0.04$ & $0.12\pm0.03$\vspace{1mm}\\
$\log \epsilon_\mathrm{Yb}$     & \ybii   & $0.85\pm0.11$ &      & 0.86 &      &      & $-$0.01 & \ph0.03 & $0.85\pm0.11$ & $0.92\pm0.02$\vspace{1mm}\\
$\log \epsilon_\mathrm{Lu}$     & \luii   & $0.10\pm0.09$ &      & 0.12 &      &      & $-$0.02 & \ph0.00 & $0.10\pm0.09$ & $0.09\pm0.02$\vspace{1mm}\\
$\log \epsilon_\mathrm{Hf}$     & \hfii   & $0.85\pm0.05$ &      & 0.88 &      &      & $-$0.03 & \ph0.00 & $0.85\pm0.05$ & $0.71\pm0.02$\vspace{1mm}\\
$\log \epsilon_\mathrm{W}$      & \wi     & $0.83\pm0.11$ &      & 1.03 &      &      & $-$0.19 & $-$0.09 & $0.83\pm0.11$ & $0.65\pm0.04$\vspace{1mm}\\
$\log \epsilon_\mathrm{Os}$     & \osi    & $1.40\pm0.05$ & 1.41 & 1.50 & 1.40 & 1.44 & $-$0.09 & $-$0.01 & $1.40\pm0.05$ & $1.35\pm0.03$\vspace{1mm}\\
$\log \epsilon_\mathrm{Ir}$     & \iri    & $1.42\pm0.07$ & 1.40 & 1.46 & 1.38 & 1.43 & $-$0.04 & \ph0.02 & $1.42\pm0.07$ & $1.32\pm0.02$\vspace{1mm}\\
$\log \epsilon_\mathrm{Au}$     & \aui    & $0.91\pm0.08$ & 0.89 & 0.93 & 0.88 & 0.93 & $-$0.02 & \ph0.02 & $0.91\pm0.08$ & $0.80\pm0.04$\vspace{1mm}\\
$\log \epsilon_\mathrm{Tl}$     & \tli    & \multicolumn{7}{c}{sunspot (Lambert et al.\ \cite{lamb2})}    &  $0.9\pm0.2$  & $0.77\pm0.03$\vspace{1mm}\\
$\log \epsilon_\mathrm{Pb}$     & \pbi    & $1.92\pm0.08$ & 1.97 & 2.05 & 1.95 & 2.00 & $-$0.14 & $-$0.06 & $1.92\pm0.08$ & $2.04\pm0.03$\vspace{1mm}\\
$\log \epsilon_\mathrm{Th}$     & \thii   & $0.03\pm0.10$ & 0.03 & 0.07 & 0.02 & 0.06 & $-$0.04 & $-$0.00 & $0.03\pm0.10$ & $0.06\pm0.03$\vspace{0.5mm}\\
\hline							  
\end{tabular}						  
\end{table*}
\afterpage{\clearpage}

\noindent where the error budget is dominated by the uncertainty in our adopted equivalent width.  Given the overall uncertainty, the photospheric abundance of Th is consistent with the meteoritic value ($\log \epsilon_{\mathrm{Th}}=0.06\pm0.03$; Lodders et al.\ \cite{lodd}).  With the HM model, the abundance would be $\log\epsilon_{\mathrm{Th}}=0.07$. 

Our 3D LTE abundance is 15$\%$ smaller than the corresponding LTE result of Caffau et al.\ (\cite{caf2}), who found $\log \epsilon_{\mathrm{Th}}=0.08\pm0.03$ with their 3D model. The 0.06\,dex difference with our LTE result can be explained by our improved, model-independent estimation of the contributions of the blends by Co\,\textsc{i} and V\,\textsc{i}.  In our opinion the stated uncertainty of Caffau et al.\  underestimates the uncertainty of the total contribution of \thii to the overall equivalent width.

\subsection{Other heavy elements}
Many other elements such as Se, Br, Kr, Te, I, Xe, Cs, Ta, Re, Pt, Hg, Tl, Bi and U are not present in the solar photospheric spectrum.  For most of them, the meteoritic abundances have to be adopted as the \textit{de facto} solar abundances. For Kr and Xe however, solar abundances can be estimated from interpolation of the theoretical $s$-process production rates.  The values for Xe and Kr in Table~\ref{table:abuns} have been taken from AGSS09. For Tl, a line of Tl\,\textsc{i} is present in the sunspot spectrum, but heavily blended.  This line has been used by Lambert et al.\ (\cite{lamb2}) to derive a very uncertain solar Tl abundance: $\log\epsilon_{\mathrm{Tl}}=0.72$--1.07.  We shall adopt $\log\epsilon_{\mathrm{Tl}}=0.9\pm0.2$.

\section{Discussion}
\label{discussion}

In some species of the heavy elements we see large differences between the abundances derived with the 3D model and the four 1D models.  The easiest way to understand these differences is to separate the elements into two groups: majority (most of the element is in this ionisation stage, in most cases the once ionised species) and minority (mostly neutral species of elements with relatively low ionisation energy).  For majority elements with weak, low-excitation lines, the 3D effect and the differences between the various model atmospheres are small (of order 10$\%$ or lower), as these lines form rather deep in the photosphere and are not strongly sensitive to temperature.  In this group we find \altnbii, \altcdi, the rare Earths, \althfii, \altiri, \aui and \altthii. It is clear that the lines of these tracers are not very sensitive to the temperature structure of the model atmosphere: in the rather deep layers where these lines are formed, the mean temperature structures of the different models are quite similar.  However, for majority elements with stronger lines such as \altzni, \altsrii, \altyii, \zrii and \altbaii, we see that the 3D effects and dependence on the model atmosphere often become much larger.  These lines are formed significantly higher in the photosphere, where the mean temperature structure of the model atmospheres differ far more.  

For the minority elements \altcui, \altgai, \altgei, \altrbi, \altsri, \altmoi, \altrui, \altrhi, \altpdi, \altagi, \altsni, \altwi, \osi and \altpbi, all lines are extremely temperature-sensitive.  This is why we observe large differences in abundance results from these lines with different model atmospheres.  These differences however depend subtly on the characteristics of each line, and its mean optical depth of formation. We caution that such lines are likely sensitive to departures from LTE, which in most cases have been explored theoretically neither in 1D nor 3D. 

\begin{table}[htb]
\centering
\caption{Recommended present-day solar photospheric abundances for the heavy elements Cu to Th, compared with oft-used solar abundance compilations: AG89 (Anders \& Grevesse \cite{ag89}), GS98 (Grevesse \& Sauval \cite{gs98}), AGS05 (Asplund et al.\ \cite{AGS05}), AGSS09 (Asplund et al.\ \cite{asp8}), LPG09 (Lodders et al.\ \cite{lodd}). Preferred values are from this work except where noted.}
\label{table:compilations} 
\begin{tabular}{l@{\hspace{2mm}}l@{\hspace{3mm}}c@{\hspace{2mm}}c@{\hspace{3mm}}c@{\hspace{3mm}}c@{\hspace{2mm}}c@{\hspace{2mm}}c}
\hline \phantom{1}Z & el. & Preferred & AG89 & GS98 & AGS05 & AGSS09 & LPG09 \\
\hline
\hline
  29 &  Cu& $    4.18\pm 0.05 $\phantom{\tablefootmark{1}}             &    4.21 &    4.21 &    4.21 &    4.19 &    4.21\\
  30 &  Zn& $    4.56\pm 0.05 $\phantom{\tablefootmark{1}}             &    4.60 &    4.60 &    4.60 &    4.56 &    4.62\\
  31 &  Ga& $    3.02\pm 0.05 $\phantom{\tablefootmark{1}}             &    2.88 &    2.88 &    2.88 &    3.04 &    2.88\\
  32 &  Ge& $    3.63\pm 0.07 $\phantom{\tablefootmark{1}}             &    3.41 &    3.41 &    3.58 &    3.65 &    3.58\\
  36 &  Kr& $    3.25\pm 0.06 $\tablefootmark{\textcolor{BrickRed}{1}} &         &         &    3.28 &    3.25 &    3.28\\
  37 &  Rb& $    2.47\pm 0.07 $\phantom{\tablefootmark{1}}             &    2.60 &    2.60 &    2.60 &    2.52 &    2.60\\
  38 &  Sr& $    2.83\pm 0.06 $\phantom{\tablefootmark{1}}             &    2.90 &    2.97 &    2.92 &    2.87 &    2.92\\
  39 &   Y& $    2.21\pm 0.05 $\phantom{\tablefootmark{1}}             &    2.24 &    2.24 &    2.21 &    2.21 &    2.21\\
  40 &  Zr& $    2.59\pm 0.04 $\phantom{\tablefootmark{1}}             &    2.60 &    2.60 &    2.59 &    2.58 &    2.58\\
  41 &  Nb& $    1.47\pm 0.06 $\phantom{\tablefootmark{1}}             &    1.42 &    1.42 &    1.42 &    1.46 &    1.42\\
  42 &  Mo& $    1.88\pm 0.09 $\phantom{\tablefootmark{1}}             &    1.92 &    1.92 &    1.92 &    1.88 &    1.92\\
  44 &  Ru& $    1.75\pm 0.08 $\phantom{\tablefootmark{1}}             &    1.84 &    1.84 &    1.84 &    1.75 &    1.84\\
  45 &  Rh& $    0.89\pm 0.08 $\phantom{\tablefootmark{1}}             &    1.12 &    1.12 &    1.12 &    0.91 &    1.12\\
  46 &  Pd& $    1.55\pm 0.06 $\phantom{\tablefootmark{1}}             &    1.69 &    1.69 &    1.69 &    1.57 &    1.66\\
  47 &  Ag& $    0.96\pm 0.10 $\phantom{\tablefootmark{1}}             &    0.94 &    0.94 &    0.94 &    0.94 &    0.94\\
  48 &  Cd& $    1.77\pm 0.15 $\phantom{\tablefootmark{1}}             &    1.86 &    1.77 &    1.77 &         &    1.77\\
  49 &  In& $    0.80\pm 0.20 $\tablefootmark{\textcolor{BrickRed}{2}}  &    1.66 &    1.66 &    1.60 &    0.80 &    1.50\\
  50 &  Sn& $    2.02\pm 0.10 $\phantom{\tablefootmark{1}}             &    2.00 &    2.00 &    2.00 &    2.04 &    2.00\\
  54 &  Xe& $    2.24\pm 0.06 $\tablefootmark{\textcolor{BrickRed}{1}} &         &         &    2.27 &    2.24 &    2.27\\
  56 &  Ba& $    2.25\pm 0.07 $\phantom{\tablefootmark{1}}             &    2.13 &    2.13 &    2.17 &    2.18 &    2.17\\
  57 &  La& $    1.11\pm 0.04 $\phantom{\tablefootmark{1}}             &    1.22 &    1.17 &    1.13 &    1.10 &    1.14\\
  58 &  Ce& $    1.58\pm 0.04 $\phantom{\tablefootmark{1}}             &    1.55 &    1.58 &    1.58 &    1.58 &    1.61\\
  59 &  Pr& $    0.72\pm 0.04 $\phantom{\tablefootmark{1}}             &    0.71 &    0.71 &    0.71 &    0.72 &    0.76\\
  60 &  Nd& $    1.42\pm 0.04 $\phantom{\tablefootmark{1}}             &    1.50 &    1.50 &    1.45 &    1.42 &    1.45\\
  62 &  Sm& $    0.95\pm 0.04 $\phantom{\tablefootmark{1}}             &    1.00 &    1.01 &    1.01 &    0.96 &    1.00\\
  63 &  Eu& $    0.52\pm 0.04 $\phantom{\tablefootmark{1}}             &    0.51 &    0.51 &    0.52 &    0.52 &    0.52\\
  64 &  Gd& $    1.08\pm 0.04 $\phantom{\tablefootmark{1}}             &    1.12 &    1.12 &    1.12 &    1.07 &    1.11\\
  65 &  Tb& $    0.31\pm 0.10 $\phantom{\tablefootmark{1}}             &   -0.10 &   -0.10 &    0.28 &    0.30 &    0.28\\
  66 &  Dy& $    1.10\pm 0.04 $\phantom{\tablefootmark{1}}             &    1.10 &    1.14 &    1.14 &    1.10 &    1.13\\
  67 &  Ho& $    0.48\pm 0.11 $\phantom{\tablefootmark{1}}             &    0.26 &    0.26 &    0.51 &    0.48 &    0.51\\
  68 &  Er& $    0.93\pm 0.05 $\phantom{\tablefootmark{1}}             &    0.93 &    0.93 &    0.93 &    0.92 &    0.96\\
  69 &  Tm& $    0.11\pm 0.04 $\phantom{\tablefootmark{1}}             &    0.00 &    0.00 &    0.00 &    0.10 &    0.14\\
  70 &  Yb& $    0.85\pm 0.11 $\phantom{\tablefootmark{1}}             &    1.08 &    1.08 &    1.08 &    0.84 &    0.86\\
  71 &  Lu& $    0.10\pm 0.09 $\phantom{\tablefootmark{1}}             &    0.76 &    0.06 &    0.06 &    0.10 &    0.12\\
  72 &  Hf& $    0.85\pm 0.05 $\phantom{\tablefootmark{1}}             &    0.88 &    0.88 &    0.88 &    0.85 &    0.88\\
  74 &   W& $    0.83\pm 0.11 $\phantom{\tablefootmark{1}}             &    1.11 &    1.11 &    1.11 &    0.85 &    1.11\\
  76 &  Os& $    1.40\pm 0.05 $\phantom{\tablefootmark{1}}             &    1.45 &    1.45 &    1.45 &    1.40 &    1.45\\
  77 &  Ir& $    1.42\pm 0.07 $\phantom{\tablefootmark{1}}             &    1.35 &    1.35 &    1.38 &    1.38 &    1.38\\
  79 &  Au& $    0.91\pm 0.08 $\phantom{\tablefootmark{1}}             &    1.01 &    1.01 &    1.01 &    0.92 &    1.01\\
  81 &  Tl& $    0.90\pm 0.20 $\tablefootmark{\textcolor{BrickRed}{3}} &    0.90 &    0.90 &    0.90 &    0.90 &    0.95\\
  82 &  Pb& $    1.92\pm 0.08 $\phantom{\tablefootmark{1}}             &    1.85 &    1.95 &    2.00 &    1.75 &    2.00\\
  90 &  Th& $    0.03\pm 0.10 $\phantom{\tablefootmark{1}}             &    0.12 &         &         &    0.02 &        \\  
\hline
\end{tabular}
\tablefoot{\tablefoottext{1}{AGSS09}; \tablefoottext{2}{Vitas et al.\ (\cite{vita})}; \tablefoottext{3}{Lambert et al.\ (\cite{lamb2})}}
\end{table}

\section{Comparison with previous solar abundance compilations}
\label{compilations}

Table \ref{table:compilations} compares our recommended present-day photospheric abundances for the Sun with those advocated by previous, commonly-used compilations of the solar chemical composition. In most cases the differences are relatively minor ($\pm 0.05$\,dex), although our Ga ($+0.14$\,dex compared with Lodders et al.\ \cite{lodd} for example), Rb ($-0.13$\,dex), Sr ($-0.09$), Ru ($-0.09$\,dex), Rh ($-0.23$\,dex), Pd ($-0.09$\,dex),  In ($-0.70$\,dex), Ba ($+0.08$\,dex), W ($-0.28$\,dex), Au ($-0.10$\,dex) and Pb ($-0.08$\,dex) abundances show greater variations relative to recent compilations. In many of these cases the previous recommended values were based on very old analyses, often with outdated atomic data and poor treatment of blends, whereas here we carry out new 3D-based analyses for these elements. 

Compared with AGSS09 there have only been minor adjustments ($\pm 0.02$) of the recommended solar abundances for most elements. Notable exceptions for which larger differences are present are Rb ($-0.05$\,dex), Sr ($-0.04$\,dex), Ba  ($+0.07$\,dex), Ir ($+0.04$\,dex) and Pb ($+0.17$\,dex). In most cases, these changes have been driven by improved transition probabilities and/or the appearance of (new) NLTE calculations.  

We will comment on the agreement between photospheric and meteoritic abundances in more detail in future work, so we restrict ourselves here to some brief comments.  Ignoring the noble gases (in which CI chondritic meteorites are heavily depleted), the photospheric excesses of $0.10$\,dex or more that we see relative to the CI chondrites for Rb, Hf, W, Ir and Au might be due to unidentified blends and/or continuum placement.  For Rh, Pd, Ag and Pb, which are under-abundant in the Sun compared to meteorites, one might postulate unidentified or underestimated NLTE effects: the photospheric lines we have used of \altrhi, \altpdi, \agi and \pbi are rather low-excitation lines of minority species, so are expected to be prone to NLTE effects.

\section{Conclusions}
\label{conclusions}

In this paper we have redetermined the solar abundances of almost every element from Cu to Th, using a new, highly realistic 3D hydrodynamic model of the solar atmosphere.  In most cases we have based our determination on a full reanalysis of the relevant lines in the solar spectrum, including the latest atomic data and extremely demanding line selection.  We have accounted for departures from LTE wherever such corrections exist. In some cases, where other authors have already performed very detailed and careful analyses based on full spectrum synthesis, we have computed the net impact on the relevant lines of using our 3D model, and adjusted the previous 1D results accordingly.  Together with the results of Papers I and II, the results we present here make up the first fully complete, detailed and homogeneous analysis of the medium-and-heavy-element ($Z>8$) composition of the Sun.

\begin{acknowledgements}
We are deeply indebted to Maria Bergemann (\altsri, \srii and \altbaii) and Yoichi Takeda (\altzni) for providing NLTE corrections for some of our lines on request.  We also thank Chris Sneden, Jim Lawler, Emile Bi\'emont and Jean-Fran\c{c}ois Wyart for providing their recent works prior to publication, Glenn Wahlgren for helpful discussions regarding our \aui line, Remo Collet, Regner Trampedach and Wolfgang Hayek for other discussions, and the referee for constructive feedback.  NG, PS and MA variously thank the Max Planck Institut f\"ur Astrophysik, Garching, the Centre Spatial de Li\`ege, the Department of Astrophysics, Geophysics and Oceanography, University of Li\`ege and Mount Stromlo Observatory for support and hospitality during the production of this paper.  We acknowledge further support from the Royal Belgian Observatory (NG), IAU Commission 46, the Lorne Trottier Chair in Astrophysics, the (Canadian) Institute for Particle Physics, the Banting Fellowship scheme as administered by the Natural Science and Engineering Research Council of Canada, the UK Science \& Technology Facilities Council (PS) and the Australian Research Council (MA).
\end{acknowledgements}

\Online

\onecolumn
\setcounter{table}{0}
\begin{center}
\topcaption{\label{table:lines} Lines retained in this analysis: atomic and solar data, line weightings, LTE abundance results for the 5 models used in this analysis, NLTE corrections to the LTE result (when available), and the corresponding 3D+NLTE abundance result}
\tablefirsthead{%
  \hline
  $\lambda$ & & E$_\mathrm{exc}$ & $\log$ $gf$ & $gf$ & $W_\lambda$ & Wt.  &\multicolumn{5}{c}{LTE Abundances}& $\Delta_{\rm NLTE}$ & 3D \\
  (nm) & & (eV) & & ref. & (pm) &  & 3D & $\langle 3{\rm D}\rangle$ & HM & \marcs & {\sc miss} & (3D) & NLTE\\
  \hline
   & & & & & & & & & & & & & \\
  }
\tablehead{%
  \hline 
  $\lambda$ & & E$_\mathrm{exc}$ & $\log$ $gf$ & $gf$ & $W_\lambda$ & Wt.  &\multicolumn{5}{c}{LTE Abundances}& $\Delta_{\rm NLTE}$ & 3D \\
  (nm) & & (eV) & & ref. & (pm) &  & 3D & $\langle 3{\rm D}\rangle$ & HM & \marcs & {\sc miss} & (3D) & NLTE\\
  \hline
  \multicolumn{14}{r}{continued.}\\
  \hline
   & & & & & & & & & & & & & \\
  }
\tabletail{%
  \hline
  \multicolumn{14}{r}{continued on next page}\\
  \hline
}
\tablelasttail{\hline}
\begin{mpsupertabular}{r@{}c c c c r c c c c c c r c}
\multicolumn{14}{c} {\cui} \\

  510.5541 & & 1.390 & $-1.516$ &  1 &  9.300  &  1  &  4.152  & 4.135  & 4.217  & 4.087   & 4.141    & $+0.020$ & 4.172  \\
  521.8201 & & 3.820 & $+0.264$ &  1 &  5.250  &  1  &  4.214  & 4.185  & 4.228  & 4.125   & 4.207    & $+0.020$ & 4.234  \\
  522.0070 & & 3.820 & $-0.616$ &  1 &  1.450  &  1  &  4.220  & 4.223  & 4.264  & 4.182   & 4.251    & $+0.010$ & 4.230  \\
  793.3130 & & 3.790 & $-0.372$ &  2 &  2.800  &  1  &  4.187  & 4.176  & 4.216  & 4.133   & 4.196    & $-0.050$ & 4.137  \\
  809.2634 & & 3.820 & $-0.045$ &  2 &  4.200  &  1  &  4.169  & 4.142  & 4.182  & 4.094   & 4.157    & $-0.050$ & 4.119  \\
   & & & & & & & & & & & & & \\

\multicolumn{14}{c} {\zni} \\
  472.2159 & & 4.030 & $-0.380$ &  3 &  6.800  &  1  &  4.687  & 4.580  & 4.603  & 4.511   & 4.602    & $-0.032$ & 4.655  \\
  481.0534 & & 4.080 & $-0.160$ &  3 &  7.500  &  1  &  4.649  & 4.539  & 4.560  & 4.466   & 4.559    & $-0.039$ & 4.610  \\
  636.2347 & & 5.800 & $+0.140$ &  4 &  2.150  &  1  &  4.535  & 4.519  & 4.524  & 4.469   & 4.545    & $-0.014$ & 4.521  \\
 1105.4280 & & 5.800 & $-0.330$ &  4 &  1.400  &  1  &  4.485  & 4.475  & 4.481  & 4.435   & 4.500    & $-0.018$ & 4.467  \\
 1305.3650 & & 6.650 & $+0.320$ &  4 &  1.900  &  2  &  4.548  & 4.549  & 4.558  & 4.507   & 4.575    & $-0.007$ & 4.541  \\
   & & & & & & & & & & & & & \\

\multicolumn{14}{c} {\gai} \\
  417.2053 & & 0.100 & $-0.337$ &  5 &  5.220  &  1  &  3.023  & 3.003  & 3.085  & 2.956   & 3.019    &   &   \\
   & & & & & & & & & & & & & \\

\multicolumn{14}{c} {\gei} \\
  326.9489 & & 0.890 & $-1.080$ &  6 &  4.350  &  1  &  3.634  & 3.547  & 3.624  & 3.511   & 3.566    &   &   \\
   & & & & & & & & & & & & & \\

\multicolumn{14}{c} {\rbi} \\
  780.0268 & & 0.000 & $+0.144$ &  7 &  0.670  &  1  &  2.530  & 2.574  & 2.627  & 2.537   & 2.595    &   &   \\
  794.7603 & & 0.000 & $-0.164$ &  7 &  0.270  &  1  &  2.417  & 2.462  & 2.514  & 2.425   & 2.484    &   &   \\
   & & & & & & & & & & & & & \\

\multicolumn{14}{c} {\sri} \\
  460.7340 & & 0.000 & $+0.283$ &  8 &  4.450  &  2  &  2.698  & 2.688  & 2.771  & 2.644   & 2.705    & $+0.146$ & 2.844  \\
  707.0100 & & 1.830 & $-0.030$ &  9 &  0.135  &  2  &  2.680  & 2.719  & 2.774  & 2.680   & 2.743    & $+0.084$ & 2.764  \\
   & & & & & & & & & & & & & \\

\multicolumn{14}{c} {\srii} \\
 1003.6670 & & 1.810 & $-1.202$ & 10 &  6.650  &  2  &  2.963  & 2.876  & 2.887  & 2.831   & 2.886    & $-0.091$ & 2.872  \\
 1032.7360 & & 1.840 & $-0.248$ & 10 & 14.860  &  3  &  3.078  & 2.982  & 2.995  & 2.918   & 2.986    & $-0.208$ & 2.870  \\
 1091.4880 & & 1.810 & $-0.478$ & 10 & 12.360  &  1  &  2.936  & 2.829  & 2.843  & 2.772   & 2.834    & $-0.195$ & 2.741  \\
   & & & & & & & & & & & & & \\

\multicolumn{14}{c} {\yii} \\
  412.4909 & & 0.410 & $-1.500$ & 11 &  1.880  &  1  &  2.179  & 2.168  & 2.203  & 2.153   & 2.202    &   &   \\
  439.8015 & & 0.130 & $-1.000$ & 11 &  4.600  &  1  &  2.116  & 2.065  & 2.101  & 2.032   & 2.092    &   &   \\
  490.0124 & & 1.030 & $-0.101$ & 12 &  5.550  &  2  &  2.291  & 2.200  & 2.228  & 2.154   & 2.223    &   &   \\
  508.7425 & & 1.080 & $-0.170$ & 11 &  4.700  &  1  &  2.156  & 2.091  & 2.116  & 2.050   & 2.116    &   &   \\
  511.9119 & & 0.990 & $-1.360$ & 11 &  1.370  &  1  &  2.326  & 2.318  & 2.345  & 2.298   & 2.349    &   &   \\
  520.0414 & & 0.990 & $-0.570$ & 11 &  3.500  &  1  &  2.149  & 2.113  & 2.139  & 2.081   & 2.142    &   &   \\
  528.9821 & & 1.030 & $-1.850$ & 11 &  0.370  &  1  &  2.205  & 2.204  & 2.229  & 2.185   & 2.237    &   &   \\
  547.3385 & & 1.740 & $-0.954$ & 13 &  0.670  &  1  &  2.248  & 2.244  & 2.265  & 2.223   & 2.276    &   &   \\
  572.8876 & & 1.840 & $-1.054$ & 13 &  0.330  &  1  &  2.092  & 2.089  & 2.110  & 2.070   & 2.121    &   &   \\
   & & & & & & & & & & & & & \\

\multicolumn{14}{c} {\zri} \\
  424.1706 & & 0.650 & $+0.140$ & 14 &  0.270  &  1  &  2.247  & 2.346  & 2.428  & 2.320   & 2.369    &   &   \\
  454.2234 & & 0.630 & $-0.310$ & 14 &  0.430  &  1  &  2.855  & 2.955  & 3.038  & 2.927   & 2.977    &   &   \\
  468.7805 & & 0.730 & $+0.550$ & 14 &  0.850  &  1  &  2.410  & 2.498  & 2.579  & 2.469   & 2.519    &   &   \\
  480.9477 & & 1.580 & $+0.160$ & 14 &  0.145  &  1  &  2.828  & 2.894  & 2.966  & 2.862   & 2.920    &   &   \\
  481.5637 & & 0.600 & $-0.030$ & 14 &  0.275  &  1  &  2.311  & 2.416  & 2.499  & 2.389   & 2.437    &   &   \\
  538.5128 & & 0.520 & $-0.710$ & 14 &  0.185  &  1  &  2.678  & 2.793  & 2.877  & 2.766   & 2.811    &   &   \\
  644.5720 & & 1.000 & $-0.830$ & 14 &  0.090  &  1  &  2.892  & 2.994  & 3.075  & 2.963   & 3.012    &   &   \\
  710.2890 & & 0.650 & $-0.840$ & 14 &  0.060  &  1  &  2.318  & 2.442  & 2.525  & 2.416   & 2.456    &   &   \\
   & & & & & & & & & & & & & \\

\multicolumn{14}{c} {\zrii} \\
  350.5666 & & 0.164 & $-0.375$ & 15 &  5.350  &  1  &  2.666  & 2.527  & 2.564  & 2.491   & 2.551    &   &   \\
  360.7369 & & 1.236 & $-0.670$ & 15 &  1.250  &  1  &  2.522  & 2.503  & 2.532  & 2.487   & 2.537    &   &   \\
  367.1264 & & 0.713 & $-0.590$ & 15 &  3.150  &  1  &  2.480  & 2.449  & 2.483  & 2.429   & 2.481    &   &   \\
  371.4777 & & 0.527 & $-0.950$ & 15 &  2.750  &  1  &  2.542  & 2.519  & 2.554  & 2.501   & 2.551    &   &   \\
  402.4435 & & 0.999 & $-1.130$ & 16 &  1.350  &  1  &  2.659  & 2.650  & 2.682  & 2.638   & 2.684    &   &   \\
  403.4083 & & 0.802 & $-1.530$ & 15 &  0.790  &  1  &  2.594  & 2.590  & 2.624  & 2.579   & 2.624    &   &   \\
  405.0320 & & 0.713 & $-1.025$ & 15 &  2.200  &  3  &  2.571  & 2.555  & 2.589  & 2.538   & 2.589    &   &   \\
  420.8980 & & 0.713 & $-0.485$ & 15 &  4.400  &  2  &  2.627  & 2.569  & 2.602  & 2.536   & 2.599    &   &   \\
  425.8041 & & 0.559 & $-1.170$ & 15 &  2.350  &  1  &  2.585  & 2.569  & 2.603  & 2.550   & 2.602    &   &   \\
  444.2992 & & 1.486 & $-0.375$ & 15 &  2.200  &  1  &  2.608  & 2.590  & 2.618  & 2.567   & 2.624    &   &   \\
   & & & & & & & & & & & & & \\

\multicolumn{14}{c} {\nbii} \\
  319.4974 & \footnote[1]{3D correction applied to HM results of Nilsson et al.\ (\cite{nils}).} 
                  & 0.326 & $+0.120$ & 17 &         &  1  &  1.473  &        &  1.50  &         &          &   &   \\
  321.5593 & \color{BrickRed}{$^a$}  & 0.439 & $-0.235$ & 17 &         &  1  &  1.505  &        &  1.53  &         &          &   &   \\
  371.7060 & \color{BrickRed}{$^a$}  & 1.694 & $+0.030$ & 18 &         &  1  &  1.455  &        &  1.48  &         &          &   &   \\
  374.0725 & \color{BrickRed}{$^a$}  & 1.619 & $-0.307$ & 17 &         &  1  &  1.442  &        &  1.47  &         &          &   &   \\
   & & & & & & & & & & & & & \\

\multicolumn{14}{c} {\moi} \\
  550.6496 & & 1.335 & $+0.060$ & 19 &  0.450  &  1  &  1.937  & 2.011  & 2.090  & 1.983   & 2.031    &   &   \\
  553.3034 & & 1.335 & $-0.069$ & 19 &  0.270  &  1  &  1.824  & 1.907  & 1.988  & 1.877   & 1.928    &   &   \\
   & & & & & & & & & & & & & \\

\multicolumn{14}{c} {\rui} \\
  343.6736 & & 0.148 & $+0.150$ & 20 &  1.000  &  1  &  1.572  & 1.653  & 1.749  & 1.632   & 1.670    &   &   \\
  349.8945 & & 0.000 & $+0.310$ & 20 &  2.050  &  1  &  1.689  & 1.742  & 1.842  & 1.719   & 1.756    &   &   \\
  374.2280 & & 0.336 & $-0.180$ & 20 &  0.740  &  1  &  1.849  & 1.935  & 2.028  & 1.915   & 1.955    &   &   \\
  408.0600 & & 0.812 & $-0.040$ & 20 &  0.280  &  1  &  1.670  & 1.755  & 1.841  & 1.731   & 1.778    &   &   \\
  455.4517 & & 0.812 & $+0.070$ & 20 &  0.550  &  1  &  1.819  & 1.898  & 1.985  & 1.874   & 1.918    &   &   \\
  458.4443 & & 1.002 & $-0.550$ & 20 &  0.105  &  1  &  1.881  & 1.960  & 2.042  & 1.933   & 1.982    &   &   \\
   & & & & & & & & & & & & & \\

\multicolumn{14}{c} {\rhi} \\
  343.4889 & & 0.000 & $+0.450$ & 21 &  1.130  &  1  &  0.930  & 1.002  & 1.096  & 0.982   & 1.018    &   &   \\
  369.2361 & & 0.000 & $+0.150$ & 21 &  0.655  &  1  &  0.853  & 0.948  & 1.037  & 0.928   & 0.966    &   &   \\
   & & & & & & & & & & & & & \\

\multicolumn{14}{c} {\pdi} \\
  324.2701 & & 0.810 & $+0.070$ & 22 &  2.400  &  1  &  1.533  & 1.528  & 1.616  & 1.504   & 1.547    &   &   \\
  340.4581 & & 0.810 & $+0.330$ & 22 &  3.330  &  1  &  1.565  & 1.510  & 1.602  & 1.483   & 1.526    &   &   \\
   & & & & & & & & & & & & & \\

\multicolumn{14}{c} {\agi} \\
  328.0681 & & 0.000 & $-0.021$ & 23 &  3.500  &  1  &  1.044  & 1.000  & 1.093  & 0.971   & 1.015    &   &   \\
  338.2900 & & 0.000 & $-0.333$ & 23 &  2.230  &  1  &  0.869  & 0.893  & 0.986  & 0.870   & 0.911    &   &   \\
   & & & & & & & & & & & & & \\

\multicolumn{14}{c} {\cdi} \\
  508.5823 & & 3.950 & $-0.144$ & 24 &  0.073  &  1  &  1.766  & 1.760  & 1.792  & 1.728   & 1.793    &   &   \\
   & & & & & & & & & & & & & \\

\multicolumn{14}{c} {\sni} \\
  380.1025 & & 1.770 & $-0.620$ & 25 &  0.120  &  1  &  2.016  & 2.061  & 2.130  & 2.036   & 2.088    &   &   \\
   & & & & & & & & & & & & & \\

\multicolumn{14}{c} {\baii} \\
  455.4036 & & 0.000 & $+0.172$ & 26 & 18.200  &  1  &  2.236  & 2.208  & 2.241  & 2.150   & 2.226    & $-0.016$ & 2.220  \\
  585.3688 & & 0.600 & $-1.026$ & 26 &  6.300  &  1  &  2.391  & 2.250  & 2.280  & 2.203   & 2.263    & $-0.080$ & 2.311  \\
  649.6908 & & 0.600 & $-0.407$ & 27 &  9.960  &  1  &  2.372  & 2.244  & 2.271  & 2.185   & 2.249    & $-0.147$ & 2.225  \\

\multicolumn{14}{c} {\wi} \\
  400.8750 & \footnote[2]{3D correction and $gf$ update applied to HM results of Holweger \& Werner (\cite{hol}).} 
                  & 0.360 & $-0.446$ & 28 &  0.350  &  1  &  0.788  &        & 0.976  &         &          &    &  \\
  484.3846 & \color{BrickRed}{$^b$} & 0.410 & $-1.471$ & 29 &  0.035  &  1  &  0.880  &        & 1.081  &         &          &    &  \\
   & & & & & & & & & & & & & \\

\multicolumn{14}{c} {\osi} \\
  330.1559 & & 0.000 & $-0.743$ & 30 &  0.960  &  1  &  1.405  & 1.413  & 1.496  & 1.396   & 1.437    &   &   \\
   & & & & & & & & & & & & & \\

\multicolumn{14}{c} {\iri} \\
  322.0775 & & 0.350 & $-0.536$ & 31 &  0.975  &  1  &  1.418  & 1.398  & 1.455  & 1.381   & 1.429    &   &   \\
   & & & & & & & & & & & & & \\

\multicolumn{14}{c} {\aui} \\
  312.2784 & & 1.140 & $-0.950$ & 32 &  0.290  &  1  &  0.908  & 0.891  & 0.925  & 0.878   & 0.927    &   &   \\
   & & & & & & & & & & & & & \\

\multicolumn{14}{c} {\pbi} \\
  368.3480 & & 0.970 & $-0.524$ & 33 &  0.855  &  1  &  1.796  & 1.854  & 1.934  & 1.831   & 1.876    & $+0.120$ & 1.916  \\
   & & & & & & & & & & & & & \\

\multicolumn{14}{c} {\thii} \\
  401.9130 & & 0.000 & $-0.228$ & 34 &  0.314  &  1  &  0.018  & 0.021  & 0.059  & 0.012   & 0.055    & $+0.010$ & 0.028  \\
   & & & & & & & & & & & & & \\

\end{mpsupertabular}
\end{center}
\vspace{5mm}
\textbf{References:}\nopagebreak\\
\begin{minipage}[t]{0.43\linewidth}
\vspace{0pt}
\begin{enumerate}
\item Kock \& Richter (\cite{kock}), renormalised to lifetimes of Carlsson et al.\ (\cite{carl})
\item Meggers et al.\ (\cite{Meggers61}), via Bielski (\cite{biel})
\item ref.\ 4, renormalised to lifetimes of Kerkhoff et al.\ (\cite{Kerkhoff80}) 
\item Bi\'{e}mont \& Godefroid (\cite{biem8})
\item Cunningham \& Link (\cite{Cunningham67})
\item Bi\'emont et al.\ (\cite{biem2})
\item mean of lifetimes from Simsarian et al.\ (\cite{sim}) and Volz \& Schmoranzer (\cite{vol}) weighted according to uncertainties, via Morton (\cite{Morton00})  
\item Migdalek \& Baylis (\cite{Migdalek})
\item Garcia \& Campos (\cite{garc}), normalised by them to other existing lifetimes
\item relative data of Gallagher (\cite{Gallagher}), normalised with the mean of (equal uncertainty) lifetimes from Kuske et al.\ (\cite{Kuske}) and Pinnington et al.\ (\cite{Pinnington95})
\item Hannaford et al.\ (\cite{hann1})
\item ref.\ 11, renormalised with unweighted mean of lifetimes from ref.\ 11 and W\"annstr\"om et al.\ (\cite{wann})
\item ref.\ 11, renormalised using unweighted mean of lifetimes from ref.\ 11, W\"annstr\"om et al.\ (\cite{wann}) and Bi\'{e}mont et al.\ (\cite{biem7})
\item Bi\'emont et al.\ (\cite{biem3})
\end{enumerate}
\end{minipage}%
\begin{minipage}[t]{0.05\linewidth}
\vspace{0pt}
\hspace{3mm}
\end{minipage}%
\begin{minipage}[t]{0.43\linewidth}
\vspace{0pt}
\begin{enumerate}
\setcounter{enumi}{14}
\item log-scale mean of refs.\ 14 and 16
\item Ljung et al.\ (\cite{ljung})
\item Nilsson \& Ivarsson (\cite{nils2})
\item Nilsson et al.\ (\cite{nils})
\item Whaling \& Brault (\cite{whal2})
\item Wickliffe et al.\ (\cite{wick})
\item Kwiatkowski et al.\ (\cite{kwia1})
\item Xu et al.\ (\cite{xu})
\item Total doublet strength from lifetime of Carlsson et al.\ (\cite{carl2}), with the relative contribution of each line given by theoretical data of Civi\v{s} et al.\ (\cite{civis})
\item Veer et al.\ (\cite{veer})
\item Lotrian et al.\ (\cite{lotri})
\item Kurz et al.\ (\cite{Kurz})
\item Davidson et al.\ (\cite{davi})
\item log-scale mean of ref.\ 29 and Kling \& Kock (\cite{klin})  
\item den Hartog et al.\ (\cite{denh})
\item Ivarsson et al.\ (\cite{ivar})
\item Gough et al.\ (\cite{gough}), renormalised using the uncertainty-weighted mean of lifetimes from Gough et al.\ and Xu et al.\ (\cite{xu2})
\item Hannaford et al.\ (\cite{hann4})
\item Bi\'emont et al.\ (\cite{biem6})
\item Nilsson et al.\ (\cite{nils3})
\end{enumerate}
\end{minipage}
\vspace{6mm}

\newpage
\begin{center}
\topcaption{\label{table:hfs} HFS and isotopic splitting data for the lines retained in this analysis.}
\tablefirsthead{%
  \hline
   & & \multicolumn{4}{c}{Lower level} && \multicolumn{4}{c}{Upper level} \\
  \cline{3-6}
  \cline{8-11}
  $\lambda$ & Iso. & $J$ & $A$   & $B$   & HFS  && $J$ & $A$   & $B$   & HFS \\
  (nm)      &      &     & (MHz) & (MHz) & ref. &&     & (MHz) & (MHz) & ref. \\
  \hline
   & & & & & & & & & \\
  }
\tablehead{%
  \hline 
   & & \multicolumn{4}{c}{Lower level} && \multicolumn{4}{c}{Upper level} \\
  \cline{3-6}
  \cline{8-11}
  $\lambda$ & Iso. & $J$ & $A$   & $B$   & HFS  && $J$ & $A$   & $B$   & HFS \\
  (nm)      &      &     & (MHz) & (MHz) & ref. &&     & (MHz) & (MHz) & ref. \\
  \hline
  \multicolumn{11}{r}{continued.}\\
  \hline
   & & & & & & & & & \\
  }
\tabletail{%
  \hline
  \multicolumn{11}{r}{continued on next page}\\
  \hline
}
\tablelasttail{\hline}
\begin{mpsupertabular}{r c@{\hspace{10mm}}c r r ccc r r c}
\multicolumn{11}{c}{\cui: 69.2\% $^{63}$Cu ($I=\frac32$), 30.8\% $^{65}$Cu ($I=\frac32$)}\\
\multicolumn{11}{c} {Isotopic separation for 510.6\,nm from ref.\ 1} \vspace{2mm}\\
510.5541 & $^{63}$Cu & $5/2$ &     749.781  &    188.270    &  1 && $3/2$  &  194.595 &    $-$28.480  &  2 \\
510.5560 & $^{65}$Cu & $5/2$ &     804.043  &    172.980    &  1 && $3/2$  &  208.464 &    $-$26.356  &  2 \\
521.8201 & $^{63}$Cu & $3/2$ &     194.595  &  $-$28.480    &  2 && $5/2$  &          &               &    \\
521.8201 & $^{65}$Cu & $3/2$ &     208.464  &  $-$26.356    &  2 && $5/2$  &          &               &    \\
522.0070 & $^{63}$Cu & $3/2$ &     194.595  &  $-$28.480    &  2 && $3/2$  &          &               &    \\
522.0070 & $^{65}$Cu & $3/2$ &     208.464  &  $-$26.356    &  2 && $3/2$  &          &               &    \\ 
   & & & & & \hspace{6mm} & & & & \\

\multicolumn{10}{c} {\rbi: 72.2\% $^{85}$Rb ($I=\frac52$), 27.8\% $^{87}$Rb ($I=\frac32$)} \vspace{2mm}\\
780.0268 & $^{85}$Rb & $1/2$ &    1011.911  &               &  3 && $3/2$  &   25.009 &      25.039   &  4 \\ 
780.0268 & $^{87}$Rb & $1/2$ &    3417.341  &               &  5 && $3/2$  &   84.718 &      12.497   &  6 \\ 
794.7603 & $^{85}$Rb & $1/2$ &    1011.911  &               &  3 && $1/2$  &  120.720 &               &  7 \\ 
794.7603 & $^{87}$Rb & $1/2$ &    3417.341  &               &  5 && $1/2$  &  406.200 &               &  7 \\ 
   & & & & & \hspace{6mm} & & & & \\
     	           	     	        	      	       	     		 
\multicolumn{10}{c} {\yii: 100\% $^{89}$Y ($I=\frac12$)} \vspace{2mm}\\
412.4909 & $^{89}$Y  &  2    &      15.500  &               &  8 &&  3     &$-$64.400 &               &  8 \\ 
439.8015 & $^{89}$Y  &  2    &  $-$222.900  &               &  8 &&  1     &$-$108.000&               &  8 \\ 
490.0124 & $^{89}$Y  &  3    &              &               &    &&  2     &$-$22.800 &               &  8 \\ 
508.7425 & $^{89}$Y  &  4    &              &               &    &&  4     &$-$29.000 &               &  8 \\ 
511.9119 & $^{89}$Y  &  2    &  $-$149.940  &               &  9 &&  3     &$-$64.400 &               &  8 \\ 
520.0414 & $^{89}$Y  &  2    &  $-$149.940  &               &  9 &&  2     &$-$79.300 &               &  8 \\ 
528.9821 & $^{89}$Y  &  3    &              &               &    &&  2     &$-$79.300 &               &  8 \\ 
547.3385 & $^{89}$Y  &  1    &      46.560  &               & 10 &&  2     &$-$67.000 &               &  8 \\ 
572.8876 & $^{89}$Y  &  2    &   $-$26.400  &               & 10 &&  2     &$-$67.000 &               &  8 \\ 
   & & & & & \hspace{6mm} & & & & \\

\multicolumn{10}{c} {\nbii: 100\% $^{93}$Nb ($I=\frac92$)} \vspace{2mm}\\
319.4974 & $^{93}$Nb &   2   &     513.245  &               & 11 &&  3     &  614.275 &               & 11 \\ 
321.5593 & $^{93}$Nb &   4   &     989.915  &               & 11 &&  4     &  358.552 &               & 11 \\ 
371.7060 & $^{93}$Nb &   4   &      76.147  &               & 11 &&  4     &  160.389 &               & 11 \\ 
374.0725 & $^{93}$Nb &   3   &     554.616  &               & 11 &&  3     &  220.347 &               & 11 \\ 
   & & & & & \hspace{6mm} & & & & \\

\multicolumn{10}{c} {\rhi: 100\% $^{103}$Rh ($I=\frac12$)} \vspace{2mm}\\
343.4889 &$^{103}$Rh & $9/2$ &  $-$175.574  &               & 12 &&$11/2$  &          &               &    \\ 
369.2361 &$^{103}$Rh & $9/2$ &  $-$175.574  &               & 12 && $7/2$  &          &               &    \\ 
   & & & & & \hspace{6mm} & & & & \\

\multicolumn{10}{c} {\agi: 51.8\% $^{107}$Ag ($I=\frac12$), 48.2\% $^{109}$Ag ($I=\frac12$)}\\
\multicolumn{10}{c} {Isotopic separations calculated from the data of Crawford et al.\ (\cite{craw}) and Jackson \& }\\
\multicolumn{10}{c} {Kuhn (\cite{jack}), but using component idetifications of Brix et al.\ (\cite{brix}) and ref.\ 13} \vspace{2mm}\\
328.06810&$^{107}$Ag & $1/2$ &  $-$1712.560 &               & 13 && $3/2$  & $-$31.700&               & 14 \\ 
328.06827&$^{109}$Ag & $1/2$ &  $-$1976.940 &               & 13 && $3/2$  & $-$36.700&               & 14 \\ 
338.29000&$^{107}$Ag & $1/2$ &  $-$1712.560 &               & 13 && $1/2$  &$-$175.400&               & 14 \\ 
338.29018&$^{109}$Ag & $1/2$ &  $-$1976.940 &               & 13 && $1/2$  &$-$201.600&               & 14 \\ 
   & & & & & \hspace{6mm} & & & & \\

\multicolumn{10}{c} {\baii: 0.1\% $^{130}$Ba ($I=0$), 0.1\% $^{132}$Ba ($I=0$), 2.4\% $^{134}$Ba ($I=0$),}\\
\multicolumn{10}{c} {6.6\% $^{135}$Ba ($I=\frac32$), 7.9\% $^{136}$Ba ($I=0$), 11.2\% $^{137}$Ba ($I=\frac32$), 71.7\% $^{138}$Ba ($I=0$)}\\
\multicolumn{10}{c} {Isotopic separations from Wendt et al.\ (\cite{Wendt84}; 455.4\,nm), van Hove et al.\ (\cite{vanhove}; 585.4\,nm), and ref.\ 17 (649.7\,nm)} \\
455.40334&$^{130}$Ba & $1/2$ &        0.000 &   0.000       &  N && $3/2$  &    0.000 &     0.000     &  N \\ 
455.40335&$^{135}$Ba & $1/2$ &     3591.670 &               & 15 && $3/2$  &  113.000 &    59.000     & 16 \\ 
455.40340&$^{132}$Ba & $1/2$ &        0.000 &   0.000       &  N && $3/2$  &    0.000 &     0.000     &  N \\ 
455.40341&$^{137}$Ba & $1/2$ &     4018.871 &               & 15 && $3/2$  &  127.200 &    92.500     & 16 \\ 
455.40344&$^{134}$Ba & $1/2$ &        0.000 &   0.000       &  N && $3/2$  &    0.000 &     0.000     &  N \\ 
455.40347&$^{136}$Ba & $1/2$ &        0.000 &   0.000       &  N && $3/2$  &    0.000 &     0.000     &  N \\ 
455.40360&$^{138}$Ba & $1/2$ &        0.000 &   0.000       &  N && $3/2$  &    0.000 &     0.000     &  N \\ 
585.36879&$^{137}$Ba & $3/2$ &      189.730 &  44.541       & 17 && $3/2$  &  127.200 &    92.500     & 16 \\ 
585.36880&$^{138}$Ba & $3/2$ &        0.000 &   0.000       &  N && $3/2$  &    0.000 &     0.000     &  N \\ 
585.36889&$^{136}$Ba & $3/2$ &        0.000 &   0.000       &  N && $3/2$  &    0.000 &     0.000     &  N \\ 
585.36891&$^{135}$Ba & $3/2$ &      169.590 &  28.953       & 17 && $3/2$  &  113.000 &    59.000     & 16 \\ 
585.36902&$^{134}$Ba & $3/2$ &        0.000 &   0.000       &  N && $3/2$  &    0.000 &     0.000     &  N \\ 
585.36914&$^{132}$Ba & $3/2$ &        0.000 &   0.000       &  N && $3/2$  &    0.000 &     0.000     &  N \\ 
585.36927&$^{130}$Ba & $3/2$ &        0.000 &   0.000       &  N && $3/2$  &    0.000 &     0.000     &  N \\ 
649.69078&$^{137}$Ba & $3/2$ &      189.730 &  44.541       & 17 && $1/2$  &  743.700 &               & 16 \\ 
649.69080&$^{138}$Ba & $3/2$ &        0.000 &   0.000       &  N && $1/2$  &    0.000 &     0.000     &  N \\ 
649.69090&$^{136}$Ba & $3/2$ &        0.000 &   0.000       &  N && $1/2$  &    0.000 &     0.000     &  N \\ 
649.69091&$^{135}$Ba & $3/2$ &      169.590 &  28.953       & 17 && $1/2$  &  664.600 &               & 16 \\ 
649.69105&$^{134}$Ba & $3/2$ &        0.000 &   0.000       &  N && $1/2$  &    0.000 &     0.000     &  N \\ 
   & & & & & \hspace{6mm} & & & & \\

\end{mpsupertabular}
\end{center}
\vspace{5mm}
\textbf{References:}\nopagebreak\\
\begin{minipage}[t]{0.43\linewidth}
\vspace{0pt}
\begin{enumerate}
\item Fischer et al.\ (\cite{fisch})
\item Hannaford \& McDonald (\cite{hann})
\item Nez et al.\ (\cite{nez})
\item Rapol et al.\ (\cite{rapol})
\item Bize et al.\ (\cite{bize})
\item Ye et al.\ (\cite{ye})
\item Beacham \& Andrew (\cite{beach})	
\item W\"{a}nnstr\"{o}m et al.\ (\cite{wann}) 
\item Beck (\cite{Beck})
\end{enumerate}
\end{minipage}%
\begin{minipage}[t]{0.05\linewidth}
\hspace{3mm}
\end{minipage}%
\begin{minipage}[t]{0.43\linewidth}
\begin{enumerate}
\setcounter{enumi}{9}
\item Dinneen et al.\ (\cite{Dinneen})
\item Nilsson \& Ivarsson (\cite{nils2})
\item Chan et al.\ (\cite{Chan})
\item Wessel \& Lew (\cite{wess})
\item Carlsson et al.\ (\cite{carl2})
\item Trapp et al.\ (\cite{trapp})
\item Villemoes et al.\ (\cite{villemoes}) 
\item van Hove et al.\ (\cite{vanhove2})
\end{enumerate}
\end{minipage}

\twocolumn
\begin{table*}
\caption{Adopted ionisation energies $\chi_{\rm ion}$ and partition functions $U(T)$ 
for the relevant ionisation stages of the heavy elements.}
\label{table:partition} 
\begin{minipage}{\columnwidth}
\centering
\begin{tabular}{l r r r r r}
\hline
\hline Species & $E_{\rm ion}$ (eV) &  \multicolumn{4}{c}{$U(T)$} \\
               &                    & 3000\,K & 5000\,K & 8000\,K & 12000\,K \\
\hline
  \cui  &   7.726 &      2.04 &      2.34 &      3.23 &      5.63 \\
  \cuii &  20.290 &      1.01 &      1.03 &      1.33 &      2.23 \\
  \zni  &   9.394 &      1.00 &      1.01 &      1.01 &      1.35 \\
  \znii &  17.960 &      2.00 &      2.00 &      2.00 &      2.02 \\
  \gai  &   5.999 &      4.70 &      5.12 &      5.65 &      8.15 \\
  \gaii &  20.520 &      1.00 &      1.00 &      1.00 &      1.03 \\
  \gei  &   7.899 &      6.03 &      7.61 &      9.01 &     12.29 \\
  \geii &  15.940 &      3.71 &      4.41 &      4.90 &      5.27 \\
  \rbi  &   4.177 &      1.99 &      2.25 &      4.26 &     11.85 \\
  \rbii &  27.290 &      1.00 &      1.00 &      1.00 &      1.00 \\
  \sri  &   5.695 &      1.01 &      1.24 &      3.01 &     19.84 \\
  \srii &  11.030 &      2.00 &      2.15 &      2.78 &      4.26 \\
  \yi   &   6.217 &      8.84 &     11.76 &     20.43 &     37.35 \\
  \yii  &  12.220 &     10.94 &     15.79 &     22.54 &     31.80 \\
  \zri  &   6.634 &     19.93 &     32.50 &     55.44 &     89.33 \\
  \zrii &  13.130 &     29.14 &     45.59 &     67.79 &     96.11 \\
  \nbi  &   6.579 &     34.95 &     52.93 &     93.16 &    192.18 \\
  \nbii &  14.320 &     26.20 &     43.07 &     71.77 &    114.13 \\
  \moi  &   7.092 &      7.12 &      9.05 &     19.54 &     56.94 \\
  \moii &  16.160 &      6.03 &      7.72 &     15.75 &     38.08 \\
  \rui  &   7.361 &     22.45 &     34.02 &     56.98 &     97.48 \\
  \ruii &  16.760 &     15.99 &     23.97 &     36.74 &     58.12 \\
  \rhi  &   7.459 &     18.66 &     26.51 &     38.86 &     58.73 \\
  \rhii &  18.080 &     11.77 &     15.50 &     21.00 &     30.21 \\
  \pdi  &   8.337 &      1.53 &      2.94 &      5.74 &     10.30 \\
  \pdii &  19.430 &      6.68 &      7.51 &      8.50 &     10.52 \\
  \agi  &   7.576 &      1.96 &      1.96 &      2.05 &      3.42 \\
  \agii &  21.490 &      0.98 &      1.00 &      1.03 &      1.16 \\
  \ini  &   5.786 &      3.46 &      4.23 &      5.42 &     12.28 \\
  \inii &  18.870 &      1.00 &      1.00 &      1.01 &      1.05 \\
  \sni  &   7.344 &      3.36 &      5.14 &      7.07 &      9.33 \\
  \snii &  14.630 &      2.57 &      3.13 &      3.83 &      4.42 \\
  \bai  &   5.212 &      1.22 &      2.55 &      8.35 &     29.87 \\
  \baii &  10.000 &      2.78 &      4.17 &      5.97 &      7.95 \\
  \lai  &   5.577 &     14.29 &     27.04 &     58.19 &    126.50 \\
  \laii &  11.100 &     20.23 &     29.25 &     42.52 &     62.91 \\
  \cei  &   5.539 &     75.34 &    190.04 &    471.79 &    998.10 \\
  \ceii &  10.800 &     81.74 &    189.36 &    369.03 &    604.79 \\
  \pri  &   5.473 &     37.48 &    115.76 &    333.07 &    717.89 \\
\hline
\end{tabular}
\end{minipage}%
\begin{minipage}{\columnsep}
\hspace{0pt}
\end{minipage}
\begin{minipage}{\columnwidth}
\centering
\begin{tabular}{l r r r r r}
\hline
\hline Species & $E_{\rm ion}$ (eV) &  \multicolumn{4}{c}{$U(T)$} \\
               &                    & 3000\,K & 5000\,K & 8000\,K & 12000\,K \\
\hline
  \prii &  10.600 &     55.98 &    138.35 &    318.42 &    585.87 \\
  \ndi  &   5.525 &     30.35 &     92.55 &    322.69 &    828.45 \\
  \ndii &  10.700 &     49.58 &    120.48 &    315.29 &    705.33 \\
  \smi  &   5.644 &     18.22 &     38.24 &    108.47 &    299.28 \\
  \smii &  11.100 &     31.27 &     59.16 &    120.02 &    243.16 \\
  \eui  &   5.670 &      8.16 &     10.97 &     28.31 &    103.98 \\
  \euii &  11.240 &     12.49 &     15.87 &     23.96 &     42.00 \\
  \gdi  &   6.150 &     34.56 &     59.41 &    135.19 &    318.79 \\
  \gdii &  12.100 &     48.42 &     83.38 &    146.39 &    249.09 \\
  \tbi  &   5.864 &     90.06 &    165.84 &    348.95 &    703.54 \\
  \tbii &  11.500 &     53.44 &    102.16 &    181.93 &    293.33 \\
  \dyi  &   5.939 &     20.94 &     41.18 &    137.28 &    441.21 \\
  \dyii &  11.700 &     33.55 &     54.46 &    127.72 &    320.92 \\
  \hoi  &   6.022 &     18.57 &     34.40 &     96.02 &    261.06 \\
  \hoii &  11.800 &     31.09 &     49.39 &    131.82 &    424.08 \\
  \eri  &   6.108 &     16.14 &     32.36 &    102.58 &    318.00 \\
  \erii &  11.900 &     26.97 &     42.14 &     90.26 &    204.14 \\
  \tmi  &   6.184 &      7.95 &     10.60 &     28.30 &    106.31 \\
  \tmii &  12.100 &     15.42 &     17.79 &     31.30 &     76.72 \\
  \ybi  &   6.254 &      0.94 &      1.02 &      2.33 &      9.85 \\
  \ybii &  12.180 &      2.01 &      2.00 &      3.04 &      8.23 \\
  \lui  &   5.426 &      6.69 &      8.70 &     14.05 &     29.32 \\
  \luii &  13.900 &      1.02 &      1.40 &      2.79 &      5.91 \\
  \hfi  &   6.825 &      9.07 &     14.60 &     27.72 &     58.17 \\
  \hfii &  14.900 &      6.91 &     12.63 &     22.14 &     38.13 \\
  \wi   &   7.864 &      6.25 &     12.72 &     28.41 &     68.89 \\
  \wii  &  17.700 &      6.95 &     13.84 &     29.34 &     60.05 \\
  \osi  &   8.438 &     12.99 &     20.01 &     36.56 &     71.82 \\
  \osii &  18.500 &     12.12 &     18.01 &     27.13 &     40.49 \\
  \iri  &   8.967 &     14.50 &     21.10 &     33.66 &     57.06 \\
  \irii &  17.000 &     16.46 &     22.92 &     34.68 &     53.96 \\
  \aui  &   9.226 &      2.10 &      2.46 &      3.27 &      4.72 \\
  \auii &  20.500 &      0.81 &      1.12 &      1.80 &      3.17 \\
  \tli  &   6.108 &      2.00 &      2.29 &      2.95 &      5.40 \\
  \tlii &  20.430 &      1.00 &      1.00 &      1.01 &      1.02 \\
  \pbi  &   7.417 &      1.11 &      1.54 &      2.70 &      7.13 \\
  \pbii &  15.030 &      2.00 &      2.10 &      2.33 &      2.75 \\
  \thi  &   6.307 &     11.19 &     27.29 &     81.41 &    221.47 \\
  \thii &  11.900 &     14.73 &     35.99 &     88.25 &    187.33 \\
\hline
\end{tabular}
\end{minipage}
\end{table*}

\end{document}